\documentclass[twocolumn,superscriptaddress]{revtex4-1}%[onecolumn,
\usepackage{graphicx}
\usepackage{url}%\usepackage{amsmath}
\usepackage{amssymb}
\usepackage{bm}
\usepackage{xcolor}

\usepackage{dcolumn}
\usepackage[caption=false]{subfig}
\usepackage{MnSymbol}

\newcommand{\id}{\mathbf{1}}

\newcommand{\rd}{\mathrm{d}}

\newcommand{\Ho}{H}%{\hat{H}}
\newcommand{\Uo}{U}%{\hat{U}}

\newcommand{\be}{\begin{equation}}
\newcommand{\ee}{\end{equation}}
\newcommand{\bes}{\begin{eqnarray}}
\newcommand{\ees}{\end{eqnarray}}
\newcommand{\ket}[1]{{\left|  #1 \right\rangle}}
\newcommand{\kket}[1]{{\left|  #1 \right\rrangle}}
\newcommand{\bra}[1]{{\left\langle  #1 \right|}}
\newcommand{\bbra}[1]{{\left\llangle  #1 \right|}}
\newcommand{\braket}[2]{{\left\langle #1 \,| #2\right\rangle}}
\newcommand{\bbrakket}[2]{{\left\llangle #1 \,| #2\right\rrangle}}

\renewcommand{\i}{\mathrm{i}}
\newcommand{\hide}[1]{{}}

\begin{document}

\title{High-frequency expansions for time-periodic Lindblad generators}

\date{\today}

\author{Alexander Schnell}
\email{schnell@tu-berlin.de}
\affiliation{Technische Universit{\"a}t Berlin, Institut f{\"u}r Theoretische Physik, 10623 Berlin, Germany}
\affiliation{Max-Planck-Institut f\"ur Physik komplexer Systeme, 
N\"othnitzer Str.\ 38, 01187 Dresden, Germany}         
%\affiliation{Center for Theoretical Physics of Complex Systems,
%IBS, Daejeon 305-732, Korea}

\author{Sergey Denisov}
\email{sergiyde@oslomet.no}
\affiliation{Department of Computer Science, OsloMet–Oslo Metropolitan University, 0130 Oslo, Norway}
\affiliation{Max-Planck-Institut f\"ur Physik komplexer Systeme, 
N\"othnitzer Str.\ 38, 01187 Dresden, Germany}         
\affiliation{Department of Applied Mathematics, Lobachevsky University, 603950 Nizhny Novgorod, Russia}

\author{Andr\'e Eckardt}
\email{eckardt@tu-berlin.de}
\affiliation{Technische Universit{\"a}t Berlin, Institut f{\"u}r Theoretische Physik, 10623 Berlin, Germany}
\affiliation{Max-Planck-Institut f\"ur Physik komplexer Systeme, 
N\"othnitzer Str.\ 38, 01187 Dresden, Germany}         
%\affiliation{Center for Theoretical Physics of Complex Systems,
%IBS, Daejeon 305-732, Korea}

\begin{abstract}
Floquet engineering of isolated systems is often based on the concept of the effective time-independent Floquet Hamiltonian, which describes the stroboscopic evolution of a periodically driven quantum system in steps of the driving period and which is routinely obtained analytically using high-frequency expansions. 
The generalization of these concepts to open quantum systems described by a Markovian master equation of Lindblad type turns out to be non-trivial: On the one hand, already for a two-level system two different phases can be distinguished, where the effective time-independent Floquet generator (describing the stroboscopic evolution) is either again Markovian and of Lindblad type or not. On the other hand, even though in the high-frequency regime a Lindbladian Floquet generator (Floquet Lindbladian) is numerically found to exist, this behaviour is, curiously, not correctly reproduced within analytical high-frequency expansions. Here, we demonstrate that a proper Floquet Lindbladian can still be obtained from a high-frequency expansion, when treating the problem in a suitably chosen rotating frame. Within this approach, we can then also describe the transition to a phase at lower driving frequencies, where no Floquet Lindbladian exists, and show that the emerging non-Markovianity of the Floquet generator can entirely be attributed to the micromotion of the open driven system. 
%and gain thereby some understanding of the nontrivial role of the micromotion.
%These results shed light on opportunities and challenges for
%dissipative Floquet engineering. 

\end{abstract}
%\pacs{}
%\keywords{}

\maketitle

\section{Introduction}
Floquet engineering, that is the idea of  manipulating the properties of a coherent quantum system by means of strong time-periodic driving, has been successfully applied to  artificial many-body systems of ultracold atoms in optical lattices \cite{AidelsburgerEtAl11, StruckEtAl13, JotzuEtAl14, AidelsburgerEtAl15, Flaeschner16, Eckardt17,TarnowskiEtAl19,ViebahnEtAl21}.
These systems are well isolated from their environment and therefore well described by the Schrödinger equation. 
However, with the recent progress in the  engineering of quantum materials as well as complex photonic many-body systems \cite{Koch2012,Nori2014}, also the control of these systems via periodic forcing becomes an interesting and promising perspective. They are typically interacting with an environment, which introduces dissipation to the system's dynamics; see, e.g., Ref.~\cite{Koch2012}.
It is, therefore, desirable to extend the concept of Floquet engineering to \emph{open} quantum systems. 
In this context, two questions are of interest. The first one concerns the properties of the non-equilibrium steady state that such open periodically modulated systems approach in the long-time limit  \cite{BreuerEtAl00, KetzmerickWustmann10, VorbergEtAl13,
ShiraiEtAl14, SeetharamEtAl15,  DehghaniEtAl15, IadecolaEtAl15,VorbergEtAl15, ShiraiEtAl16, 
LetscherEtAl17, SchnellEtAl18, ChongEtAl2018, QinEtAl18}.
The second question, which will be discussed in this paper, concerns the \emph{transient} dynamics of these systems. Here, analogously to the Floquet engineering of isolated quantum systems, one can ask, whether it is possible to find %(and control) 
an effective time-independent description of the stroboscopic dynamics of the system %in steps of the driving period 
\cite{SchnellEtAl18FL,HaddadfarshiEtAl15,ReimerEtAl18,RestrepoEtAl17,DaiEtAl16,DaiEtAl2017,HotzSchaller21,Szczygielski21,GundersonEtAl21,MizutaEtAl21}.

While for a closed system the stroboscopic dynamics %(observed at integer multiples of the driving period) 
can always be recast into an effective coherent evolution governed by a 
time-independent  Floquet Hamiltonian \cite{EckardtAnisimovas15,Eckardt17}, 
%we investigate 
it is not obvious whether such a mapping %to an effective time-homogeneous evolution 
exists for an open periodically modulated system. 
More specifically, when considering the Markovian evolution described by Lindblad-type master equations, the question is whether the stroboscopic dynamics 
can be described by an effective time-\emph{independent} Floquet generator of the Lindblad type (henceforth addressed as \emph{Floquet Lindbladian}). 
The existence of such Floquet Lindbladians has implicitly  been assumed in recent works \cite{HaddadfarshiEtAl15,RestrepoEtAl17,DaiEtAl16,DaiEtAl2017}.
However, in Ref.~\cite{SchnellEtAl18FL} it was  shown that, already for a simple two-level system, there is no guarantee that such an operator exists. Namely, extensive parameter regions were found, where it does not exist, while in other extensive parameter regions, including the high-frequency limit, it does exist. 

The high-frequency regime plays an important role for Floquet engineering of isolated systems. On the one hand, it is appealing because in this regime unwanted heating via resonant excitations is suppressed \cite{Eckardt17,EckardtAnisimovas15,WeinbergEtAl15,BilitewskiCooper15b,WinterspergerEtAl20}. %\cmt{Add further references!}. 
On the other hand, %a second reason is that here 
it is possible to calculate the Floquet Hamiltonian by using  systematic high-frequency expansions, such as the Magnus expansion \cite{BlanesEtAl09}, and thus to analytically predict the properties of the Floquet system. It is, therefore, very natural to generalise such high-frequency expansions to open systems, as it has been done in various recent papers \cite{HaddadfarshiEtAl15,RestrepoEtAl17,DaiEtAl16,DaiEtAl2017,MizutaEtAl21,IkedaEtAl21}. However, it was observed that the corresponding expansions usually do not provide time-independent  Floquet generators of Lindblad type \cite{HaddadfarshiEtAl15,ReimerEtAl18,MizutaEtAl21}. Below we will demonstrate this failure of the Magnus expansion for the model used in Ref.~\cite{SchnellEtAl18FL}, despite the fact that the Floquet Lindbladian 
%(extracted from exact integration of the problem) 
was explicitly shown to exist. %Thus, one may ask the question, whether 

In this paper, we address the question, whether it is possible to construct a high-frequency expansion that is consistent with respect to the expected Lindblad-type stroboscopic evolution of the model. 
For this purpose, we compare four different approaches. Firstly, they are distinguished by the expansion technique they are based on, (i) a %Floquet-
Magnus expansion \cite{BlanesEtAl09}
%of the Floquet generator %(which comprises information about both the spectrum of the stroboscopic generator as well as of the time-periodic micromotion) 
or (ii) a van-Vleck-type high-frequency expansion~\cite{EckardtAnisimovas15}. %of the effective generator and the micromotion operator (each containing information only about the spectrum and the micromotion, respectively). 
Secondly, they differ by the reference frame, in which the model system is treated, i.e., either (a) in the direct frame or (b) in a suitably chosen rotating frame. We find that it is the appropriately chosen rotating reference frame [approach (b)], which allows to compute Lindblad-type Floquet generators in the high-frequency limit for our model. As a second major result, we  find that the break-down of the existence of a Floquet  Lindbladian, which was found by using the procedure described in Ref.~\cite{SchnellEtAl18FL}, can be related to the micromotion of the system~\cite{EckardtAnisimovas15}. This becomes apparent when performing the van-Vleck type high-frequency expansion [approach (ii)] in the rotating frame. %\cmt{Mention that the latter can be seen using van Vleck.}

The remaining part of this paper is organized as follows. In Sec.~\ref{sec:Floquet-Lind-intro} %and \ref{sec:driven-qbit-model} 
we summarize the results of Ref.~\cite{SchnellEtAl18FL} by
%Sec.~\ref{sec:Floquet-Lind-intro} 
outlining the general concept of the Floquet Lindbladian and
%In Sec.~\ref{sec:driven-qbit-model} we 
applying this concept to a driven two-level system. In Sec.~\ref{sec:ext-space}
we introduce the Magnus expansion, as well as the extended Floquet Hilbert space for the open system and generalize  the related 
van-Vleck high-frequency expansion  to open quantum systems. In Sec.~\ref{sec:FL-Highfreq} we study the problems that 
arise, when the high-frequency expansions are performed in the direct frame of reference.  
In Sec.~\ref{sec:FL-Highfreq-rotfr} we show that both the Magnus and the van-Vleck high-frequency expansions provide a valid Lindbladian  in the high-frequency limit, 
 when applied in the rotating frame that we introduce in Sec.~\ref{sec:Rotfr}. Moreover, we discuss the non-trivial role played by the micromotion. 

\section{The Floquet Lindbladian}
\label{sec:Floquet-Lind-intro}
In order to make the considerations self-consistent, we start by briefly sumarizing the main findings of Ref.~\cite{SchnellEtAl18FL}, where  the existence of the Floquet Lindbladian  is discussed. 

\subsection{Definition of the Floquet Lindbladian and the problem of its existence}
We consider the time-dependent Markovian master equation \cite{BreuerEtAl09,RivasEtAl2014,BreuerEtAl16RMP,DeVegaAlonsoRMP17}
\be\label{eq:tdm}
\partial_t {\rho} = \mathcal{L}(t)\rho = -{i}[H(t),\rho] + \mathcal{D}(t)\rho,
\ee
for the system's density operator $\rho$, described by a time-periodic Lindbladian generator
$\mathcal{L}(t)=\mathcal{L}(t+T)$.  In this work we set $\hbar=1$, therefore all energies are given in units of frequency.  The Lindbladian is characterized by a Hermitian time-periodic
Hamiltonian $H(t)$ and a dissipator 
\be
\mathcal{D}(t)\rho = 
\sum_i \gamma_i(t) \big[L_i(t)\rho L_i^\dag(t)-\frac{1}{2}\{L_i^\dag(t)L_i(t),\rho\}\big],
\ee
with jump operators $L_i(t)$ and non-negative rates $\gamma_i(t)$,
which both, in general, are time periodic with the same period $T$.
Note that
the time-dependent variation of $\mathcal{L}(t)$ may be due to a time-periodic modulation of the coherent evolution, governed by the Hamiltonian $H(t)$, and/or due to a time-periodic
modulation of the dissipative channels, represented by the rates $\gamma_i (t) \geqslant 0$ and the
jump operators $L_i(t)$.
This time-local form  guarantees that the corresponding   
evolution -- for any time $t$ --- can described with a completely positive (CP) and trace preserving (TP) map~\cite{BreuerEtAl09}.  %\textcolor{blue}{
Following the terminology of Ref.~\cite{Wolf2008}, such an evolution is  called \emph{time-dependent Markovian} \cite{Wolf2008}. Correspondingly, the evolution generated by a time-\emph{in}dependent Lindbladian is termed  \emph{Markovian}.  We follow this
nomenclature (note that there are also alternative terminologies, e.g., time-dependent and time-independent Markovian evolutions can be  combined together and simply called "Markovian"~\cite{RivasEtAl2014}).%}

%\textcolor{blue}{
The time-dependent Markovian evolution generated by time-dependent Lindbladians is the subject of our analysis. %Note that the  non-negativity of the rates is only a sufficient condition and 
Note that the non-negativity of the rates is only a sufficient condition to produce an evolution in the form of a CPTP map  for any time $t$.
There are cases when the rates can acquire negative values but the resulting map  nevertheless remains completely positive and trace preserving ~\cite{AddisEtAl14,SiudzinskaChruscinski2020}. %\cite{RivasEtAl2014,BreuerEtAl16RMP,DeVegaAlonsoRMP17}. 
We also consider such Lindbladians as relevant evolution generators; important is that the corresponding stroboscopic maps [see Eq.~\eqref{eq:oceo}] belong to the CPTP class.%}

Let us briefly outline the Lindblad master equation for the time-homogeneous case \cite{BreuerPetruccione}. 
A %completely positve and trace preserving (CPT) evolution or 
quantum dynamical semigroup is an evolution $\mathcal{P}(t, t_0)$ %\in L(L(\mathcal{H}))$ 
of the density matrix $\varrho$ %\in L(\mathcal{H})$ 
in a Hilbert space $\mathcal{H}$,
\begin{align} 
\varrho(t) = \mathcal{P}(t,t_0) \varrho(t_0), 
%\quad \text{and henceforth we denote} \quad  \mathcal{P}(t) = \mathcal{P}(t,0),
\end{align}
where henceforth we use the shorthand $\mathcal{P}(t) = \mathcal{P}(t,0)$. The semi-group
%that is motivated by the constraints that a `physical' evolution of the density matrix 
should obey several constraints: It is continuous, $\lim_{t \to 0^+}\mathcal{P}(t)\varrho = \varrho$, 
trace preserving, $\mathrm{Tr}(\mathcal{P}(t)\varrho)=\mathrm{Tr}(\varrho)$, has the semigroup property,  $\mathcal{P}(t+s)=\mathcal{P}(t)\mathcal{P}(s)$, i.e.~the evolution has no memory of its history (it is Markovian),
and is completely positive, $\mathcal{P}(t)\otimes \id \geq 0$, where $\id$ is the identity on the  space~$L(\mathcal{H})$ of linear operators acting in Hilbert space $\mathcal{H}$.  

As it was shown by Gorini, Kossakowski and Sudarshan~\cite{GoriniEtAl76} and Lindblad~\cite{Lindblad1976}, the superoperator $\mathcal{L}$ that generates this semi-group, i.e.
\begin{align} 
\partial_t{\rho}(t) = \mathcal{L}\rho(t),  \text{ or equally } \mathcal{P}(t)=\exp(\mathcal{L}t),
\end{align}
has to be of the form  %Eq.~\eqref{eq:Lindblad-general-basis},
%\begin{align} 
%\mathcal{L} = -\frac{i}{\hbar}[H,\cdot ] +\sum_{i,j=1}^{N^2-1} d_{ij} \big[A_i\cdot A_j^\dagger-\frac{1}{2}\{A_j^\dag A_i,\cdot \}\big],
%\label{eq:Lindblad-timeindep}
%\end{align}
\begin{align}
\mathcal{L} = -i \left[ H, \cdot \right] +  \sum_{i,j=1}^{N^2-1} d_{ij} \left(A_i \cdot A_j^\dagger - \frac{1}{2} \left\lbrace A_j^\dagger A_i, \cdot \right\rbrace  \right),
\label{eq:Lindblad-general-basis}
\end{align}
(henceforth refereed to as the \emph{Lindblad form}), where $H$ is a Hermitian operator (Hamiltonian), $\lbrace A_i\rbrace$ is a Schmidt-Hilbert basis in $L(\mathcal{H})$ ($\mathrm{dim}(\mathcal{H})=N$) and $d \geqslant 0$ is a Hermitian and positive semidefinite  \emph{Kossakowski matrix}. The corresponding jump operators $L_i$ and  rates $\gamma_i$  can be found by diagonalizing the Kossakowski matrix.

Let us turn now to the superoperator $\mathcal{P}(t)$ describing the map that is generated by a time-dependent Lindbladian  $\mathcal{L}(t)$, as in Eq.~\eqref{eq:tdm}, %via
% $\partial_t{\rho}(t) = \mathcal{L}(t) \rho(t)$. 
which formally yields 
\begin{align} 
\mathcal{P}(t)=\mathcal{T}\exp\left(\int_0^t\mathrm{d}t\mathcal{L}(t)\right).
\end{align}
where $\mathcal{T}$ is the time-ordering operator.
By definition, the map $\mathcal{P}(t)$ is 
completely positive and trace preserving, i.e.~it is a quantum channel \cite{Holevo2012}. 
%But, in general, the semigroup property will not
%hold, therefore the evolution need not be (time-independent) Markovian.

%However, 
Since the evolution is time periodic, it is interesting  to consider the stroboscopic dynamics, given by the one-cycle evolution map \cite{szczygielski2014,HartmannEtAl17}
\be\label{eq:oceo}
\mathcal{P}(T)=\mathcal{T}\exp\bigg[\int_0^T\!\rd t\, \mathcal{L}(t) \bigg].
\ee
The repeated application of it describes the stroboscopic evolution of the system, i.e.~for all $\rho(0)$ one has
\begin{align}
\rho(nT) = \mathcal{P}(T)^n \rho(0).
\end{align}

In analogy to the case of a closed system  \footnote{Note that, despite the fact that the Floquet generator of an isolated system is Hermitian and can, thus, be considered a Floquet \emph{Hamiltonian}, its properties can be rather different from those of a Hamiltonian of an undriven system. Namely, for generic (interacting non-integrable) systems, the eigenstates of the Floquet Hamiltonian are expected to be superpositions of states at all energies corresponding to infinite-temperature ensembles in the sense of eigenstate thermalisation \cite{LazaridesEtAl14b,DAlessioRigol14}.}, %a Floquet Lindbladian is a time-independent Lindblad superoperator $\mathcal{L}_F$ for which
we can now formally define a Floquet generator, i.e.\ a time-independent superoperator $\mathcal{K}$, such that
%\begin{align}
%\mathcal{P}(T) = \exp\left({\mathcal{L}_F T}\right).
%\end{align}
\begin{align}
\mathcal{P}(T) =  \exp\left({\mathcal{K} T}\right) \quad\text{  or   }\quad  \mathcal{K} =\frac{ \log(\mathcal{P})}{T} 
\label{eq:generator-cand}
\end{align}
%The aim is to find a time-independent Markovian evolution, generated by the Floquet Lindbladian $\mathcal{L}_F$,
%that coincides with $\mathcal{P}(t)$ at stroboscopic instances of time.
for the open driven system described by  Eq.~\eqref{eq:tdm}.
As it was discussed in Ref.~\cite{SchnellEtAl18FL}, it is not guaranteed that this Floquet generator $\mathcal{K}$ is of Lindblad form. However, if it is of Lindblad form, we will call it \emph{Floquet Lindbladian} and write
\begin{equation}
\mathcal{L}_F = \mathcal{K}.
\end{equation}
%that such a Lindblad operator exists.
%Although, by using the complex matrix logarithm one is always able to find a general superoperator $\mathcal{K}$ such that
%this superoperator $\mathcal{K}$ is not necessarily of Lindblad form, i.e.~the corresponding evolution $\exp\left({\mathcal{K} t}\right)$ is not necessarily a quantum dynamical semigroup anymore.

%\textcolor{blue}{
At first glance, it may appear counter-intuitive that the effective generator $\mathcal{K}$ is not of the Lindblad form. The map $\mathcal{P}(T)$ is time-dependent Markovian~\cite{WolfEtAl08} and therefore is CP-divisible~\cite{Wolf2008,RivasEtAl2014,BreuerEtAl16RMP,DeVegaAlonsoRMP17,ChakrabortyEtAl19}. I.e., for any $t$ and $t'$, $0 < t',t < T$, the map can be split as  $\mathcal{P}(t)=\mathcal{P}(t, t')\mathcal{P}(t')$, with $\mathcal{P}(t, t')$ being a CPTP map. Here, as a result of time-inhomogeneity, $\mathcal{P}(t,t’)$ is not simply a function of the time difference $t-t’$. 
%However, maps obtained with this division for the same duration $\Delta t = t - t'$ but for different values of $t', t$, in general case will be different thus being explicitly dependent on  starting time $t'$. In short, the evolution generated by a time-dependent Lindbladian is not time-homogeneous.
%}
%\textcolor{blue}{
The set of dynamical maps that are time-dependent Markovian  is larger than the set of Markovian maps \cite{WolfEtAl08}. Hence, by implementing a time-dependent protocol, one may
end up with a CPTP map that can only be obtain with a time-independent generator of a non-Lindblad form.  %The non-Markovianity of this effective generator stems from the ``memory" of the precise time-dependence of the protocol used to realize the map. 
Therefore, the existence of a Floquet-Lindbladian is not guaranteed.%}

%\textcolor{blue}{
Whether the Floquet-generator is of the Lindblad form or not is relevant for Floquet engineering. Namely, if it is of the Lindblad form, the stroboscopic evolution can be interpreted as the result of a time-independent Lindblad-type master equation, which is just monitored stroboscopically. If, in turn, no Floquet Lindbladian exists, the stroboscopic evolution, despite being Markovian by construction, cannot be interpreted as a stroboscopically monitored continuous time-independent Markovian process.%}

%For Floquet engineering on the other hand, the goal %is to engineer a desired time-independent %Lindbladian operator, hence those driving protocols %that give rise to a non-Lindbladian effective %generator have to be excluded.

%Let us elaborate a bit on this point.
Note that, due to the multi-branch structure of the  complex logarithm, there is a whole family of Floquet superoperators $\mathcal{K}_{\mathbf{x}}$, labeled by a set of integers ${\mathbf{x}} = \{x_1,...,x_{n_c}\}$  that specifies a particular branch of the logarithm, where $n_c$ is the number of complex conjugated pairs in the spectrum of $\mathcal{P}(T)$. In order to find a Floquet Lindbladian or refute its existence, we have to check whether at least one of these candidates $\mathcal{K}_{\mathbf{x}}$  is of the Lindblad form. Details on this procedure can be found in Appendix \ref{sec:app-Wolf-method}. In short, the test is  checking two conditions, which require that $\mathcal{K}_{\mathbf{x}}$ has to (i) preserve Hermiticity and (ii) has to be conditionally completely positive \cite{WolfEtAl08}.
Also note that, given that one has extracted operator $\mathcal{K}_{\mathbf{x}}$ from the matrix logarithm, one can always recast it in a quasi-Lindblad form, formally given by Eq.~\eqref{eq:Lindblad-general-basis}, with some operator $H$ and Kossakowski matrix $d$. The implementation of the test is then equivalent to testing $H$ for Hermiticity and $d$ for positive semi-definiteness, $d \geqslant 0$. We will use this test when performing the high-frequency expansions in Sections~\ref{sec:FL-Highfreq} and \ref{sec:FL-Highfreq-rotfr}.

If there is no set of integers ${\mathbf{x}}$ such that condition (i) and (ii) are fulfilled, then no Floquet Lindbladian exists. In this situation, it is instructive to quantify the distance from Markovianity for the non-Lindbladian generator $\mathcal{K}_{\mathbf{x}}$, by picking the branch giving the minimal distance. For this purpose, we compute the measure for non-Markovianity proposed by Wolf \emph{et al.}~\cite{WolfEtAl08}. This measure is based on adding a noise term $\mu\mathcal{N}$ of strength $\mu$ to 
the generator and determining the minimal strength required to make at least one of the candidates Lindbladian, i.e.
%Here
%
%  In the region where there is no such generator, we use a distance measure that was also suggested in Ref.~\cite{WolfEtAl08}
%to measure the distance $\mu_\mathrm{min}$ to Markovianity, i.e.~the distance of the closest candidate $\mathcal{K}$ to a valid physical generator. This distance
%is obtained by determining the minimum amount of noise  $\mu_\text{min}$  that is needed to render at least one of the candidate channels Linbladian, i.e.
\begin{align}
\mu_{\mathrm{min}}=\min_{\mathbf{x} \in \mathbb{Z}^{n_c}} \min \left\lbrace \mu \geq 0 {\Big\vert} \begin{array}{l}   \mathcal{K}_{\mathbf{x}} + \mu \mathcal{N}  \text{ is a valid} \\
\text{Lindblad generator}\end{array} \right\rbrace.
\end{align}
Here, $\mathcal{N}$ is the generator of the depolarizing channel 
$\exp(T\mu \mathcal{N})\rho = e^{-\mu T} \rho + [1 - e^{-\mu T}] \frac{\mathbf{1}}{N}$.

%\textcolor{blue}
{
Various other measures for non-Markovianity have been proposed in the literature \cite{ChruscinskiEtAl11,ChruscinskiManiscalco14}. %\textbf{[Chruscinski, D., A. Kossakowski, and A. Rivas, 2011, Phys. Rev. A
%83, 052128, D. Chruscinski and S. Maniscalco
%Phys. Rev. Lett. 112, 120404]}. 
Besides the one introduced above, in Ref.~\cite{SchnellEtAl18FL} we also calculated a measure that qualifies the violation of the positivity of the Choi representation of the map \cite{RivasEtAl10} %\textbf{[A. Rivas, S. F. Huelga, and M. B. Plenio, Phys. Rev. Lett. 105, 050403 (2010)]} 
and found that for our specific model (up to a factor of 1/2) it coincides with the measure of Ref.~\cite{WolfEtAl08}.
However, while these measures might provide different values for the distance from Markovianity in the regions where no Floquet Lindbladian exists, all of them will classify the same regions in parameter space as Markovian (those where the Floquet generator can be brought into the Lindblad form). Thus, the phase diagram will be independent of the chosen Markovianity measure.}

 %It is an open question whether this is a feature of this particular model or the reason for this is of a more fundamental sort. %Perhaps, other measures should also be tried.
%However,  it is evident that a particular choice  cannot change the ``yes/no'' partition of the parameter space, which is the main aspect of our studies. Henceforth we will quantify non-Markoviniaty with the measure given by Eq.~(11).}

\subsection{Model
%: Driven-dissipative two-level system
}
\label{sec:driven-qbit-model}

\begin{figure}%Markovian
	\centering
	\includegraphics[width=0.9\columnwidth]{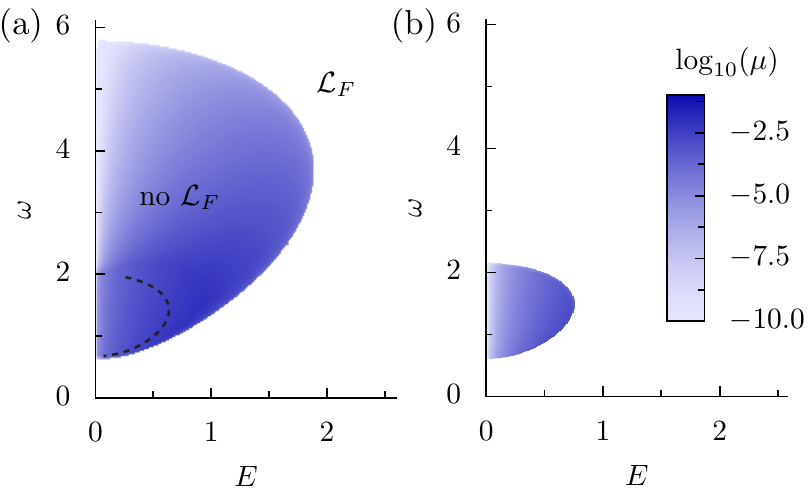}
	\caption[Distance to Markovianity from effective generator $\mathcal{K}$ of the stroboscopic dynamics for the two-level system model]{Distance to Markovianity $\mu_{\mathrm{min}}$ of the Floquet generator $\mathcal{K}$ obtained from the one-cycle evolution superoperator 
		as a function of driving strength $E$ and frequency $\omega$, for weak dissipation $\gamma =0.01$ and two
		driving phases (a) $\varphi = 0$  and (b) $\varphi = \pi/2$. In the white
		region, where $\mu_\text{min}=0$, it is of Lindblad type, so that a Floquet Lindbladian $\mathcal{L}_F$ exists. 
		On the dashed line the Floquet map $\mathcal{P}(T)$ pair of eigenvalues coincide when crossing the negative real semi-axis.}
	\label{fig:phases-exact}
\end{figure}

To illustrate the problem,  we consider a driven two-level system described by the master equation $\partial_\tau \varrho(\tau) = \mathcal{L}(\tau) \varrho(\tau)$ with time-periodic Lindbladian generator
$%\begin{align}
\mathcal{L}(\tau) = -{i} [H(\tau), \cdot] + \kappa \, \left(\sigma_- \cdot \sigma_+ - \frac{1}{2} \lbrace\sigma_+ \sigma_-, \cdot \rbrace \right),
%\label{eq:Lindblad-two-level system}
$ %\end{align} 
with 
$ %\begin{align}
H(\tau) = \frac{\Delta}{2} \sigma_z + \mathcal{E} \cos(\Omega \tau- \varphi)\, \sigma_x.
%\end{align}
$
Here $\sigma_x$, $\sigma_z$, and $\sigma_-$ are standard Pauli and lowering operators.
%Using the level splitting $\Delta$ and $1/\Delta$ as units for energy and time, %(so that henceforth $\Delta=\hbar=1$), 
After introducing $t= \tau\Delta$, i.e.~using $1/\Delta$ as unit for time, we find  $\partial_t \varrho(t) = \mathcal{L}(t) \varrho(t)$ with
\begin{align}
 \mathcal{L}(t)= -{i} [H(t), \cdot] + \gamma \, \left(\sigma_- \cdot \sigma_+ - \frac{1}{2} \lbrace\sigma_+ \sigma_-, \cdot \rbrace \right),
\label{eq:Lindblad-two-level system}
\end{align}
and
\begin{align}
H(t) = \frac{1}{2} \sigma_z + {E} \cos(\omega t- \varphi)\, \sigma_x.
\end{align}
This model is characterized by four dimensionless parameters: the relative dissipation strength $
\gamma=\kappa/\Delta$ as well as the relative strength $E = \mathcal{E}/\Delta$, frequency $\omega=\Omega/\Delta$ and phase $\varphi$ of the driving.

In Fig.~\ref{fig:phases-exact} we present the distance from Markovianity for the effective time-independent Floquet generator of our model, obtained using the procedure described in the previous section.  Note that the spectrum of a CPTP map is invariant under complex conjugation. Thus for the two-level system we have at most one pair of 
complex eigenvalues and, therefore, have to check a single integer $x$ labelling the branches of the operator logarithm. 
If we find a branch $x_0$ with a generator of the Lindblad form, then this would be our Floquet Lindbladian $\mathcal{L}_F=\mathcal{K}_{x_0}$. In Fig.~\ref{fig:phases-exact}(a), we mark the region where  such a branch was found and therefore the Floquet Lindbladian exists with white color.
%in the parameter space $(E, \omega)$.
%Let us now discuss our results. 
In the region where no such branch exists, we plot the distance from Markovianity
$\mu_\text{min}$ for the closest branch.
%of the effective generator of the one-cycle evolution superoperator. %versus driving 
%amplitude $E$ and frequency $\omega$. 
For  weak dissipation $\gamma=0.01$ and $\varphi=0$,
%The 
%extended 
%blue lobe, where $\mu_\text{min}>0$, corresponds to
%an extended 
%a phase, where the Floquet Lindbladian 
%$\mathcal{L}_F$ does not exist. 
an extended non-Lindbladian phase is  surrounded by a Lindbladian 
phase (white region) where $\mu_\text{min}=0$ so that $\mathcal{L}_F$ can be constructed.
%It contains also the $\omega$ axis, corresponding to the trivial undriven limit $E=0$. 
%Note that only 
%for a fine-tuned set of parameters, lying on the dashed line in Fig.~\ref{fig:phases-exact}(a), 
%$\mathcal{P}(T)$ possesses negative eigenvalues. However, they come in a degenerate pair, such that
%the 
%construction of a Floquet Lindbladian is not hindered by condition (i). 

For sufficiently large and small driving frequencies $\omega$ as well as for zero driving ($E=0$) and in the regime of strong driving amplitudes $E$, a Floquet Lindbladian is found to exist. Only for intermediate driving frequencies $\omega$ and sufficiently small (but finite) driving strengths $E$, a lobe-shaped region exists, where the Floquet generator is not markovian, i.e. not of Lindblad-type. 

\begin{figure}
	\centering
	\subfloat[$\varphi=0.1\pi$]{
		%\begin{minipage}{0.3\textwidth}
		\includegraphics[scale=0.6]{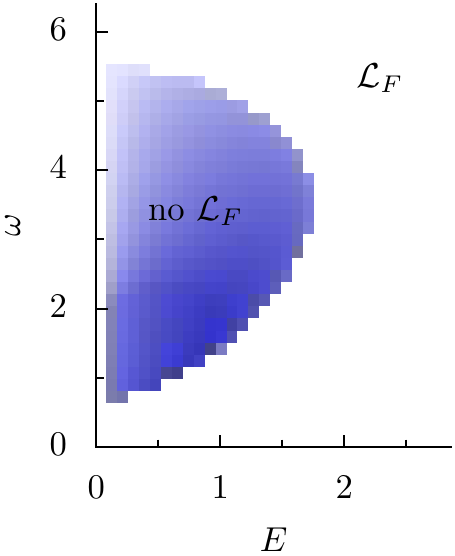}
		%\end{minipage}
	}
	\subfloat[$\varphi=0.2\pi$]{
		%\begin{minipage}{0.3\textwidth}
		\includegraphics[scale=0.6]{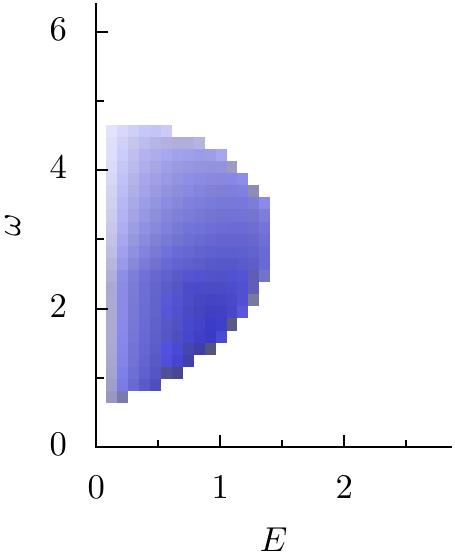}
		%\end{minipage}
	}\\
	\subfloat[$\varphi=0.3\pi$]{
		%\begin{minipage}{0.3\textwidth}
		\includegraphics[scale=0.6]{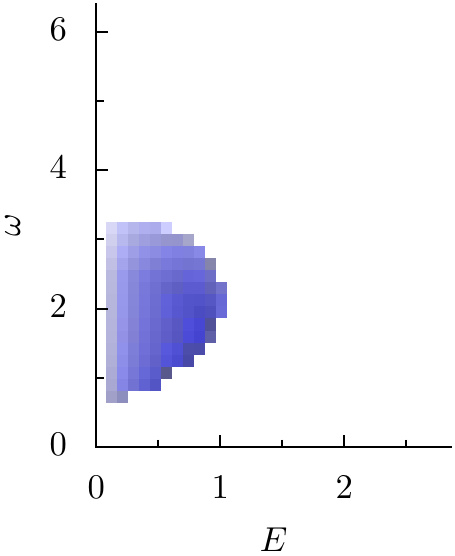}
		%\end{minipage}
	}
	\subfloat[$\varphi=0.4\pi$]{
		%\begin{minipage}{0.3\textwidth}
		\includegraphics[scale=0.6]{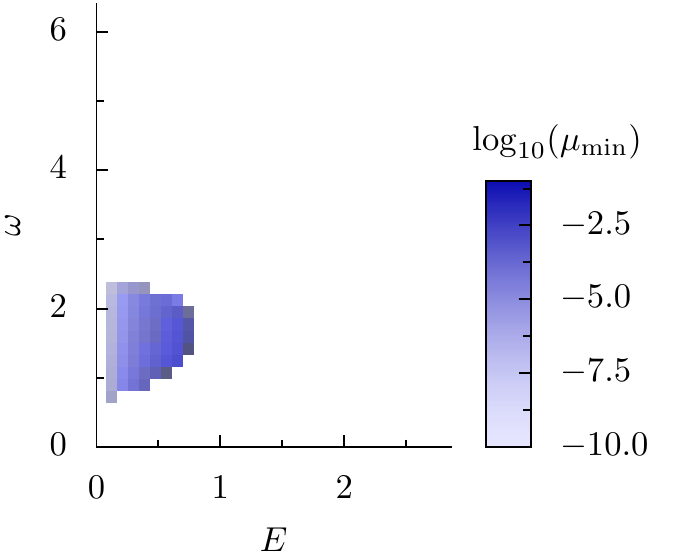}
		%\end{minipage}
	}
	\caption{Distance to Markovianity $\mu_{\mathrm{min}}$ %of the exact effective generator $\mathcal{K}$ as in Fig.~\ref{fig:phases-exact} 
	for different values of the driving phase $\varphi$. Other parameters are as in Fig.~\ref{fig:phases-exact}.}
	\label{fig:Phases-func-phi}
\end{figure}

Figure~\ref{fig:phases-exact}(b) shows the phase diagram for another driving phase, $\varphi=\pi/2$. Remarkably, 
compared to $\varphi=0$, Fig.~\ref{fig:phases-exact}(a), the non-Lindbladian phase covers now a much smaller region
of the parameter space. 
In Fig.~\ref{fig:Phases-func-phi} we plot the same phase diagram again, but for multiple intermediate values of the driving phase $\varphi$ and observe how the non-Lindbladian region continuously changes its shape with driving phase and appears to be smallest for $\varphi=\pi/2$.
The phase boundaries therefore depend on the driving phase or, in other words, on when during the driving period we monitor the stroboscopic evolution of the system. 

In the coherent case ($\gamma=0$ for our model), we can decompose the time evolution operator of a Floquet system from time $t_0$ to time $t$ like (see, e.g., Ref.\ \cite{EckardtAnisimovas15})
\begin{align}
\Uo(t,t_0)=U_F(t)\exp[-i(t-t_0)H_\text{eff}]U_F^\dag(t_0),
\label{eq:def-eff-Hamil}
\end{align}
where $\Uo_F(t)=\Uo(t+T)$ is a
unitary operator describing the time-periodic \textit{micromotion} of the Floquet states of the system and 
$H_\text{eff}$ is a time-independent effective Hamiltonian. The Floquet Hamiltonian $\Ho^F_{t_0}$, defined 
via $U(t_0+T,t_0)=\exp(-i T\Ho^F_{t_0})$ so that it describes the stroboscopic evolution of the system at 
times $t_0$, $t_0+T$, \ldots, is for general $t_0$ then given by (see, e.g., Ref.~\cite{EckardtAnisimovas15})
\begin{equation}
 H^F_{t_0} =\Uo_F(t_0)\Ho_\text{eff}\Uo_F^\dag(t_0).
 \end{equation} 
%where we used  the notation $\Ho_F=\Ho^F_0$ for $t_0=0$. 
Thus the operator $H^F_{t_0}$ depends on the micromotion via a $t_0$-dependent unitary rotation. 
However, in the dissipative system the micromotion will no longer be captured with a unitary operator.
This explains why the effective time-independent generator of the stroboscopic evolution can change its character in a nontrivial fashion, e.g.\ from Lindbladian to non-Lindbladian form, as a function of $t_0$ (or, equivalently, of the driving phase $\varphi$). 
%A quantitative discussion of this is found in Sec.~\ref{sec:FL-Highfreq-rotfr}, where we aim to extract the micromotion operator for the dissipative system.
In Sec.~\ref{sec:FL-Highfreq-rotfr}, we will present strong evidence for the fact that the breakdown of `Lindbladianity' of the Floquet generator is entirely due to the impact of the micromotion operator.

The fact that the Floquet generator for the stroboscopic evolution $\mathcal{K}$ is found to be of Lindblad form in the high-frequency regime (Fig.~\ref{fig:phases-exact}), suggests that it is possible to analytically approximate this Floquet Lindbladian by using systematic high-frequency expansions. However, in the literature it was found that one of the most conventional high-frequency expansions, the Magnus expansion~\cite{BlanesEtAl09}, generally does \emph{not} produce a valid Lindblad generator~\cite{ReimerEtAl18, HaddadfarshiEtAl15}. Below, in Sec.~\ref{sec:FL-Highfreq} we show that this is also the case, when directly applying the Magnus expansion to our model (\ref{eq:Lindblad-two-level system}). We will then show how a high-frequency expansion that is consistent with the phase diagrams of Fig.~\ref{fig:phases-exact} can still be obtained by conducting it in a suitably chosen rotating frame. In the rotating frame, it even explains the transition to the non-Lindbladian phase as a consequence of the non-unitary micromotion, when the frequency is lowered.

\hide{
TODO: Vllt verschieben;

However, a straight-forward definition of the rotating frame is only due to the fact that in our model system, Eq.~\eqref{eq:Lindblad-two-level system}, there is a driving of
the coherent part of the evolution only. If we were to start from a model where also the dissipative part of the
evolution is driven, such a definition is not obvious. Therefore, in the second part of the chapter we show that,
when the driving is a single harmonic, one can equally arrive at the same result of Section \ref{sec:Magnus-rotfr} by
performing an expansion in the extended Hilbert space, which does not rely on any definition of the rotating frame at all. 
In Section \ref{sec:ext-space} we  first introduce the notion
of this extended space. In Section \ref{sec:expan-ext-sp} the expansion, which is a perturbative treatment of the constant part of the Lindbladian, is then performed
and we show that for our model system we recover  the exact same results as in Section \ref{sec:Magnus-rotfr}.}

\section{High-frequency expansions and extended Floquet space for open quantum systems}%{Obtaining the Floquet-Lindbladian in extended space}
\label{sec:ext-space}
A standard tool to extract the Floquet Hamiltonian in the high-frequency limit
is the Magnus expansion \cite{BlanesEtAl09}. 
%It has originally been developed as a means to find fundamental solutions of an ordinary linear differential equation (of a general type) with time-dependent coefficients. 
%As such, we apply it to our  two-level system model to extract the Floquet generator. 
%The Magnus expansion is a high-frequency expansion of the dynamics that is commonly used
%for systems with coherent dynamics governed by a time-periodic Hamiltonian $H(t)$. 
In line with what has been developed in the literature \cite{ReimerEtAl18, HaddadfarshiEtAl15}
we apply the Magnus expansion to the special case of a time-periodic Lindblad superoperator. For a two-level system it takes the general form
\begin{align}
\mathcal{L}(H, {d}) = -i \left[H, \cdot \right] + \sum_{n,m} d_{nm} \, \left(\sigma_n \cdot \sigma_m - \frac{1}{2} \left\lbrace\sigma_m \sigma_n, \cdot \right\rbrace \right),
\label{eq:Lind-gen}
\end{align}
with traceless Hamiltonian $H$ governing the coherent evolution and Kossakowski matrix ${d}$ governing the dissipative component of the evolution. Recall that for the
evolution to be physical, i.e.~completely positive and trace-preserving, the Kossakowski matrix has to be positive semi-definite, ${d} \geqslant 0$. With this notation, commutators of Lindblad superoperators can be evaluated by using the general expressions for the commutators of two general two-level system Lindblad superoperators (see Appendix \ref{sec:app-comm-lind}).

%Despite this success, for open quantum systems, however, it was observed in the literature 
%that the Magnus expansion typically produces non-Lindbladian terms in low orders \cite{ReimerEtAl18, HaddadfarshiEtAl15}.
%These terms are non-Lindbladian as the corresponding superoperator cannot be brought into Lindblad form.  

As an alternative approach to compute the Floquet generator of an open system, we will also work out a non-Hermitian version of van-Vleck degenerate perturbation theory in the Floquet space of time-periodic density matrices. This extended Floquet state space is given by the product space of the original state space of density matrices with that of time-periodic functions. This approach is a generalisation of the method described in Ref.~\cite{EckardtAnisimovas15} for isolated driven quantum systems.  It has the advantage that it clearly isolates the effect of the micromotion. Namely, it gives rise to an effective generator that is independent of the driving phase. Combining this object with a driving-phase dependent micromotion operator, then provides the Floquet generator for the stroboscopic evolution.

%Instead of performing time integration, the stroboscopic map $\mathcal{P}(T)$ and therefore also the 
%candidate $\mathcal{K}$ for the Floquet Lindbladian can also be obtained by solving an eigenvalue equation
%in the extended Hilbert space (extended by the space of time-periodic functions), much like in the case of the coherent evolution.

%If this diagonalization is only performed perturbatively up to a given order in inverse frequency, we find a high-frequency expansion that we want to call van-Vleck high-frequency  expansion \cite{AnisimovasEtAl15}. %(despite the fact that it may vary slightly from the conventional Floquet-Magnus expansion in the literature; a discussion of this subtle difference is found in Ref.~\cite{AnisimovasEtAl15}).
%We will see that the problem of an non-Lindbladian leading order that we found for the Magnus expansion is also present in this approach.

\subsection{The Magnus expansion}
Because the Lindblad superoperator is time periodic, we can expand it in the Fourier series,
\begin{align}
\mathcal{L}(t) &= \sum_{n\in \mathbb{Z}} e^{i\omega n t} \mathcal{L}_n.
\label{eq:lindbl-four}
\end{align}
The Magnus expansion  \cite{BlanesEtAl09} is a general  high-frequency expansion for linear differential equations with periodic  coefficients.
Therefore it can be directly applied to our problem.
%The Magnus expansion is a high frequency expansion that 
It gives rise to one candidate $\mathcal{K}$ for $\mathcal{L}_F$.
Let us denote this expansion of the generator by
\begin{align}
\mathcal{K}_\mathrm{Mag} &= \sum_{n=1}^{\infty} \mathcal{K}^{(n)},
\end{align}
which we approximate by truncating the series after some order $k$, giving 
\begin{align}
\mathcal{K}_\mathrm{Mag,k} &= \sum_{n=1}^{k} \mathcal{K}^{(n)}.
\end{align}
%Then, it holds that
The leading coefficients read
\begin{align}
\mathcal{K}^{(1)} &= \frac{1}{T} \int_0^T\mathrm{d}t \mathcal{L}(t)  = \mathcal{L}_0
\label{eq:Magnus-zeroth}\\
\mathcal{K}^{(2)} &= \frac{1}{2T} \int_0^T\mathrm{d}t \int_0^t\mathrm{d}t' \left[\mathcal{L}(t), \mathcal{L}(t') \right] \\
&= i \sum_{n=1}^{\infty} \frac{\left[\mathcal{L}_n, \mathcal{L}_{-n}\right]+ \left[\mathcal{L}_0, \mathcal{L}_n - \mathcal{L}_{-n}\right]}{n \omega},\\
\begin{split}
\mathcal{K}^{(3)} &= \frac{1}{6T} \int_0^T\mathrm{d}t \int_0^t\mathrm{d}t'\int_0^{t'}\mathrm{d}t'' \bigl( \left[\mathcal{L}(t), \left[\mathcal{L}(t'), \mathcal{L}(t'')\right] \right] \\
&\qquad \qquad +  \left[\mathcal{L}(t''), \left[\mathcal{L}(t'), \mathcal{L}(t)\right] \right] \bigr).
\end{split}
\label{eq:sec-ord-Magnus}
\end{align}
For an expression of the third-order contribution in terms of the Fourier components of $\mathcal{L}(t)$ see Appendix~\ref{sec:app-discrep-magnus}.

\subsection{Floquet space}
Since $\mathcal{L}(t)$ is periodic, we can apply Floquet's theorem to Eq.~\eqref{eq:tdm} and  find that the 
fundamental solutions of Eq.~\eqref{eq:tdm} are Floquet states of the form
\begin{equation}
\varrho_{a}(t) = e^{-i\Omega_{a}t} \, \Phi_a(t)
\label{eq:timeev-rho}
\end{equation}
where index $a$ runs over all $N^2$ fundamental solutions, 
%(the Hilbert space $\mathcal{H}$ has dimension $N$), 
with complex numbers $\Omega_{a}$ (replacing the quasienergies in the case of an isolated system) and time-periodic operators $\Phi_a(t) =\Phi_a(t + T)$ (replacing the Floquet modes).
Note that the representation in Eq.~\eqref{eq:timeev-rho} is not unique,  namely by setting
\begin{align}
\Omega_{a} & \longrightarrow \Omega_{a}  + m \omega, \qquad m \in \mathbb Z\\
\Phi_a(t) & \longrightarrow \mathrm{e}^{i m \omega t} \Phi_a(t) 
\end{align}
we could find an equivalent representation of Eq.~\eqref{eq:timeev-rho}, that will later appear as a (seemingly) independent solution in the Floquet space formalism.

We can expand the time-periodic operators $\Phi_a$ in a Fourier series
\begin{align}\label{eq:FEFM}
%\mathcal{L}(t) &= \sum_{n\in \mathbb{Z}} e^{i\omega n t} \mathcal{L}_n,\\
\Phi_a(t) &= \sum_{n\in \mathbb{Z}} e^{i\omega n t} \, \Phi_{a,n}.
\end{align}
Plugging both Fourier expansions, Eq.~\eqref{eq:lindbl-four} and Eq.~\eqref{eq:FEFM}, into Eq.~\eqref{eq:tdm}, we find
\begin{equation}
\sum_n  (-i \Omega_{a}  + i \omega n) \,  \Phi_{a,n}  e^{i\omega n t}
	= \sum_{k,m} \mathcal{L}_k  \Phi_{a,m} \, e^{i\omega (k+m) t}. 
\end{equation}
Recall that the $\mathcal{L}_n$ are superoperators that act on the $\Phi_{a,n}$, which are linear operators on $\mathcal{H}$, 
$\Phi_{a,n} \in L(\mathcal{H})$. 

By comparing the prefactors of the exponential functions, we find an eigenvalue equation in the `extended' Hilbert space $ L(\mathcal{H}) \otimes \mathcal{T}$, where $\mathcal{T}$ shall denote the space of time-periodic functions with period $T$. It reads
\begin{equation}
\Omega_{a} \Phi_{a,n} = \sum_{m} (i \mathcal{L}_{n-m} + \delta_{nm} \, m \omega \, \mathbf{1}) \,  \Phi_{a,m} = \sum_{m} \bar{\mathcal{Q}}_{nm}  \,  \Phi_{a,m},
\label{eq:extended-space}
\end{equation}
where $\bar{\mathcal{Q}}$ is the extended-space representation of the superoperator,
\begin{align}
Q(t) = i\mathcal{L}(t)-i\partial_t.
\end{align}
This superoperator is the generalization of the quasienergy operator $H(t)-i \partial_t$ found for isolated systems
to the open system.

Similar to the case of isolated systems, Eq.~\eqref{eq:extended-space} possesses a transparent block structure
%\begin{align}
%\Omega_a \begin{pmatrix}
%\dots \\
%{\Phi_{a,-1}} \\ {\Phi_{a,0}} \\
%{\Phi_{a,1}} \\ \dots
%\end{pmatrix} = 
%\begin{pmatrix}
%\dots & \\
%& i\mathcal{L}_{1} &  i\mathcal{L}_{0} - \omega \,  \mathbf{1} &  i\mathcal{L}_{-1}& i\mathcal{L}_{-2}&  i\mathcal{L}_{-3}\\
%& i\mathcal{L}_{2}  & i\mathcal{L}_{1}  & i\mathcal{L}_{0} & i\mathcal{L}_{-1} &  i\mathcal{L}_{-2}\\
%& i\mathcal{L}_{3} & i\mathcal{L}_{2} & i\mathcal{L}_{1} &i\mathcal{L}_{0} + \omega \,  \mathbf{1} &  i\mathcal{L}_{-1}\\
%&&& && &\dots
%\end{pmatrix}
%\begin{pmatrix}
%\dots \\
%{\Phi_{a,-1}} \\ {\Phi_{a,0}} \\
%{\Phi_{a,1}} \\ \dots
%\end{pmatrix},
%\end{align}
\begin{align}
\Omega_a \begin{pmatrix}
\dots \\
{\Phi_{a,-1}} \\ {\Phi_{a,0}} \\
{\Phi_{a,1}} \\ \dots
\end{pmatrix} = 
\begin{pmatrix}
\dots & \\
 &  i\mathcal{L}_{0} - \omega \,  \mathbf{1} &  i\mathcal{L}_{-1}& i\mathcal{L}_{-2}\\
  & i\mathcal{L}_{1}  & i\mathcal{L}_{0} & i\mathcal{L}_{-1} \\
 & i\mathcal{L}_{2} & i\mathcal{L}_{1} &i\mathcal{L}_{0} + \omega \,  \mathbf{1} \\
&&& &\dots
\end{pmatrix}
\begin{pmatrix}
\dots \\
{\Phi_{a,-1}} \\ {\Phi_{a,0}} \\
{\Phi_{a,1}} \\ \dots
\end{pmatrix},
\label{eq:ext-sp-matrixform}
\end{align}
however the entries in the vectors are now operators and the entries
in the matrix are nonhermitian (but Hermiticity-preserving) superoperators.

%The problem may be solved numerically by representing the matrices $\Phi_{a,n}$ as ($N^2$-dimensional) vectors,  the $\mathcal{L}_n$ as ($N^2 \times N^2$) matrices 
%and truncating the sum at some cutoff index. In principle now this problem will have an extensive number (growing with the cutoff) of eigenvalues $\Omega_{a,r}$, but it suffices to consider
%the $N^2$ solutions in the strip $-\omega/2 \leq \mathrm{Re} \, {\Omega_a} <  \omega/2$. 

\subsection{van-Vleck  high-frequency expansion}
\label{sec:Floq-Magn-dissi}
The aim of the van-Vleck  high-frequency expansion is to find a rotation $\bar{\mathcal{D}}$ that block diagonalizes the problem in the extended space,
\begin{align}
\bar{\mathcal{Q}}' = \bar{\mathcal{D}}^{-1}\bar{\mathcal{Q}}\bar{\mathcal{D}},
\label{eq:trafo-orig-extspace}
\end{align}
such that
\begin{align}
\bar{\mathcal{Q}}'_{nm} = \delta_{nm} (i \mathcal{K}_{\mathrm{eff}} + \, m \omega \, \mathbf{1}).
\label{eq:Quasiop-diagonal}
\end{align}
This transformation to a block-diagonal form is desired, since Eq.~\eqref{eq:ext-sp-matrixform} is block diagonal for a time-independent generator.
As we will see, this transformation therefore leads into a frame where the dynamics is governed by the time-independent generator $\mathcal{K}_{\mathrm{eff}}$.
%Note that the essence of this rotation $\bar{\mathcal{D}}$ is 
However, in contrast to the closed system, $\bar{\mathcal{Q}}$ is not necessarily Hermitian, so the
rotation $\mathcal{D}$ is in general not a unitary transformation. Still, the spectrum $\Omega_a$ is of course invariant under this transformation.

In analogy to the coherent case~\cite{EckardtAnisimovas15},
it suffices to take into account time-periodic transformations $\mathcal{D}(t) = \sum_{n}e^{i\omega n t} \mathcal{D}_n$,
therefore in extended space the operator $\bar{\mathcal{D}}_{nm}$
may only depend on the difference of the phonon indices $\bar{\mathcal{D}}_{nm}=\mathcal{D}_{n-m}$. 
First of all, we observe that for two time-local time-periodic superoperators, 
\begin{align}
\mathcal{A}(t)= \sum_{n\in \mathbb{Z}} e^{i\omega n t} \mathcal{A}_n \quad \text{and} 
\quad \mathcal{B}(t)= \sum_{n\in \mathbb{Z}} e^{i\omega n t} \mathcal{B}_n,
\end{align}
the product of both operators in the time domain
\begin{align}
\mathcal{C}(t)&=\mathcal{A}(t) \mathcal{B}(t)= \sum_{n,m\in \mathbb{Z}} e^{i\omega (n+m) t} \mathcal{A}_n \mathcal{B}_m\\
&= \sum_{n,m\in \mathbb{Z}} e^{i\omega n t} \mathcal{A}_{n-m} \mathcal{B}_m,
\end{align}
leads in the extend space to 
\begin{align}
\bar{\mathcal{C}}_{nm}& = \mathcal{C}_{n-m} = \sum_{k\in \mathbb{Z}}\mathcal{A}_{n-m-k} \mathcal{B}_{k} \\
&= \sum_{k\in \mathbb{Z}}\mathcal{A}_{n-k} \mathcal{B}_{k-m}=(\bar{\mathcal{A}}\bar{\mathcal{B}})_{nm}.
\end{align}
Therefore, products in the time domain directly translate into products in the 
extended space and vice versa. As a result, the inverse transformation $\bar{\mathcal{D}}^{-1}$ in the extended space is just the representation 
of the inverse transformation in time,
\begin{align}
\mathcal{D}^{-1}(t) = \sum_{n}e^{i\omega n t} (\mathcal{D}^{-1})_n \quad \text{with} \quad  \mathcal{D}^{-1}(t)  \mathcal{D}(t)= 1 ,
\end{align}
i.e.~we have $(\bar{\mathcal{D}}^{-1})_{nm}=(\mathcal{D}^{-1})_{n-m}$.
%Let us begin by discussing the nature 

Thus, the transformation in Eq.~\eqref{eq:trafo-orig-extspace} becomes
%\begin{align}
$\Phi_a'(t) = {\mathcal{D}}^{-1}(t)\Phi_a(t)$, and therefore $\varrho'(t) = {\mathcal{D}}^{-1}(t)\varrho(t)$.
The equation of motion in the  transformed frame reads
\begin{align} \nonumber
\partial_t {\varrho}'(t) &= (\partial_t \mathcal{D}^{-1}(t)) \varrho(t) +\mathcal{D}^{-1}(t) \partial_t\varrho(t) \\ 
&\equiv {\mathcal{L}}'(t) {\varrho}'(t) 
\end{align}
Thus, much like to the coherent case, this transformation 
is equivalent to
\begin{align}
\label{eq:gaugetrafo-micromotion}
{\mathcal{L}}'(t) [\cdot] &= (\partial_t \mathcal{D}^{-1}(t))\mathcal{D}(t) \cdot + \mathcal{D}^{-1}(t) \mathcal{L}(t)[\mathcal{D}(t) \cdot].%\\
%&= -\mathcal{L}_d(t) \cdot   + \Lambda(t) \mathcal{L}_d(t)[\Lambda(t)^{-1} \cdot] + %\Lambda(t) \mathcal{L}_0[\mathcal{D}(t) \cdot] 
\end{align}
resembling a gauge transformation.
%Thus, much like in the coherent case, this transformation 
%is equivalent to a (non-unitary) gauge transformation.

As pointed out already in the literature \cite{DaiEtAl2017} and in analogy 
to the closed system, Eq.~\eqref{eq:def-eff-Hamil}, the effective generator $\mathcal{K}_\text{eff}$ appearing in Eq.~\eqref{eq:Quasiop-diagonal} fulfills
\begin{align}
\mathcal{P}(t,t_0)=\mathcal{D}(t)\exp[(t-t_0)\mathcal{K}_\text{eff}]\mathcal{D}^{-1}(t_0).
\end{align}
It is the time-independent generator describing the evolution in a ``rotating frame of reference''. 
However, since the dynamics is dissipative, the  time-periodic ``micromotion'' operator $\mathcal{D}(t)$ that describes this 
transformation, is generally not unitary anymore. 
Defining a general Floquet generator $\mathcal{K}_{t_0}$ via
\begin{align}
\mathcal{P}(t_0+T,t_0)=\exp\left(\mathcal{K}_{t_0}T\right),
\end{align}
so that $\mathcal{K}=\mathcal{K}_{0}$ corresponds to the Floquet generator defined by Eq.~(\ref{eq:generator-cand}) for the case of $t_0=0$, it can be expressed in terms of the effective generator $\mathcal{K}_\text{eff}$ and the micromotion operator $\mathcal{D}(t)$, 
\begin{align}
\mathcal{K}_{t_0}=\mathcal{D}(T+t_0)\mathcal{K}_\text{eff}\mathcal{D}^{-1}(t_0).
\label{eq:Floquet-Lind-t0}
\end{align}
Since the micromotion superoperator $\mathcal{D}(t_0)$ is generically non-unitary, it is possible that it maps a Lindbladian effective generator $\mathcal{K}$ to a non-Lindbladian Floquet generator $\mathcal{K}_{t_0}$. This explains the driving-phase dependence (which is equivalent to a dependence on $t_0$) observed in Fig.~ \ref{fig:phases-exact}. Moreover, below we find strong evidence suggesting that $\mathcal K_\text{eff}$ is always of Lindblad type, so that the non-Markovianity of $\mathcal{K_{t_0}}$, as it is found in the non-Lindbladian lobes of Fig.~\ref{fig:phases-exact}, must entirely entirely be to the micromotion captured by $\mathcal{D}(t_0)$. 

In Ref.~\cite{DaiEtAl2017} a high-frequency expansion for both the effective generator $\mathcal{K}_\mathrm{eff}$ and the micromotion superoperator $\mathcal{D}(t)$ were derived. 
Here we present an alternative derivation of such a high-frequency expansion by applying van-Vleck-type degenerate perturbation theory the extended Floquet space. 
Genealizing the reasoning of Ref.~\cite{EckardtAnisimovas15} to the non-Hermitian problem of the open system, we decompose $\mathcal{Q}$ into an unperturbed blockdiagonal part $\bar{\mathcal{Q}}_0$ and a perturbation $\bar{\mathcal{V}}$ that can also contain block-off-diagonal terms, 
\begin{align}
	\bar{\mathcal{Q}} = \bar{\mathcal{Q}}_0 + \lambda \bar{\mathcal{V}},
\end{align}
with $ (\bar{\mathcal{Q}}_0)_{nm} = \delta_{nm} m\omega \mathbf{1} $. 
Applying van Vleck perturbation theory, we obtain (Appendix \ref{sec:app-deg-perttheory})
\begin{eqnarray}
\mathcal{K}_\text{eff} &=& \sum_{n=1}^{\infty} \mathcal{K}^{(n)}_\text{eff}
\\
\mathcal{D}(t)&=& \exp(\mathcal G(t)) \quad  \text{ with }  \quad \mathcal G(t) = \sum_{n=1}^{\infty} \mathcal {G}^{(n)}(t), 
\end{eqnarray}
where (see also  \cite{DaiEtAl2017})
\begin{align}
	\mathcal{K}^{(1)}_\text{eff} &= \mathcal{L}_0, 
	\label{eq:Keff-zeroth}
	\\
	\mathcal{K}^{(2)}_\text{eff} &=  i \sum_{n=1}^{\infty} \frac{\left[\mathcal{L}_n, \mathcal{L}_{-n}\right]}{n\omega}, \\
	\mathcal{K}^{(3)}_\text{eff} &=  -\sum_{n\neq 0} \left( \frac{\left[\mathcal{L}_n, \left[\mathcal{L}_{0}, \mathcal{L}_{-n}\right]\right]}{2 n^2\omega^2}  + \sum_{ \substack{m\neq0,\\ m\neq n}} \frac{\left[\mathcal{L}_m, \left[\mathcal{L}_{n-m}, \mathcal{L}_{-n}\right]\right]}{3nm\omega^2}\right).
	\label{eq:Keff-2}
\end{align}
and
\begin{align}
\mathcal G^{(1)}(t) &= -i  \sum_{n\neq 0} e^{i n\omega t} \frac{\mathcal{L}_n}{n\omega},\\
\mathcal G^{(2)}(t) &= -  \sum_{n\neq 0} e^{in\omega t} \left( \frac{\left[\mathcal{L}_0,\mathcal{L}_n\right]}{n^2\omega^2} + \sum_{\substack{m \neq 0,\\ m \neq n}}  \frac{\left[\mathcal{L}_{n-m},\mathcal{L}_m\right]}{2mn\omega^2} \right).
\end{align}
These expressions take exactly the same structure as those found for isolated systems \cite{EckardtAnisimovas15}.

\section{High-frequency expansion: Direct frame}
\label{sec:FL-Highfreq}
%After having discussed high frequency expansions in the rotating frame, let us now investigate what these expansions yield when directly applied to the problem, without integrating out the driving term.
Let us now apply both types of high-frequency expansion described in the previous section to our model system. Although a Lindblad-type Floquet generator is found numerically to exist in the high-frequency regime, this behaviour is not reproduced by both the Magnus and the van-Vleck-type expansion when directly applied to the model \eqref{eq:Lindblad-two-level system}.

%Here we explicitly demonstrate that, even though for our model system a Floquet-Lindbladian was shown to exist in the high-frequency regime,  the existence of the Floquet Lindbladian is guaranteed in the high-frequency limit (as we have seen in Section \ref{sec:driven-qbit-model}), as we show in Sec.~\ref{sec:Magnus}, the Magnus expansion does not yield a valid Lindbladian Floquet generator %(i.e.~it is not a valid Lindbladian) 
%on the dominating order of the high-frequency expansion.
%
%A second high-frequency expansion that we  consider is the van-Vleck-type perturbation theory in the extended Hilbert space that we introduced in Sec.~\ref{sec:Floq-Magn-dissi}. As we show in Sec.~\ref{sec:Floq-Magn-dirframe}, for the open system the problem of an non-Lindbladian leading order is shared by this expansion.

 \subsection{Emergence of non-Lindbladian terms in the Magnus expansion}
\label{sec:Magnus}

%As already mentioned, 
Let us compute the leading terms of the Magnus expansion for the effective Floquet generator for the two-level system defined in Eq.~\eqref{eq:Lindblad-two-level system} with driving phase $\varphi=0$. The Fourier-expansion of our model yields three non-vanishing terms, 
\begin{align}
\mathcal{L}_0 &= -i \left[\frac{\sigma_z}{2}, \cdot\right] + \gamma \, \left(\sigma_- \cdot \sigma_+ - \frac{1}{2} \lbrace\sigma_+ \sigma_-, \cdot \rbrace \right) 
\intertext{ and }  \mathcal{L}_1 &= \mathcal{L}_{-1} = -i \frac{E}{2} \left[ \sigma_x, \cdot\right].
\end{align}
The second order of the expansion drops out, $\mathcal{K}^{(2)}=0$, (as well as all other even orders). Using Eq.~\eqref{eq:sec-ord-Magnus}, up to the third order we, therefore, find
\begin{align}
\mathcal{K}_\mathrm{Mag, 3} =  \mathcal{L}_0  +   \frac{2}{\omega^2} \left[\mathcal{L}_0, \left[\mathcal{L}_{0}, \mathcal{L}_{1}\right]\right] - \frac{1}{\omega^2} \left[\mathcal{L}_1, \left[\mathcal{L}_{0}, \mathcal{L}_{1}\right]\right].%+ \mathcal{O}(1/\omega^4).
\end{align}
%(Note to us: This is what comes out for the second expression by Vegas et al that we present in the appendix as well as if one calculates the terms by hand. However when
%one uses Mintert's expression, one finds a different prefactor. I therefore presume that Mintert's expression is wrong.)

By using the general expressions for the commutator of two general two-level system Lindblad superoperators that we present in Appendix \ref{sec:app-comm-lind}, 
we compute
\begin{align}
%\begin{split}
\left[\mathcal{L}_{0}, \mathcal{L}_{1}\right]  = \mathcal{L}(H_a, {d_a}),
\end{align}
with
\begin{align}
\text{ with } H_a = \frac{E}{2} \sigma_y  \quad\text{ and }\quad  d_a= \gamma E \begin{pmatrix}
0 & 0  & -i \\
0 & 0 & -1 \\
i & -1 & 0
\end{pmatrix},
\end{align}
where $\mathcal{L}(H, {d})$ is defined in Eq.~\eqref{eq:Lind-gen}.
Similarly, we find 
\begin{align}
\left[\mathcal{L}_{0},  \left[\mathcal{L}_{0}, \mathcal{L}_{1}\right] \right]  = \mathcal{L}(H_b, {d}_b), 
\end{align}
with 
\begin{align}
 H_b =-\frac{E}{2} \sigma_x  
\quad \text{ and }\quad {d}_b = 2 \gamma E \begin{pmatrix}
0 & 0  & 1\\
0 & 0 & -i \\
1& i & 0
\end{pmatrix},
\end{align}
as well as
\begin{align}
\left[\mathcal{L}_{1},  \left[\mathcal{L}_{0}, \mathcal{L}_{1}\right] \right]  = \mathcal{L}(H_c, {d}_c), 
\end{align}
with
\begin{align}
 H_c = \frac{E^2}{2} \sigma_z+ \mathcal{O}(\gamma^2)
 \text{ and }{d}_c = \gamma E^2 \begin{pmatrix}
0 & i  & 0\\
-i & 2 & 0 \\
0& 0 & -2
\end{pmatrix}+ \mathcal{O}(\gamma^2).
\label{eq:commut-L_101}
\end{align}
Altogether, in third order Magnus expansion (and first order in $\gamma$), the Floquet generator is approximated by
\begin{align}
\mathcal{K}_\mathrm{Mag,3} 
&= \mathcal{L}(H_\mathrm{Mag,3} ,{d}_\mathrm{Mag,3} ), 
\label{eq:FloqLind-Magnus}
\end{align}
with
\begin{align}
H_\mathrm{Mag,3}  = -\frac{\varepsilon}{\omega} \sigma_x  + \frac{1}{2}\left(1-\varepsilon^2\right) \sigma_z
\end{align}
and
\begin{align}
 {d}_\mathrm{Mag,3}  = \gamma \begin{pmatrix}
1 & i  (1-\varepsilon^2) & 4\varepsilon/\omega\\
-i(1-\varepsilon^2) & 1-2\varepsilon^2 & -4i\varepsilon/\omega  \\
4\varepsilon/\omega& 4i\varepsilon/\omega & 2\varepsilon^2
\end{pmatrix},
\end{align}
where $\varepsilon={E}/{\omega}$. 

\begin{figure}
	\centering
	\subfloat[]{
		\begin{minipage}{0.5\columnwidth}
			\includegraphics[scale=0.65]{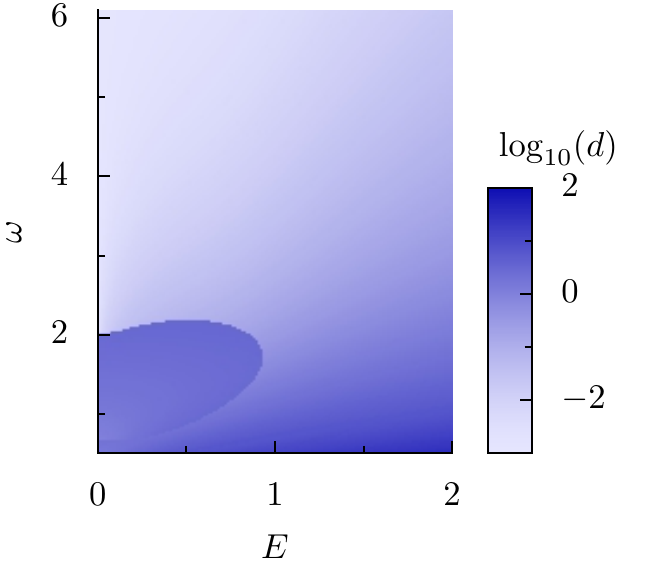}
	\end{minipage}}
	%\hspace{1cm}
	\subfloat[]{
		\begin{minipage}{0.4\columnwidth}
			\vspace{0.1cm}
			\includegraphics[scale=1.]{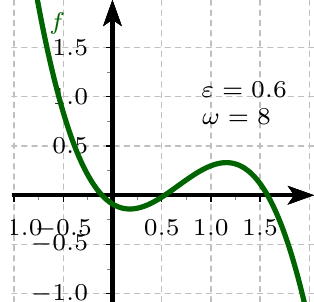}
			\vspace{0.1cm}
	\end{minipage}}
	\caption[Distance of the leading order Magnus expansion $\mathcal{K}_\mathrm{Mag,3}$ to the exact generator and sketch of the characteristic polynomial of coefficient matrix]{(a)~Distance (the Frobenius norm) $d=\vert\vert \mathcal{K}_\mathrm{Mag,3} - \mathcal{K}_{x_0}\vert \vert_F$  between the generator 
		$\mathcal{K}_\mathrm{Mag,3}$ obtained with the  third-order Magnus expansion in the direct frame and the exact generator $ \mathcal{K}_{x_0} \in \mathrm{log}(\mathcal{P}(T))/T$
		from  branch closest to a Lindblad generator.
		(b)~Typical graph $f(\lambda)$ of the characteristic polynomial of  matrix ${d}_\mathrm{Mag,3}/\gamma$ of the
		third order Magnus expansion $\mathcal{K}_\mathrm{Mag,3}$ of the Floquet generator. The matrix ${d}_\mathrm{Mag,3}$ therefore has one negative eigenvalue for all parameters $\varepsilon, \omega$ and $\gamma>0$.}
	\label{fig:char-pol-Magnus}
\end{figure}

The matrix distance of the matrix representation of the superoperator %of this approximation
 $\mathcal{K}_\mathrm{Mag,3}$ to the matrix representation of  the exact Floquet generator $\mathcal{K}$ is shown in
Fig.~\ref{fig:char-pol-Magnus}(a).
Note that although for high frequencies, $\omega \to \infty$, this distance approaches zero,  for any finite $\gamma \neq 0$ the generator $\mathcal{K}_\mathrm{Mag,3}$ is not
a valid Lindbladian generator in the whole region of the parameters.
This can be seen from the characteristic polynomial of its dissipator matrix ${d}_\mathrm{Mag,3} $, which apart from the prefactor $\gamma$ reads
\begin{align}
f(\lambda)&=\mathrm{det}({d_\mathrm{Mag,3}}/\gamma- \lambda \mathbf{1}) \\
&= - \lambda^3 + 2 \lambda^2 - \lambda \left(4\varepsilon^2-5\varepsilon^4-\frac{32\varepsilon^2}{\omega^2}\right)-2\varepsilon^6.	
\label{eq:char-polyn-magnus}
\end{align}
As illustrated in Fig.~\ref{fig:char-pol-Magnus}(b), for $\lambda \rightarrow -\infty$ we have $f(\lambda) \rightarrow \infty$, but at the same time one finds $f(0)=-2\varepsilon^6 < 0$. Therefore there will always be a 
negative eigenvalue $\lambda$ and the Kossakowski matrix ${d}_\mathrm{Mag,3}$ is not positive semi-definite. As a result, the third-order Magnus approximation of the Floquet generator $\mathcal{K}_\mathrm{Mag,3}$ is not of Lindblad form. This is unsatisfactory, since the Floquet generator has been shown to be of Lindblad form numerically in the limit of large driving frequencies.  

%In Section \ref{sec:Magnus-rotfr} below we  will show 
%that the this problem of a non-Lindbladian leading order does not occur, when we first transform into a co-rotating frame,
%and then perform the Magnus expansion. This shows that when performing high-frequency expansions for dissipative systems, some frames of reference can be more favorable for the expansion than others.

As was already pointed out in the literature \cite{HaddadfarshiEtAl15}, the negative eigenvalue emerges due to the fact that  the characteristic polynomial
has terms that are of higher order than $1/\omega^2$ up to which the Magnus expansion was performed. It is indeed expected, that
the characteristic polynomial is correct only up to this order,
\begin{align}
f(\lambda)= - \lambda^3 + 2 \lambda^2 -  4\varepsilon^2 \lambda, % + \mathcal{O}\left(\frac{1}{\omega^4}\right),
\label{eq:char-pol-w2}
\end{align}
and that the next higher order will only be revealed after evaluating the Magnus expansion up to fourth order and so on.
Note that if we only take into account the terms up to order $1/\omega^2$, Eq.~\eqref{eq:char-pol-w2}, indeed, the characteristic polynomial only has nonnegative
eigenvalues, so one could argue that complete positivity is only violated in orders higher than $1/\omega^2$. 
However, if one would want to find  a generator that is a valid Lindbladian in this order $1/\omega^2$, there is no well-defined procedure on how to modify the terms in the dissipator matrix ${d}$, such that its characteristic polynomial is exactly the one in
Eq.~\eqref{eq:char-pol-w2}.

The problem of an non-Lindbladian generator $\mathcal{K}_\mathrm{Mag}$ is not originating from a wrong choice of branch for $\mathcal{K}_\mathrm{Mag}$. We have also checked the other branches of $\mathcal{K}_\mathrm{Mag}$ numerically and they also do not yield a valid Lindbladian generator. In the high-frequency limit $\omega\to \infty$, we generally expect that it suffices to investigate the principal branch. This is because for the high-frequency expansion $\mathcal{K}_\mathrm{Mag}(\omega)$ one has (cf.~Appendix \ref{sec:app-Wolf-method})
\begin{align}
\mathcal{K}_{\mathrm{Mag}, \mathbf{x}}(\omega) = \mathcal{K}_\mathrm{Mag}(\omega)+i \omega  \sum_{c=1}^{n_c}  x_c \left( P_c(\omega) -   P_{c*}(\omega)\right).
\end{align}
In the high-frequency limit, the principal branch $\mathcal{K}_\mathrm{Mag}(\omega)$ converges to the diabatic (or rotating-wave) Lindbladian $\mathcal{K}_\mathrm{Mag}(\omega)\rightarrow \mathcal{L}_0$, therefore all the projectors will also converge, $P_c(\omega)\rightarrow P_c(\infty) $. As long as
\begin{align}
	 \Phi_{\perp} (P_c(\infty) - P_{c*}(\infty))^\Gamma\Phi_{\perp} \neq 0
\end{align}
the matrices $V_c$ in the Markovianity test, Eq.~\eqref{eq:Fl-matrix-cond}, will scale linearly 
with $\omega$ in that limit. Therefore, for $\omega\to\infty$ all matrices $V_\mathbf{x}(\omega)$
for branches  different from $\mathbf{x}=0$ will diverge, leaving only the principal branch as a candidate.

\subsection{Non-Lindbladian terms in the van-Vleck  high-frequency expansion}
\label{sec:Floq-Magn-dirframe}

Let us now  investigate the effective generator $\mathcal{K}_\text{eff}$ using the van-Vleck  high-frequency expansion. Since again the second order vanishes, it provides in third-order high-frequency approximation and first order with respect to~$
\gamma$
\begin{align}
\mathcal{K}_{\mathrm{eff},3} =  \mathcal{L}_0 - \frac{1}{\omega^2} \left[\mathcal{L}_1, \left[\mathcal{L}_{0}, \mathcal{L}_{1}\right]\right].%+ \mathcal{O}(1/\omega^4).
\end{align}
Employing Eq.~\eqref{eq:commut-L_101}, 
we obtain
\begin{align}
\mathcal{K}_\mathrm{eff,3} 
= \mathcal{L}(H_\mathrm{eff,3} , {d}_\mathrm{eff,3}),
\label{eq:EffLind-Magnus}
\end{align}
with
\begin{align}
H_\mathrm{eff,3} =  \frac{1}{2}\left(1-\varepsilon^2\right) \sigma_z, 
\end{align}
and
\begin{align}
{d}_\mathrm{eff,3} = \gamma \begin{pmatrix}
1 & i  (1-\varepsilon^2) & 0\\
-i(1-\varepsilon^2) & 1-2\varepsilon^2 & 0  \\
0 & 0 & 2\varepsilon^2
\end{pmatrix}.
\end{align}
Here we may directly read off one eigenvalue of ${d}_\mathrm{eff,3}/\gamma$
\begin{align}
\lambda_3 = 2 \varepsilon^2.
\end{align}
The other eigenvalues follow from solving
\begin{align}
0  = \tilde f(\lambda) =  \lambda^2 - 2(1-\varepsilon^2) \lambda -\varepsilon^4.	
\end{align}
Again, $\tilde f(0)=-\varepsilon^4 < 0$ while asymptotically $\tilde f$ is positive, therefore there must be 
one negative eigenvalue, and also the effective generator is non-Lindbladian.
Thus, the van-Vleck  high-frequency expansion shares the problems of the Magnus expansion that it does not provide an effective generator of Lindblad form in the high-frequency limit.

%n the rotating frame, however, also the Floquet-Magnus expansion will give rise to a physical generator, as we will see in Section \ref{sec:rotfr-extsp}.

%In Section \ref{sec:expan-ext-sp}, we present an alternative high-frequency expansion %that is  a perturbative treatment of the constant part $\mathcal{L}_0$
%in the extended space. This high-frequency expansion does not yield such problems as it %produces generators that are physical in first and second order of the expansion. 

\section{Rotating frame of reference}
\label{sec:Rotfr}

When considering Floquet engineering in the high-frequency limit, we know from isolated systems that often the regime of strong driving, with the driving amplitude comparable to $\omega$ (which is large compared to other relevant system parameters), is of special interest, since here the driving leads to a noticeable modification of the system properties. A prominent example is coherent destruction of tunneling \cite{GrossmannEtAl91,EckardtEtAl05b, EckardtEtAl09}%\cmt{Refs!}
, occurring when the amplitude of the energy modulation between two tunnel-coupled states is equal to about $2.4 \omega$. To, nevertheless, be able to treat this regime using high-frequency expansions, typically a gauge transformation to a rotating frame of reference is performed, before conducting the high-frequency expansion. This frame is defined so that it integrates out the strong driving term, corresponding to the transition to the interaction picture with the driving term playing the role of the unperturbed Hamiltonian. Comparing the results of a high-frequency expansion in the original frame with those obtained in the rotating frame, the terms of the latter correspond to a partial resummation of infinitely many terms of the previous. Namely, while in the original frame, the $n$th order contains powers of the driving amplitude $\le n$, each order of the rotating-frame expansion can contain arbitrary powers of the driving amplitude. The rotating frame expansion is, thus, non-perturbative with respect to the driving amplitude. 

We will now perform such a transformation to a rotating frame also for the open quantum system. However, differently from the case of isolated systems, it will now not only improve the convergence properties of the high-frequency expansion for strong driving. Rather remarkably, it also ensures that the leading orders of the expansion give rise to approximations to the Floquet generator that can be of Lindblad type. Thus, the problem discussed in the previous section, namely that the Magnus and the van-Vleck expansion do not provide Lindblad-type generators when directly applied to our model system, is cured when conducting the high-frequency expansions in the rotating frame of reference.

\subsection{Rotating frame of reference}
\label{sec:Magnus-rotfr}

We decompose the time-dependent Lindbladian
into its time-average and a driving term,
\begin{align}
\mathcal{L}(t) =  \mathcal{L}_0 + \mathcal{L}_d(t),
\end{align}
with
\begin{align}
 \mathcal{L}_d(t) = \sum_{n\neq 0} e^{i n \omega t} \mathcal{L}_n.
\end{align}
Let us, for the sake of simplicity, assume that $\mathcal{L}_d(t)$ commutes with itself at different times, 
\begin{align}
\left[\mathcal{L}_d(t), \mathcal{L}_d(t')\right] =0, \quad \forall t,t',
\end{align}
which is equivalent to 
$\left[\mathcal{L}_n, \mathcal{L}_m\right] =0$, $\forall n,m\neq 0$.
In analogy to the  coherent case of isolated systems,  we consider the transformation generated by the driving term, %Eq.~\eqref{eq:trafo-rotfr}
\begin{align}
\tilde{\varrho}(t) = \Lambda^{-1}(t) \varrho(t),
\end{align}
with
\begin{align}
\Lambda^{-1}(t)= \exp\left(- \int_0^t \mathrm{d}t' \mathcal{L}_d(t')\right).
\end{align}
We denote operators in the rotating frame with a tilde. 
In case that only the coherent part  of the Lindbladian (i.e.\ the Hamiltonian) is driven,
$ \mathcal{L}_d(t) = -i \left[H_d(t), \cdot \right]$,
this transformation reduces to a unitary rotation of the density matrix,
\begin{align}\label{eq:trafo-rotfr}
\tilde{\varrho}(t) = U^{\dagger}(t)\varrho(t) U(t),  
\end{align}
with
\begin{align}
U(t)= \exp\left(-i \int_0^t \mathrm{d}t' H_d(t')\right).
\end{align}
The equation of motion in the rotating frame reads
\begin{align}
\partial_t \tilde{\varrho}(t) &= (\partial_t \Lambda^{-1}(t)) \varrho(t) +\Lambda^{-1}(t) \partial_t\varrho(t) \equiv \tilde{\mathcal{L}}(t) \tilde{\varrho}(t) 
\end{align}
with gauge-transformed Lindbladian
\begin{align}
\tilde{\mathcal{L}}(t) [\cdot] &= (\partial_t \Lambda^{-1}(t))\Lambda(t) \cdot + \Lambda^{-1}(t) \mathcal{L}(t)[\Lambda(t) \cdot]. %\\
%&= -\mathcal{L}_d(t) \cdot   + \Lambda(t) \mathcal{L}_d(t)[\Lambda(t)^{-1} \cdot] + %\Lambda(t) \mathcal{L}_0[\Lambda(t)^{-1} \cdot] 
\end{align}
Now, because $\mathcal L_d(t)$ commutes with itself at different times,
also $\Lambda(t)$ commutes with $\mathcal{L}_d(t)$, therefore we find
\begin{align}
\tilde{\mathcal{L}}(t) [\cdot] 
&= -\mathcal{L}_d(t) \cdot   + \Lambda^{-1}(t) \mathcal{L}_d(t)[\Lambda(t) \cdot] + \Lambda^{-1}(t) \mathcal{L}_0[\Lambda(t) \cdot] \\
&= \Lambda^{-1}(t) \mathcal{L}_0[\Lambda(t) \cdot].
\label{eq:Ltilde-rotfr}
\end{align}
By construction, we have eliminated the driving term, at the expense that the transformed static term has now acquired a 
periodic time-dependence.
%by but it comes at the expense of 
%a (possibly nonunitary) ``rotation'' of the  part of the dissipator that was static before.

From the time evolution operator in the rotating frame, $\tilde{\mathcal{P}}(t)$ (where here and in the following the initial time of the evolution is always understood to be $t=0$), we can define the Floquet Lindbladian $\tilde{\mathcal{K}}$ in the rotating frame in analogy to Eq.~\eqref{eq:generator-cand}, 
\begin{align}
\tilde{\mathcal{P}}(T) = \exp({\tilde{\mathcal{K}} T}).
\end{align}
Since for our choice of the driving term $\mathcal{L}_d(t)$, one has $\int_0^{\nu T} \mathrm{d}t \mathcal{L}_d(t)=0$, $\nu\in \mathbb{N}_0$, the transformation $\Lambda(t)$ becomes the identity at stroboscopic times $t=\nu T$. Thus, at stroboscopic times the rest frame and the rotating frame coincide, so that
\begin{align}
\tilde{\varrho}(\nu T) = \varrho(\nu T)
\end{align}
as well as 
\begin{align}
\tilde{\mathcal{P}}(\nu T) =  \mathcal{P}(\nu T).
\end{align}
In particular, one has  $ \tilde{\mathcal{P}}(T) = \mathcal{P}(T)$, which implies that 
\begin{align}
\tilde{\mathcal{K}}= \mathcal{K}.
\end{align}
Note that this is not true for a general choice of $\mathcal{L}_d(t)$, e.g.,~if $\mathcal{L}_d(t)$ does not commute with itself at 
different times. 

\subsection{Explicit transformation for our model system}
\label{sec:rotfr-Ln-expl}
We now work out the transformation to the rotating frame for our model system [Eq.~\eqref{eq:Lindblad-two-level system}]. Since only the Hamiltonian is driven, the transformation is unitary,  
\begin{align}
\tilde{\varrho}(t) = U^{\dagger}(t) \varrho(t) U(t),
\end{align}
where 
\begin{align}
U(t)= \exp\left(-i \chi(t)\sigma_x\right), \quad\text{with}\quad \chi(t) = \frac{E}{\omega} \sin(\omega t).
\end{align}
We again consider driving phase $\varphi=0$ only.
We find
\begin{align}
\begin{split}
\tilde{\mathcal{L}}(t) [\cdot] %&= -i \left[-\partial_t\chi(t)\sigma_x, \cdot\right]  -i \left[\frac{1}{2} \tilde\sigma_z(t)+ E \cos(\omega t)\sigma_x, \cdot\right] + \gamma \, \left(\tilde\sigma_-(t) \cdot \tilde\sigma_+(t) - \frac{1}{2} \lbrace\tilde\sigma_+(t) \tilde\sigma_-(t), \cdot \rbrace \right)\\
=&  -i \left[\frac{1}{2}\tilde\sigma_z(t), \cdot\right] \\&+ \gamma \, \left(\tilde\sigma_-(t) \cdot \tilde\sigma_+(t) - \frac{1}{2} \lbrace\tilde\sigma_+(t) \tilde\sigma_-(t), \cdot \rbrace \right).
\end{split}
\label{eq:Lindbl-rotfr}
\end{align}
%Here let us denote,
%\begin{align}
%\int_0^t \mathrm{d}t' H_d(t') =
 %\int_0^t\mathrm{d}t' E \cos(\omega t')   \sigma_x= \frac{E}{\omega} \sin(\omega t) %\sigma_x\equiv \chi(t) \sigma_x,
%\end{align}
Here the rotated Pauli operators read
\begin{align}
\begin{split}
\tilde\sigma_z(t) &= U^{\dagger}(t) \sigma_z U(t) \\ &= \cos(2\chi(t)) \sigma_z + \sin(2\chi(t)) \sigma_y,
\end{split}\\
\begin{split}
\tilde\sigma_\pm(t) &= U^{\dagger}(t) \sigma_\pm U(t) \\&= \sigma_x \pm i \left[ \cos(2\chi(t)) \sigma_y - \sin(2\chi(t)) \sigma_z\right].
\end{split}
\end{align}

%\subsection{Fourier components of the transformed Lindbladian}
In order to perform the high-frequency expansions in the rotating frame, let us now determine the Fourier components
of the transformed Lindbladian $\tilde{\mathcal{L}}(t)$, Eq.~\eqref{eq:Lindbl-rotfr}. Using the definition $z=2E/\omega$, we may rewrite the Fourier transform
\begin{align}
&\mathcal{F}_n[\cos(2\chi(t))]\equiv \frac{1}{T} \int_0^T \cos(2\chi(t)) e^{-i n \omega t}\mathrm{d}t \\
&= \frac{1}{T} \int_0^T \frac{1}{2} \left(e^{i z\sin(\omega t)} + e^{-i z\sin(\omega t)}\right) e^{-i n \omega t}\mathrm{d}t \\
&= \frac{1}{2}  \left[J_n(z) + J_{-n}(z) \right] = e_n J_n(z).
\end{align}
Here $J_n(z)$ is the $n$-th Bessel function of the first kind, we have used $J_{-n}(z)=(-1)^n J_{n}(z)$ and defined
\begin{align}
e_n =\left\lbrace \begin{array}{cc} 1, & n \text{ even,}\\ 0, & n \text{ odd,} \end{array} \right. \quad\text{and}\quad o_n =\left\lbrace \begin{array}{cc} 0, &  n \text{ even,}\\ 1, & n \text{ odd.} \end{array} \right.
\end{align}
Similarly, we find
\begin{align}
\mathcal{F}_n[\sin(2\chi(t))]&= -i o_n J_n(z), \\
\mathcal{F}_n[\sin(2\chi(t))\cos(2\chi(t))]&= -i \frac{o_n}{2} J_n(2z),\\
\mathcal{F}_n[\cos(2\chi(t))^2]&= \frac{1}{2}\left[\delta_{n0}+ e_n J_n(2z)\right],\\
\mathcal{F}_n[\sin(2\chi(t))^2]&= \frac{1}{2}\left[\delta_{n0}- e_n J_n(2z)\right],
\end{align}
so that the Fourier components of the Lindblad generator in the rotating frame read
\begin{align}
\tilde{\mathcal{L}}_n &= \mathcal{L}(H_n, {d}_n), 
\label{eq:L-fourier-rotfr}
\end{align}
with
\begin{align}
 H_n = \frac{J_n(z)}{2} \left(e_n \sigma_z  - i o_n \sigma_y \right)
\end{align}
and
\begin{align}
{d}_n&= \gamma
\begin{pmatrix}
\delta_{n0} & i e_n J_n(z) & -o_n J_n(z)\\
-i e_n J_n(z)  & \frac{\delta_{n0}+e_n J_n(2z)}{2} & \frac{i}{2} o_n J_n(2z) \\
o_n J_n(z) & \frac{i}{2} o_n J_n(2z) &  \frac{\delta_{n0}-e_n J_n(2z)}{2}
\end{pmatrix}.
\end{align}
(Note that each of the individual Fourier-components $\tilde{\mathcal{L}}_n$ can be brought to Lindblad form simply by the multiplication with a suitable phase factor.)

\section{High-frequency expansion: Rotating frame}
\label{sec:FL-Highfreq-rotfr}

Let us now perform both types of high-frequency expansion in the rotating frame of reference. 

\subsection{Magnus expansion in the rotating frame}
\label{sec:Magnus-rotfr-two-level system}
%For our model system, 
%Here we perform the Magnus expansion of the generator in the rotating frame and find a generator that is a valid Lindbladian
%already in the lowest order of the high-frequency expansion, in the strong contrast to the high-frequency expansions in the direct frame (that we addressed  in Section~\ref{sec:Magnus}).
%Whether and how this procedure can be generalized to more complex systems and for systems 
%where the dissipative part of the generator is driven remains an open question.

\subsubsection{First order Magnus expansion in the rotating frame}
\label{sec:Magnus-rotfr-0}

\begin{figure}
	\centering
	\subfloat[]{
		%\begin{minipage}{0.3\textwidth}
		\includegraphics[scale=0.65]{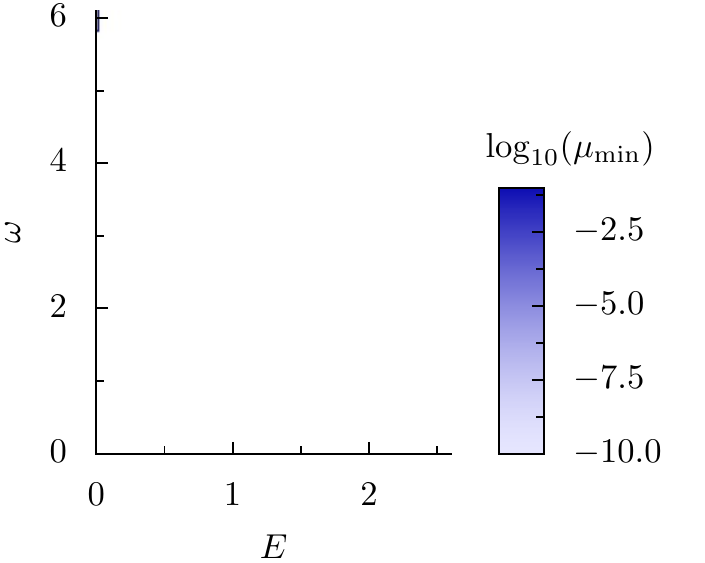}
		%\end{minipage}
	}
	\subfloat[]{
		%\begin{minipage}{0.3\textwidth}
		\includegraphics[scale=0.65]{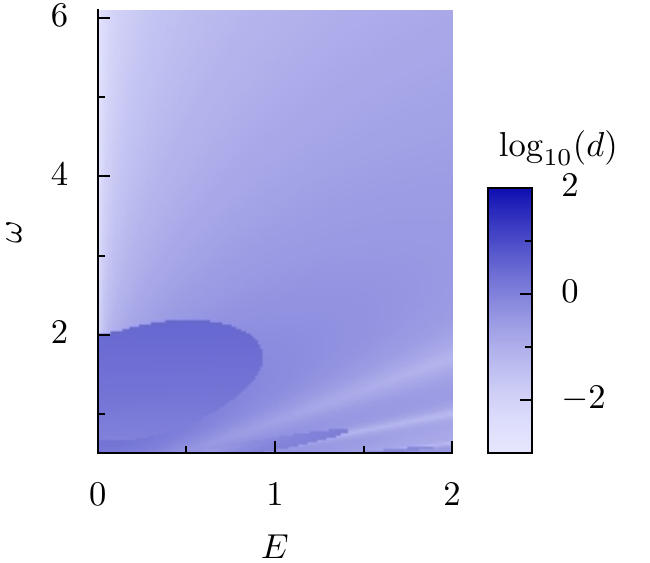}
		%\end{minipage}
	}
	
	\caption{(a)~Distance to  Markovianity $\mu_\mathrm{min}$  of the Floquet generator $\tilde{\mathcal{K}}_{\mathrm{Mag},1}$($=\tilde{\mathcal{K}}_{\mathrm{eff},1}=\tilde{\mathcal{K}}_{\mathrm{vV},1}$) obtained with the first-order Magnus expansion
		in the rotating frame for the same model and parameter $\gamma=0.01$ as in Fig.~\ref{fig:phases-exact}(a). The generator is a valid Lindbladian for  for all parameters $(E, \omega)$. (b)~Distance (the Frobenius norm) $d=\vert\vert \tilde{\mathcal{K}} - \mathcal{K}_{x_0}\vert \vert_F$ between the generator 
		$\tilde{\mathcal{K}}$ obtained by the first-order Magnus expansion $\tilde{\mathcal{K}}_{\mathrm{Mag},1}$
		in the rotating frame and  the candidate $ \mathcal{K}_{x_0} \in \mathrm{log}(\mathcal{P}(T))/T$
		for the Floquet-Lindbladian $\mathcal{L}_F$ of branch ${x_0}$, which is 
		closest to the valid Lindbladian generator.}
	\label{fig:dist-Mag-rotfr}
\end{figure}

The lowest order of the Magnus expansion in the rotating frame reads
\begin{align}
\tilde{\mathcal{K}}_{\mathrm{Mag,1}} &= \tilde{\mathcal{L}}_0 = \mathcal{L}(H_{\mathrm{Mag,1}} , {d}_{\mathrm{Mag,1}}),
\label{eq:FloqLind-rotframe}
\end{align}
with
\begin{align}
 H_{\mathrm{Mag,1}} = \frac{J_0(z)}{2} \sigma_z
 \end{align}
 and
 \begin{align}
{d}_{\mathrm{Mag,1}} &= \gamma \begin{pmatrix}
1 & i  J_0(z) & 0\\
-i J_0(z)  & \frac{1}{2}[1+ J_0(2z)] & 0\\
0& 0 &  \frac{1}{2}[1- J_0(2z)]
\end{pmatrix}.\end{align}
where, again,  $z=2E/\omega$.
Note that for $z \to 0$, i.e.~for $E\to 0$ or $\omega \to \infty$ (such that $J_0(z)\to 1$) %and $J_0(2z)\to 1$ 
we recover the static Hamiltonian and dissipator, as expected.
In Fig.~\ref{fig:dist-Mag-rotfr}(b) we plot the distance of the matrix representation of the superoperator of this approximation $\tilde{\mathcal{K}}$ to  the exact Floquet generator and see a much better agreement than what one finds for the lowest order in the direct frame [cf.~Fig.~\ref{fig:char-pol-Magnus}(a)], especially for smaller values of $\omega$. This is expected because the transformation to the rotating frame integrates out the driving term which corresponds to a partial resummation of infinitely many orders in $E /\omega$, here entering via the nonlinear function Bessel function $J_0$. In the direct frame, however, the leading order correction in the Magnus expansion only captures terms up to order $(E /\omega)^2$.

The eigenvalues of the coefficient matrix ${d}_{\mathrm{Mag,1}}$ read 
\begin{align}
\lambda_{1/2} &= \gamma \left[\mu(z) \pm \sqrt{\mu(z)^2+J_0(z)^2- \frac{1}{2}[1+ J_0(2z)] } \right],\label{eq:eigval-1stord-1} \\ 
\lambda_3 &= \frac{\gamma}{2} [1- J_0(2z)],
\end{align}
with $\mu(z)= [3+ J_0(2z)] / 4$.
The corresponding generator is a valid Lindbladian generator only if all three eigenvalues are non-negative. This is generally the
case, since
\begin{align}
&J_0(z)^2 - \frac{1}{2}[1+ J_0(2z)] \\
&=  J_0(z)^2  - \frac{1}{2} \sum_{k \in \mathbb{Z}} J_k(z) ^2  - \frac{1}{2}  \sum_{k\in \mathbb{Z}} J_{k}(z)J_{-k}(z)\\
%&= J_0(z)^2 - \frac{1}{2}  \sum_{k\in \mathbb{Z}} J_{2k}(z)^2 + \frac{1}{2}  \sum_{k\in \mathbb{Z}} J_{2k+1}(z)^2- \frac{1}{2}  \\
&= J_0(z)^2 -  \sum_{k\in \mathbb{Z}} J_{2k}(z)^2 = -  \sum_{k\neq 0} J_{2k}(z)^2
\leq  0,
\end{align}
In the first step we have used the identity $J_n(y+z)= \sum_{k \in \mathbb{Z}} J_k(y) J_{n-k}(z)$ and that $1= \sum_{k \in \mathbb{Z}} J_k(z) ^2$. %$=  \sum_{k \in \mathbb{Z}} J_{2k}(z) ^2 +  \sum_{k \in \mathbb{Z}} J_{2k+1}(z) ^2 $.
This shows that the values that the square root in Eq.~\eqref{eq:eigval-1stord-1} takes will be smaller than $\mu(z)$. Therefore,
the first order expansion in the rotating frame 
produces a nontrivial generator $\tilde{\mathcal{K}}_{\mathrm{Mag,1}}$  that is a valid Lindbladian for all parameter values [Fig.~\ref{fig:dist-Mag-rotfr}(a)].

When comparing the result that we obtain in the rotating frame, Eq.~\eqref{eq:FloqLind-rotframe},
to the one that we obtain when directly performing the Magnus expansion, Eq.~\eqref{eq:FloqLind-Magnus}, we 
find that by expanding the Bessel function to second order, $J_0(z) \approx  1-z^2/4$, by using $z=2\varepsilon$ we recover the terms
$\propto \varepsilon^2$ in Eq.~\eqref{eq:FloqLind-Magnus}, while the terms $\propto \varepsilon/\omega$ will be found in the next order of the rotating-frame Magnus expansion. 

%However, if one eliminates the terms $\propto \varepsilon/\omega$ in Eq.~\eqref{eq:FloqLind-Magnus}, only keeping the terms $\propto \varepsilon^2$,
%still the resulting generator is not a valid Lindbladian. Thus, to find a valid Lindbladian generator one needs to know the higher order terms in $1/\omega$, but those are hard to extract in the direct frame 
%because orders higher than three in the Magnus expansion
%are cumbersome to compute.

\subsubsection{Second order Magnus expansion in the rotating frame}
\label{sec:Magnus-rotfr-1}
The second order term of the rotating-frame Magnus expansion  reads
\begin{align}
\tilde{\mathcal{K}}^{(2)} &= i \sum_{n>0} \frac{\left[\tilde{\mathcal{L}}_n, \tilde{\mathcal{L}}_{-n}\right]+ \left[\tilde{\mathcal{L}}_0, \tilde{\mathcal{L}}_n - \tilde{\mathcal{L}}_{-n}\right]}{n\omega} 
=  \sum_{n>0} 2 o_n  \frac{\left[\tilde{\mathcal{L}}_0, i \tilde{\mathcal{L}}_n \right]}{n\omega},
\end{align}
where in the second step we have used that for the Fourier components in Eq.~\eqref{eq:L-fourier-rotfr} we have $\tilde{\mathcal{L}}_{-n} = (-1)^n \tilde{\mathcal{L}}_{n}$.

By employing the general expressions derived in Appendix \ref{sec:app-comm-lind}, 
%It is straight-forward to compute  
we find that for odd $n$ 
\begin{align}
&\left[\tilde{\mathcal{L}}_0, i \tilde{\mathcal{L}}_n \right] = \mathcal L(H_n, {d}_n),  
\end{align}
with
\begin{align}
 H_n = -\frac{J_0(z)J_n(z)}{2} \sigma_x, 
\end{align}
and
\begin{align}
 \quad {d}_n = \frac{\gamma}{2} \begin{pmatrix}
0 &  0 & f_n(z)\\
0  &0 & -4i J_0(z)J_n(z) \\
f_n(z) & 4i J_0(z)J_n(z) & 0
\end{pmatrix},
\end{align}
where $f_n(z) = J_n(z)[1+J_0(2z)] +J_n(2z)J_0(z)$. Moreover, we ignored terms of second or higher order in  $\gamma$. Thus, up to second order, the Magnus expansion in the
rotating frame reads
\begin{align}
\tilde{\mathcal{K}}_{\mathrm{Mag,2}} = \mathcal{L}(H_{\mathrm{Mag,2}}, {d}_{\mathrm{Mag,2}}),
\label{eq:FloqLind-rotframe-ord1}
\end{align}
with 
\begin{align}
 H_{\mathrm{Mag,2}} =J_0(z) \left[ \frac{1}{2} \sigma_z - {\frac{\nu(z)}{\omega}} \sigma_x \right]
\end{align}
and 
 \onecolumngrid
 \begin{align}
 {d}_{\mathrm{Mag,2}} = \gamma \begin{pmatrix}
1 & i  J_0(z) &  \frac{1}{\omega} [{\nu(z)(1+J_0(2z))+J_0(z) \nu(2z)}]\\
-i J_0(z)  & \frac{1}{2}[1+ J_0(2z)] & - \frac{4i}{\omega}J_0(z) \nu(z)\\
\frac{1}{\omega}[{\nu(z)(1+J_0(2z))+J_0(z) \nu(2z)}]&  \frac{4i}{\omega} J_0(z) \nu(z)&  \frac{1}{2}[1- J_0(2z)]
\end{pmatrix},
\end{align}
\twocolumngrid 
\noindent where we have introduced $\nu(z) = \sum_{n>0} [o_n J_n(z)/n]$. Since in leading order order $\nu(z) \simeq z/2$, we also recover the terms $\propto \varepsilon/\omega$ in Eq.~\eqref{eq:FloqLind-Magnus}. 

\begin{figure}
	%\centering
	\subfloat[]{
	\includegraphics[scale=0.59]{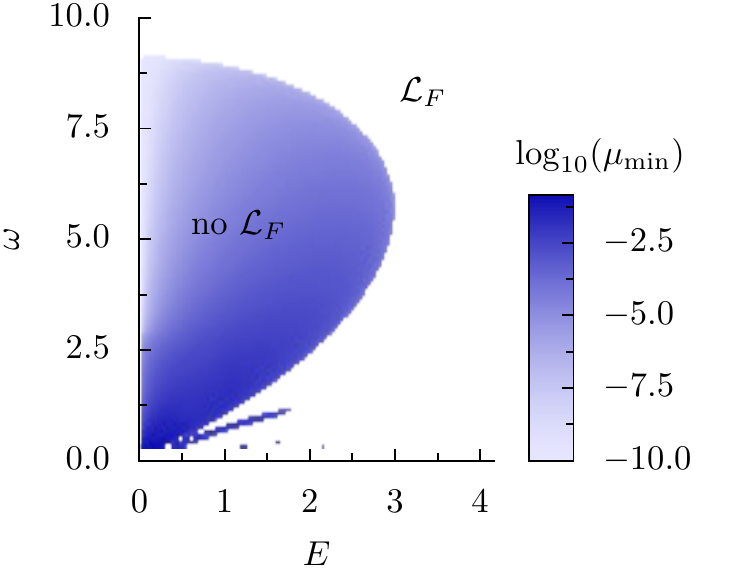}
	}\hspace{-0.3cm}
	\subfloat[]{
		%\begin{minipage}{0.3\textwidth}
		\includegraphics[scale=0.59]{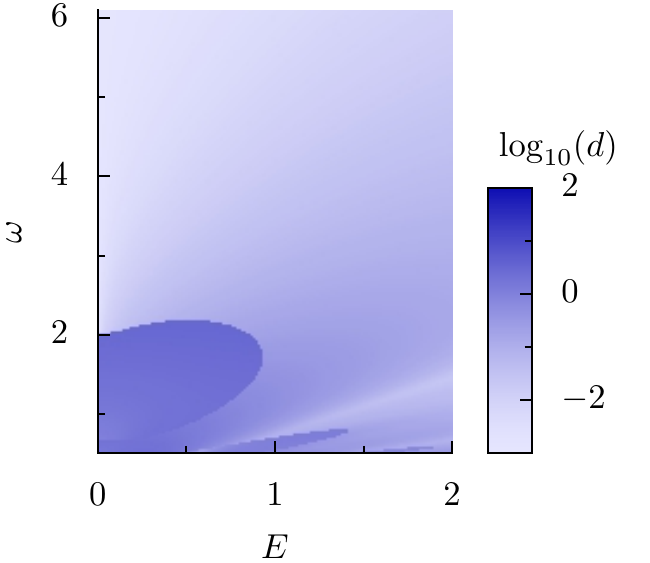}
		%\end{minipage}
	}
	\caption{(a)~Distance to  Markovianity $\mu_\mathrm{min}$  of the Floquet generator $\tilde{\mathcal{K}}_{\mathrm{Mag},2}$ obtained with the second-order Magnus expansion in the rotating frame
		for the same model and parameter $\gamma=0.01$ as in Fig.~\ref{fig:phases-exact}.  Note that we only calculate distances for
	    $\omega\geq 0.3$, values below this are drawn in white. (b)~Matrix distance $d$ 
of the candidate $\tilde{\mathcal{K}}_{\mathrm{Mag},2}$ to the candidate $\mathcal{K}$ obtained from the logarithm of $\mathcal{P}(T)$.}
	\label{fig:rotfr-1st-order}
\end{figure}

In Fig.~\ref{fig:rotfr-1st-order}(b) we show the distance of the matrix representation of the superoperator of $\tilde{\mathcal{K}}_{\mathrm{Mag,2}}$ to  the exact Floquet generator and see a small improvement compared to the first-order result in Fig.~\ref{fig:dist-Mag-rotfr}(b). 
However, the distance from Markovianity, which is plotted in Fig.~\ref{fig:dist-Mag-rotfr}(a), acquires qualitatively different behaviour in second order. While the Floquet generator was always Markovian (i.e.\ of Lindblad form) in first order, in second order we can now distinguish parameter regions, where it is of Lindblad type, from others, where it is not. Remarkably, the map shown in Fig.~\ref{fig:rotfr-1st-order}(a) resembles very much the exact phase diagram of Fig.~\ref{fig:phases-exact}(a). Namely, we can clearly observe a lobe-shape region, where the Floquet generator is non-Markovian. While this region is larger than in the exact phase diagram, the transition between Lindbladian and non-Lindbladian Floquet generator is qualitatively captured correctly by the Floquet-Mangus expansion. Only at very low frequencies, where we cannot expect the high-frequency expansion to provide meaningful results, we find as an artifact a thin non-Markovian stripe, which is not present in the exact phase diagram. 

%In contrast to the lowest order of the expansion, in this order the Floquet generator is not a valid Lindbladian generator for all parameters $E, \omega$.
%Similar to Section~\ref{sec:driven-qbit-model}, we check whether the Floquet generator is a valid Lindblad generator by testing for conditional complete positivity and
%in case that this fails we compute the distance $\mu_\mathrm{min}$  to Markovianity.
%We show this distance in Fig.~\ref{fig:rotfr-1st-order}(a) and in this order it resembles  already  very much the structure that one
%obtains by directly computing the logarithm of the map $\mathcal{P}(T)$. Nevertheless, the ear-shaped structure extends to larger values of $E$ and $\omega$ than it does for the exact generator.

\subsection{Van-Vleck high-frequency expansion in the rotating frame}
\label{sec:rotfr-extsp}

After having seen that, starting from the rotating frame of reference, the Magnus expansion qualitatively reproduces the exact 
phase diagram, let us now also evaluate the leading orders of the van-Vleck expansion. Different, however, from the previous 
section, where we were able to derive analytic expressions for the Magnus expansion, here calculations get quite involved and 
so we treat this expansion numerically. 
For this purpose, it is convenient to first discuss the action of the transformation to the rotating frame, $\Lambda(t)$ in the extended Floquet space.

%
%
%
%As we have seen in Section \ref{sec:ext-space}, the aim of the van-Vleck  high-frequency expansion 
%is to find a transformation $\mathcal{D}(t)$ that `removes' the micromotion leaving 
%an evolution with the time-independent effective generator $\mathcal{K}_\mathrm{eff}$.
%This transformation $\mathcal{D}$  can be regarded as  a generalized gauge transformation much like the generalized rotating-frame
%transformation $\Lambda$. It is therefore
%instructive to represent $\Lambda$ in the extended Hilbert space. By doing so, we obtain
%a more systematic approach to the high-frequency expansion in the rotating frame.

\subsubsection{Floquet-space formalism}

%As Eq.~\eqref{eq:gaugetrafo-micromotion} suggests, 
Both the rotating-frame transformation $\Lambda(t)$ and the micromotion  $\mathcal{D}(t)$
are generalized gauge transformations. Instead of finding $\mathcal{D}(t)$
directly, however, we may first perform a transformation to the rotating frame, $
\tilde{\varrho}(t) = \Lambda^{-1}(t) \varrho(t)$, and then find the micromotion transformation there.
%\begin{align}
%\tilde{\mathcal{P}}(t,t_0)=\tilde{\mathcal{D}}(t)\exp[(t-t_0) {\mathcal{L}}_\text{eff}]\tilde{\mathcal{D}}^{-1}(t_0).
%\end{align}
%By using that ${\mathcal{P}}(t,t_0)=\Lambda(t)\tilde{\mathcal{P}}(t,t_0)\Lambda^{-1}(t_0)$ we find that
%\begin{align}
%{\mathcal{P}}(t,t_0)=\Lambda(t)\tilde{\mathcal{D}}(t)\exp[(t-t_0){\mathcal{L}}_\text{eff}]\tilde{\mathcal{D}}^{-1}(t_0)\Lambda^{-1}(t_0).
%\end{align}
Since $\Lambda(t)$ is periodic, we have
\begin{align}
\Lambda(t)= \sum_{n} e^{in\omega t} \Lambda_n, 
\end{align}
so we also may represent it in extended space $\bar{\Lambda}_{nm}= \Lambda_{n-m}$.
Note that this representation is only possible since we assume the driving term $\mathcal{L}_d(t)$ to commute with itself at 
different times, so that no time-ordering is needed. As a result $\Lambda(t)$ is a time-local, and thus also time-periodic,  superoperator. %\cmt{I reformulatd the last sentence. Please check it!}

As a result, in the rotating frame the generalized quasi-energy operator reads
 \begin{align}
\bar{ \tilde{\mathcal{Q}}}=\bar{\Lambda}^{-1} \bar{{\mathcal{Q}}}\bar{\Lambda}.
 \end{align}
Like in the direct frame,  the goal is to find a transformation  $\tilde{\mathcal{D}}$ such that
  \begin{align}
 \bar{{\mathcal{Q}}}' =\bar{ \tilde{\mathcal{D}}}^{-1}\bar{ \tilde{\mathcal{Q}}}\bar{ \tilde{\mathcal{D}}}
 \end{align}
 where $\bar{{\mathcal{Q}}}'$ is block diagonal.
 
 With respect to the original frame of reference, the micromotion operator is given by the combination 
 \begin{align}
 {\mathcal{D}}(t)=\Lambda(t)\tilde{\mathcal{D}}(t).
 \end{align}
From this expression, we can once more directly see that for strong driving the high-frequency expansion in the direct frame will at least have a slow convergence only.
% In this limit, it is expected that $\bar{ \tilde{\mathcal{Q}}}$ will only be slightly disturbed form being block diagonal $\bar{ \tilde{\mathcal{Q}}} \approx \bar{ \mathcal{Q}}_0+ \mathcal{O}(\omega^0)$ with $\bar{\mathcal{Q}}_0 \propto \omega^1$, therefore
% the  contribution of $\tilde{\mathcal{D}}(t)$ to $ {\mathcal{D}}(t)$ will be small only.
Namely, the transformation $\Lambda$ involves a summation of infinitely many terms in $E/\omega$.
%, so $\Lambda$ is not perturbative  in the driving strength.
% However, a high-frequency expansion in the direct frame tries to capture the dynamics of $\tilde{\mathcal{D}}(t)$ and $\Lambda(t)$ on an equal footing 
% from which problems are arising.
 %and thus will be  dominated by the dynamics of the rotating frame transformation $\Lambda(t)$.
 
In the regular (non-extended) superoperator-space, the rotating-frame quasienergy operator, reads
 \begin{align}
 \tilde{\mathcal{Q}}(t) = i\tilde{\mathcal{L}}(t) - i \partial_t .
\end{align}
Here the time-periodic Lindbladian generator in the rotating frame, $\tilde{\mathcal{L}}$, is given by Eq.~\eqref{eq:Ltilde-rotfr}. 
Its Fourier-components $\tilde{\mathcal{L}}_{n}$ are directly related to its Floquet-space representation, 
 \begin{align}
\bar{\tilde{\mathcal{L}}} = \bar{\Lambda}^{-1} \bar{{\mathcal{L}_0}}\bar{\Lambda} \quad \text{ i.e. }  \quad  \bar{\tilde{\mathcal{L}}}_{nm}= \tilde{\mathcal{L}}_{n-m} = \sum_{k} \Lambda^{-1}_{n-k} \mathcal{L}_0 \Lambda_{k-m}, 
\end{align}
which allows for their efficient numerical calculation. To this end let us determine the coefficients $\Lambda_n$. In Appendix \ref{sec:app-rotfr-fourier} we show that for 
driving terms of the form
\begin{align}
\mathcal{L}_d(t) = \phi(t) \mathcal{L}_d', 
\label{eq:scalar-drivingterm-Lindbl}
\end{align}
with scalar function $ \phi(t) = \sum_{m\neq 0} e^{im\omega t} \phi_m$,
one finds the explicit Floquet-space expression:
\begin{align}
\label{eq:Lambda_n-expl}
\bar{ \Lambda} 
&= \prod_{m\neq0} \bar{f}^{(m)}\left(\frac{\phi_m \mathcal{L}_d'}{im\omega}\right)\bar{g}^{(m)}\left(\frac{\phi_m \mathcal{L}_d'}{im\omega}\right).
%\label{eq:Lambda-rotfr-trafo-expl}
\end{align}
Here we have introduced $\bar{f}^{(m)}_{nl}= f^{(m)}_{n-l}$, $\bar{g}^{(m)}_{nl}= g^{(m)}_{n-l}$ as well as
\begin{align}
f^{(m)}_n(x) &= 
\left\{\begin{array}{cc} 
J_{k}(x) &\text{ if } n = k m, k \in \mathbb{Z},\\
0  &  \text{ else. }
\end{array} \right. \\
 g^{(m)}_n(x) 
&= 
\left\{\begin{array}{cc} 
e^{-x} I_{k}(x) &\text{ if } n = k m, k \in \mathbb{Z},\\
0  &  \text{ else, }
\end{array} \right.
\end{align}
with Bessel functions of first kind, $J_k$, and modified Bessel functions of first kind, $I_k$,. 
%taken at some matrix argument (which may be evaluated most easily by performing a spectral decomposition of  $\mathcal{L}_d'$).
Since $\Lambda^{-1}(t)$ is directly obtained from $\Lambda(t)$  by setting $\phi(t) \rightarrow -\phi(t)$, we find $\bar \Lambda^{-1}$ from Eq.~\eqref{eq:Lambda_n-expl}
by setting  $\phi_m \rightarrow -\phi_m$.

For our example system we have
\begin{align}
\phi(t) = {2} \cos(\omega t), \qquad  \mathcal{L}_d' = \mathcal{L}_1 = \mathcal{L}_{-1} = -i \left[\frac{E}{2} \sigma_x, \cdot\right]
\end{align}
From Eq.~\eqref{eq:Lambda_n-expl} (or an explicit calculation) we find
\begin{align}
\Lambda_n = J_{n}\left( \frac{2 \mathcal{L}_1}{i \omega}\right),
\end{align}
which finally yields 
\begin{align}
\label{eq:L_n-matrx-expl}
\tilde{\mathcal{L}}_{n} = \sum_{k} J_{n-k}\left( -\frac{2 \mathcal{L}_1}{i \omega}\right)\mathcal{L}_0 J_{k}\left( \frac{2 \mathcal{L}_1}{i \omega}\right). 
\end{align}
By translating superoperators into $N^2\times N^2$-dimensional matrices
as shown in Appendix \ref{sec:app-matrep-two-level system}, we therefore have an alternative procedure to  the one we 
obtained in Section \ref{sec:rotfr-Ln-expl} to
calculate the operators $\tilde{\mathcal{L}}_{n}$ and from this the van-Vleck  high-frequency  expansion.
An explicit calculation of $\tilde{\mathcal{L}}_{n}$ using this matrix representation is given in Appendix \ref{sec:app-two-level system-expl}. [Plugging this result into the first order of the Magnus expansion, Eq.~\eqref{eq:Magnus-zeroth}, one recovers 
 $\tilde{\mathcal{K}}_\mathrm{Mag,1}$ of Section \ref{sec:Magnus-rotfr-0}.]

Equation \eqref{eq:L_n-matrx-expl} is a good starting point for numerical
investigations, because it can be evaluated easily, after having represented the superoperators $\mathcal{L}_0, \mathcal{L}_1$ by $N^2\times N^2$-dimensional matrices.
From the expressions in Section \ref{sec:Floq-Magn-dissi} we can then compute the terms of the van-Vleck high-frequency  expansion in the rotating frame. 
We compute both the approximate effective generator,  $\tilde{\mathcal{K}}_{\mathrm{eff},n}=\sum_{k=1}^{n} \tilde{\mathcal{K}}_\mathrm{eff}^{(k)}$, as well as the approximate micromotion operator $\tilde{\mathcal{D}}_n(t)= \exp(\sum_{k=1}^{n} \mathcal G_k(t))$. Note that for the latter, the expansion of the exponent is truncated, rather than that of the full exponential function. For isolated systems, this makes sure that the micromotion operator is unitary also in finite orders of the approximation \cite{EckardtAnisimovas15}. Combining both approximations, we can compute the $n$th order approximation to the Floquet generator 
\begin{align}
\tilde{\mathcal{K}}_{\mathrm{vV}, n} = \tilde{\mathcal{D}}_{n-1}(0)  \tilde{\mathcal{K}}_{\mathrm{eff},n} \tilde{\mathcal{D}}_{n-1}^{-1}(0).
\label{eq:vVFG}
\end{align}
Here we only need to consider the micromotion correction up to the order of $n-1$, since all terms contained in $\tilde{\mathcal{K}}_{\mathrm{eff},n} $ are of order one or higher. The approximation (\ref{eq:vVFG}) is generally different from the one obtained from the truncated Magnus expansion in the rotating frame. If, instead, we had expanded and truncated $\tilde{\mathcal{D}}_n(t)$ directly, rather than its exponent, we would have recovered the Magnus approximation. 

\subsubsection{First order van-Vleck  high-frequency expansion in the rotating frame}
Note that from comparing Eq.~\eqref{eq:Magnus-zeroth} to Eq.~\eqref{eq:Keff-zeroth} we learn that in the leading first order (i.e.\ zeroth order in $1/\omega$), the van-Vleck high-frequency expansion
of the effective generator $\tilde{\mathcal{K}}_{\mathrm{eff}}$ and the Magnus expansion $\tilde{\mathcal{K}}_{\mathrm{Mag}}$ coincide, 
$\tilde{\mathcal{K}}_{\mathrm{eff,1}}=\tilde{\mathcal{K}}_{\mathrm{Mag, 1}}$,
therefore in first order also the effective generator exists for all parameter values.
%We will have a more detailed discussion of the van-Vleck high-frequency expansion in the rotating frame in the following Sec.~\ref{sec:rotfr-extsp}.
Additionally, in leading (zeroth) order the micromotion operator is simply the identiy,  $\tilde{\mathcal{D}}_{0}(0)=\mathbf{1}$, so that in leading (first) order, the Floquet generator is equal to the effective generator, 
\begin{equation}
\tilde{\mathcal{K}}_{\mathrm{vV}, 1} = \tilde{\mathcal{K}}_{\mathrm{eff},1}= \tilde{\mathcal{K}}_\mathrm{Mag,1}.
\end{equation}
%Furthermore, % as we discussed in Section \ref{sec:Magnus-rotfr-0} already, 
%on first order one also finds that the effective generator coincides with the corresponding order of the Magnus expansion,
Thus, in the rotating frame for the first-order van-Vleck Floquet generator $\tilde{\mathcal{K}}_{\mathrm{vV}, 1}$ both the distance to Markovianity as well as the distance from the exact Floquet generator are identical to the ones shown in Fig.~\ref{fig:dist-Mag-rotfr}.  In particular,  $\tilde{\mathcal{K}}_{\mathrm{vV}, 1}$ is of Lindblad type in the whole parameter plane $(E, \omega)$ [cf.~Fig.~\ref{fig:dist-Mag-rotfr}(a)].  

\subsubsection{Second order van-Vleck  high-frequency expansion in the rotating frame}
From the second order on, the truncated van-Vleck expansion for the Floquet generator, $\tilde{\mathcal{K}}_{\mathrm{vV}, n}$,   
deviates both from the effective generator $\tilde{\mathcal{K}}_{\mathrm{eff},n}$ and from the truncated Magnus expansion of the Floquet generator, $\tilde{\mathcal{K}}_{\mathrm{Mag},n}$.
However, since for our model we have $\tilde{\mathcal{L}}_{-n} = (-1)^n \tilde{\mathcal{L}}_{n}$ and, therefore, 
$[\tilde{\mathcal{L}}_{n}, \tilde{\mathcal{L}}_{-n}]=0$, the second-order contribution to the effective generator vanishes, so that 
\begin{equation}
\tilde{\mathcal{K}}_{\mathrm{eff,2}}=\tilde{\mathcal{K}}_{\mathrm{eff,1}}. 
\end{equation}
Thus, the only new contribution to the Floquet generator 
\begin{equation}
\tilde{\mathcal{K}}_{\mathrm{vV}, 2} = \tilde{\mathcal{D}}_1(0)\tilde{\mathcal{K}}_{\mathrm{eff,2}}\tilde{\mathcal{D}}_1^{-1}(0)
= \tilde{\mathcal{D}}_1(0)\tilde{\mathcal{K}}_{\mathrm{eff,1}}\tilde{\mathcal{D}}_1^{-1}(0)
\end{equation}
 stems from the micromoton operator $\tilde{\mathcal{D}}_1(0)$. 
 
\begin{figure}
	\centering
	\subfloat[]{\includegraphics[scale=0.6]{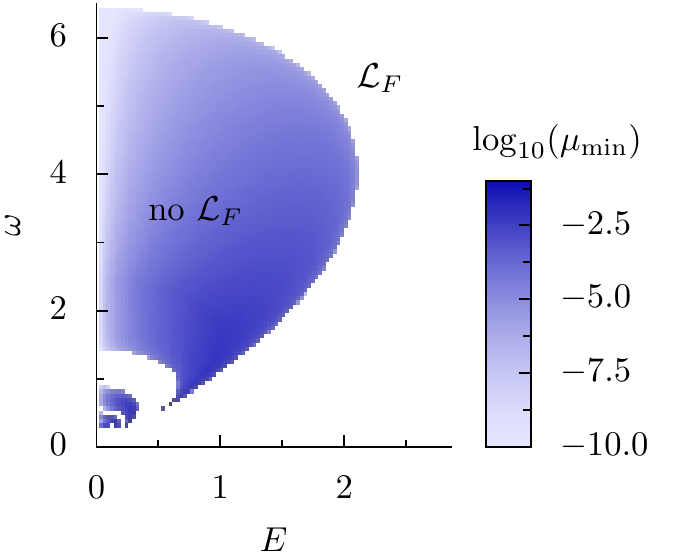}}
	%\hspace{1cm}
	\subfloat[]{
		\includegraphics[scale=0.6]{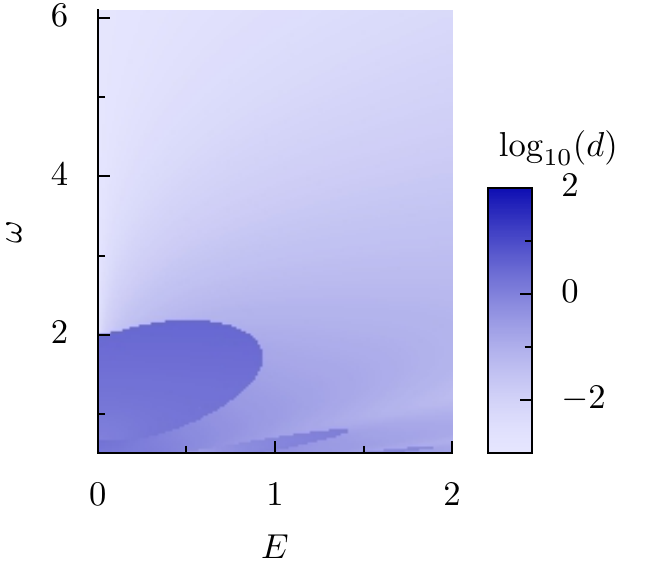}
	}
	\caption[Distance to Markovianity and distance to the exact generator for the second order $\tilde{\mathcal{K}}_{\mathrm{vV},2}$ of the van-Vleck  high-frequency  expansion in the rotating frame]{(a)~Distance to  Markovianity $\mu_\mathrm{min}$  of the Floquet generator ${\mathcal{K}}$ obtained with the second-order van-Vleck  high-frequency  expansion $\tilde{\mathcal{K}}_{\mathrm{vV},2}$ 
		in the rotating frame, where we do not expand the exponential in $\tilde{\mathcal{D}}(t)=\exp(\tilde{\mathcal{G}}(t))$.
	We present the same model and parameter $\gamma=0.01$ as in Fig.~\ref{fig:phases-exact}(a). Note that we only calculate distances for
	$\omega\geq 0.3$, values below this are drawn in white. 
	 (b)~Distance $d$ 
of the candidate $\tilde{\mathcal{K}}_{\mathrm{vV},2}$ to the exact candidate $\mathcal{K}$ obtained from the logarithm of $\mathcal{P}(T)$.}
	\label{fig:rotfr-1st-order-withD}
\end{figure}

In Fig.~\ref{fig:rotfr-1st-order-withD}, we plot the distance from Markovianity (a) [as well as the distance from the exact Floquet 
generator for $\tilde{\mathcal{K}}_{\mathrm{vV,2}}$ (b)].  Apart from some artifacts at very low frequencies, we find a lobe-shaped non-Markovian region, where no Floquet-Lindbladian can be found. Thus, like the Magnus expansion, also the 
van-Vleck expansion explains the structure of the exact phase diagram shown in Fig.~\ref{fig:phases-exact}(a). However, the phase boundaries obtained within the second-order van-Vleck approximation [Fig.~\ref{fig:rotfr-1st-order-withD}(a)] are closer to the exact ones [Fig.~\ref{fig:phases-exact}(a)] than those obtained with the Magnus expansion [Fig~\ref{fig:dist-Mag-rotfr}(a)].

Apart from providing a quantitatively better approximation to the exact results, the van-Vleck expansion has another (and more important) advantage compared to the Magnus expansion. Namely, it disentangles effects that result from the micromotion, which are contained in  $\tilde{\mathcal{D}}(t_0)$, from those contained in the $t_0$-independent effective generator $ \tilde{\mathcal{K}}_{\mathrm{eff}}$. Since $\tilde{\mathcal{K}}_{\mathrm{eff}, 2}$ is Markovian in the whole parameter plane $(E, \omega)$, we can now clearly see that for our model system the origin of the region with non-Markovian Floquet generator lies (entirely) in the non-unitary micromotion. While this statement is obtained from a second-order high-frequency van-Vleck expansion only, the very good agreement with the exact phase diagram, strongly suggests that this statement remains true also beyond this approximation. This is confirmed also by the third order van-Vleck approximation, which is discussed below. 
Note that the phase diagram will not be changed further, when transforming from the rotating to the direct frame of reference, because both are related by a unitary transformation for our model system, since the driving term is hermitian.

The relation of regions with non-Markovian Floquet generator with the non-unitary micromotion of the system, is consistent also 
with the strong dependence of the phase diagram on the driving phase, which is equivalent to a variation of the time $t_0$, with $t_0=\varphi T/2\pi$. Compare Figs.~\ref{fig:phases-exact}(a) and (b) corresponding to $\varphi=0$ and $\varphi=\pi/2$, respectively or the subfigures of Fig.~\ref{fig:Phases-func-phi}. In order to explain why the non-Markovian lobe in the phase diagram is largest for $\varphi=t_0=0$ and shrinks with increasing $\varphi$, until it finds its smallest extent for $\varphi = \pi/2$ or $t_0=T/4$, let us inspect the first-order van-Vleck approximation of the micromotion operator (which describes the role of the micromotion in the second-order approximation of the Floquet generator). It reads
\begin{align}
 \tilde{\mathcal{D}}_1(t_0) &= \exp\left(-i\sum_{n\neq 0} e^{in\omega t_0} \frac{\tilde{\mathcal{L}}_n}{n\omega} \right)\\
 %&= \exp\left(-i\sum_{k=1}^{\infty} 2\cos(k\omega t_0) \frac{\tilde{\mathcal{L}}_k}{k\omega} \right),
 &= \exp\left(-i\sum_{k=1}^{\infty}2 \left[o_k \cos(k\omega t_0) + e_k  i \sin(k\omega t_0) \right] \frac{\tilde{\mathcal{L}}_k}{k\omega} \right),
 \label{eq:D_1-tilde-expl}
\end{align}
where in the second step we have employed that for our model $\tilde{\mathcal{L}}_{-n}=(-1)^n\tilde{\mathcal{L}}_n$.
When the exponent of this expression becomes small, the micromotion operator approaches the identity, which describes a 
unitary rotation that does not induce any non-Markovian behavior. The largest contribution to the exponent stems from the $k=1$ term, which vanishes precisely when $t_0=T/4$ corresponding to the driving phase $\varphi=\pi/2$ at which the non-Markovian region is smallest. Thus, the van-Vleck expansion provides analytical insight into the origin of the phase dependence of the phase diagram. 

\begin{figure}
	\centering
	\includegraphics[scale=0.8]{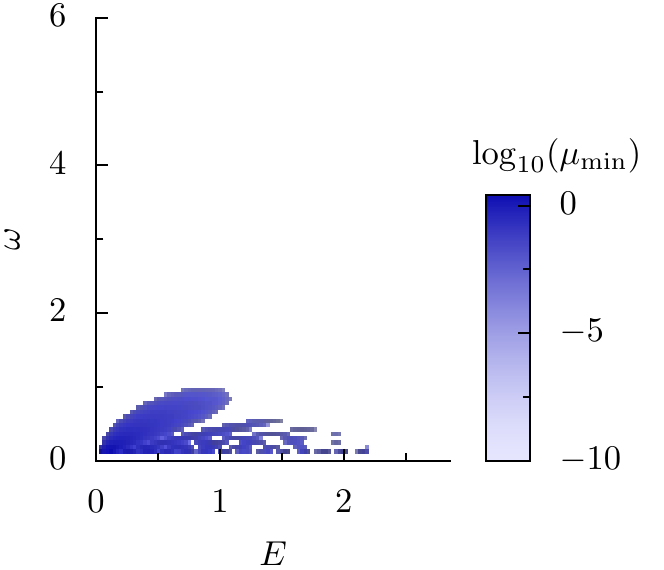}
	\caption[Distance to  Markovianity for the  effective generator $\tilde{\mathcal{K}}_{\mathrm{eff},3}$ obtained with the third-order van-Vleck  high-frequency  expansion]{Distance to  Markovianity $\mu_\mathrm{min}$  of the candidate $\tilde{\mathcal{K}}_{\mathrm{eff},3}$ for the effective generator obtained from a  third order van-Vleck  high-frequency  expansion
		in the rotating frame
		for the same model and parameter $\gamma=0.01$ as in Fig.~\ref{fig:phases-exact}.  We only calculate for
		$\omega\geq 0.1$, values below this are drawn in white.}
	\label{fig:Leff-rotfr-2}
\end{figure}

\subsubsection{Third order van-Vleck  high-frequency expansion in the rotating frame}
In order to support the conclusions drawn from the second-order van-Vleck expansion in the previous section, let us now briefly 
disucuss the third order. Calculating numerically the effective generator $\tilde{\mathcal{K}}_{\mathrm{eff},3}$, in 
Figure~\ref{fig:Leff-rotfr-2} we show the resulting distance from Markovianity. Apart from artifacts appearing at very small 
frequencies, $\tilde{\mathcal{K}}_{\mathrm{eff},3}$ is of Lindblad form essentially everywhere in the parameter plane  
$(E, \omega)$.  This confirms that non-Markovian behavior must be an effect of the micromotion.

It is interesting to see that the high-frequency expansion is able to capture the transition between the two phases and it is remarkable that rather good agreement with the exact phase diagram is found also down to quite low frequencies. But for very low frequencies, eventually also qualitative deviations from the exact result become visible. This is not surprising, since the high-frequency expansion cannot be expected to converge in this regime. For the Magnus expansion (and thus also for the van-Vleck expansion), convergence is guaranteed as long as  \cite{BlanesEtAl09,MoanNiesen2008}
\begin{align}
 \int_0^{T} \vert\vert \mathcal{L}(t)\vert \vert_2 \mathrm{d}t < \pi.
\end{align}
Here, $\vert \vert A \vert\vert_2 = \max_{\vert \vert x \vert\vert_2=1} \vert \vert Ax \vert\vert_2$
is the induced 2-norm. 

We can gain a very rough estimate for the region of convergence
by discussing the undriven limit of $E=0$ and $\gamma=0$. As shown in Appendix
\ref{sec:app-two-level system-expl}, the matrix representation of the generator then reads
$\mathcal{L}\vert_{E=0, \gamma=0}=\mathrm{diag}(0,-i,i,0)$, therefore  $\vert \vert (\mathcal{L}\vert_{E=0, \gamma=0})\vert \vert_2=1$. Thus, for $E=0$ and $\gamma=0$ we find that the Magnus expansion is only expected to converge for $\omega > 2$. For finite values of the driving strength $E$
the norm of $\mathcal{L}(t)$ will increase and thus the radius of convergence  
will decrease even further.

As a result, Figure \ref{fig:Leff-rotfr-2} shows that within the region of convergence of the Magnus expansion, $\tilde{\mathcal{K}}_{\mathrm{eff},3}$ is a valid Lindbladian. Our hypothesis, that the effective Lindbladian could exist for all parameters, is therefore not violated by the third order of the van-Vleck high-frequency expansion in the rotating frame.

%\subsection{Concluding remarks on the high-frequency expansions}
\section{Summary and outlook}
In this paper, we have studied the description of a time-periodically driven open quantum system using high-frequency expansions (Magnus- or van-Vleck-type). In particular, we have focused on the resulting approximations for the effective time-independent Floquet generator, which is defined so that it describes the stroboscopic evolution of the system in steps of the driving period. Our work is generally motivated by the interesting perspective to apply the concepts of Floquet engineering also to open quantum systems. More specifically, it was initiated by a discrepancy that arose from two observations: On the one hand we found in previous work that the Floquet generator of a simple open periodically driven Markovian two-level system is of Lindblad type in the high-frequency regime \cite{SchnellEtAl18FL}. %\cmt{ref}.
On the other hand it was pointed out that the Floquet generator resulting from a high-frequency expansion is generally not of Lindblad type \cite{ReimerEtAl18, HaddadfarshiEtAl15,MizutaEtAl21}.  We have found that high-frequency expansions can correctly describe the behaviour of the system, when applied in a rotating frame of reference. Moreover, by going beyond the leading first order, the high-frequency expansion can even explain the transition to another regime, where the Floquet generator is not of Lindblad type. By isolating the effect of the micromotion within the van Vleck approach, this transition can be attributed entirely to the properties of the non-unitary micromotion of the system, and its dependence on the driving phase can be explained. 

Our analysis emphasizes that the approach that some recent works \cite{MizutaEtAl21} take to argue about the non-existence of a Floquet Lindbladian in an interacting system on the basis of the performance of  high-frequency expansions might not be conclusive.
%Our results emphasize that the existence of the Floquet Lindbladian is a nontrivial question and that the approach that some works \cite{MizutaEtAl21} take to argue in general about the non-existence of a Lindbladian generator in interacting systems just from performing high-frequency expansions might not be suited to the problem.

We hope that our results will 
%proof useful for the description and the control of periodically driven open quantum systems and that they will 
stimulate further research of periodically driven open quantum systems. Since  we focused on a specific model, it is, for instance, a very natural question under what conditions our findings can be generalized to other models. For instance, to  the case of non-Markovian completely positive stroboscopic evolution when  time-dependent rates $\gamma_i(t)$ can become negative ~\cite{AddisEtAl14,SiudzinskaChruscinski2020}. 
Such behaviour may arise from a microscopic derivation of the equation of motion of a (Floquet) system coupled to a heat bath \cite{BastidasEtAl18, ScopaEtAl19} and  is typically neglected in the Floquet-Born-Markov secular formalism~ \cite{KohlerEtAl97,GrifoniHaenggi98}. 
%Whether our results can be used to improve this  formalism is an interesting question.
%\textcolor{blue}{
Applying our approach to such microscopically derived master equations is, therefore, another interesting perspective.%}

%\textcolor{blue}{
Finally, it is an open question, whether the observation that the origin of the non-Markovianity of the Floquet generator lies in the micromotion, which was made here based on a high-frequency expansion of a specific model, generalizes to all or a subclass of time-periodically Markovian  quantum systems.%}

%Is there a universal link between  non-Markovianity of a periodically modulated  open quantum system and the  micromotion or what are the conditions under which a rotating frame correctly capturing (non)Markovianity of the Floquet generator can be found, are intriguing issues to explore.}

\section{Acknowledgments}\label{acknowledgment}
S.D. acknowledges support by the Russian Science Foundation through Grant No. 19-72-20086 and A.S.\ and A.E.\ by the German Research Foundation (DFG) via the Research Unit FOR 2414 (project No. 277974659).  A part of this work is a result of the activity of the MPIPKS Advanced Study Group ``Open quantum systems far from equilibrium''.

\hide{

\newpage
\begin{center}
	OLD STUFF
\end{center}
\newpage

\subsection{TODO: Do we want? Second order Floquet-Magnus expansion in the rotating frame}

\subsection{Driven two-level system in the extended Hilbert space} 
In the extended Hilbert space we may represent the eigenvalue equation,  Eq.~\eqref{eq:extended-space}, for the driven dissipative two-level system as
%\onecolumngrid
\begin{align}
\Omega_a \begin{pmatrix}
\dots \\
\ket{\Phi_{a,-1}} \\ \ket{\Phi_{a,0}} \\
\ket{\Phi_{a,1}} \\ \dots
\end{pmatrix} = 
\begin{pmatrix}
\dots & \\
& A &B - \omega \,  \mathbf{1} & A\\
&& A & B & A\\
&& & A &B + \omega \,  \mathbf{1} &A\\
&&& && &\dots
\end{pmatrix}
\begin{pmatrix}
\dots \\
\ket{\Phi_{a,-1}} \\ \ket{\Phi_{a,0}} \\
\ket{\Phi_{a,1}} \\ \dots
\end{pmatrix}
\label{eq:eval-extended}
\end{align}
%\twocolumngrid
with
\begin{align}
A=\i \mathcal{L}_1  = \i \mathcal{L}_{-1}= \frac{E}{2} \begin{pmatrix}
0 & - 1 & \phantom{-}1 & 0  \\
-1 & 0 & 0 & \phantom{-}1 \\
\phantom{-}1&  0 & 0 & -1 \\
0 & \phantom{-}1 & -1 & 0
\end{pmatrix}
\text{ and } 
B=\i \mathcal{L}_0 =  \begin{pmatrix}
-4 \i \gamma & 0 & 0 & 0  \\
0 & 1 - 2 \i \gamma & 0 & 0 \\
0&  0 & -1 - 2 \i \gamma & 0 \\
4 \i \gamma & 0 & 0 & 0
\end{pmatrix}.
\end{align}
Since the matrix $B$ is not hermitian we will also need the left eigenvectors for which it holds
\begin{align}
\Omega_a \begin{pmatrix}
\dots \\
\bra{\tilde\Phi_{a,-1}} \\ \bra{\tilde\Phi_{a,0}} \\
\bra{\tilde\Phi_{a,1}} \\ \dots
\end{pmatrix}^T = 
\begin{pmatrix}
\dots \\
\bra{\tilde\Phi_{a,-1}} \\ \bra{\tilde\Phi_{a,0}} \\
\bra{\tilde\Phi_{a,1}} \\ \dots
\end{pmatrix}^T
\begin{pmatrix}
\dots & \\
& A &B - \omega \,  \mathbf{1} & A\\
&& A & B & A\\
&& & A &B + \omega \,  \mathbf{1} &A\\
&&& && &\dots
\end{pmatrix}.
\label{eq:eval-extended-left}
\end{align}

In order to extract the corresponding effective generator we have to diagonalize the problem.
To find the Floquet generator we may first use Eq.~\eqref{eq:timeev-rho} to express
\begin{align}
\ket{\varrho_{a}(T)} = e^{-\i\Omega_{a}T} \,  \ket{\Phi_a(T)} &= e^{-\i\Omega_{a}T} \,  \ket{\Phi_a(0)}
= e^{-\i\Omega_{a}T} \,  \sum_{n\in \mathbb{Z}} \ket{\Phi_{a,n}}.
\end{align}
To find the evolution for any given initial density matrix with corresponding vector $\ket{\varrho(0)}$, we
have to expand it in the Floquet states $\ket{\varrho(0)}=\sum_a c_a  \ket{\Phi_a(0)}$. The expansion coefficients are found by projection 
$c_a = \braket{\tilde\Phi_a(0)}{\varrho(0)}$ with the left Floquet states, $\bra{\tilde\Phi_a(t)} = \sum_{n\in \mathbb{Z}} e^{-\i\omega n t} \, \bra{\tilde\Phi_{a,n}}$.
Therefore it holds
\begin{equation}
\ket{\varrho(T)} =  \sum_a e^{-\i\Omega_{a}T}  \ket{\Phi_{a}(0)}\braket{\tilde\Phi_a(0)}{\varrho(0)}.
\end{equation}
This allows for identifying the one-period time-evolution superoperator $\mathcal{P}_F$,
which is defined as  $\varrho(T) =  \mathcal{P}_F{\varrho(0)}$, such that in the 
vector notation we have
\begin{equation}
\mathcal{P}_F =  \sum_a e^{-\i\Omega_{a}T}  \ket{\Phi_{a}(0)}\bra{\tilde\Phi_a(0)}.
\end{equation}
%The Floquet-Lindbladian is a superoperator for which it holds
%\begin{equation}
%	 \mathcal{P} = e^{\mathcal{L}_F T}.
%\end{equation}
As it was discussed in Section~\ref{sec:intro}, there are in principle infinitely many 
candidates of this Floquet-Lindbladian, which result from the freedom to pick a branch for 
all the eigenvalues $\Omega_{a}$ with nonzero real part. Here, let us
restrict ourselves to the principal branch
\begin{equation}
\mathcal{K}_{0} = \sum_a -\i\Omega_{a}  \ket{\Phi_{a}(0)}\bra{\tilde\Phi_a(0)}.
\end{equation}
Note again that we restrict ourselves to solutions in the strip $-\omega/2 \leq \mathrm{Re} \, {\Omega_a} <  \omega/2$.

\section{Expansion of the generator in extended space}
\label{sec:expan-ext-sp}

Here we show that for a 
driving with a single harmonic function, the expansion of the generator in the rotating frame that was performed in Section~\ref{sec:Magnus-rotfr} is equivalent 
to a perturbative expansion in the extended space, where the constant part of the Lindbladian is taken as a small perturbation to the driven part. 
Remarkably, this procedure
does not rely on a transformation to a rotating frame and should also be applicable in presence of a driving of the dissipative terms with a single harmonic.

We can use the fact that in this extended space the eigenvalue equation obeys the structure 
\begin{align}
\Omega \begin{pmatrix}
\dots \\
{\phi_{-1}} \\ {\phi_{0}} \\
{\phi_{1}} \\ \dots
\end{pmatrix} = 
\begin{pmatrix}
\dots & \\
&a& b- \omega & a\\
&& a & b & a\\
&& & a & b+ \omega  & a\\
&&& && &\dots
\end{pmatrix}
\begin{pmatrix}
\dots \\
{\phi_{-1}} \\ {\phi_{0}} \\
{\phi_{1}} \\ \dots
\end{pmatrix}
\end{align}
which for the case that $a$ and $b$ are scalar is solved 
by the transformation
\begin{align}
\kket{\phi_n} = \sum_{k\in \mathbb{Z}} J_{k-n}\left(-\frac{2a}{\omega}\right) \kket{k}.
\end{align}
Here $\kket{k}$ is the $k$-th vector of the standard basis of the extended Hilbert space and $J_k$
is the $k$-th Bessel function. The
corresponding eigenvalues are 
\begin{align}
\Omega_n = b + n \omega.
\end{align}
It turns out that  this transformation also works in the case where $A$ and $B$
are matrices, however only if these matrices do commute. 

In our case, Eq.~\eqref{eq:eval-extended}, the matrices do not commute.
However, in the limit $E, \omega \gg 1, \gamma$ the terms in the matrix $B$ can be assumed to be a small perturbation to the problem
\begin{align}
\Omega_a \begin{pmatrix}
\dots \\
\ket{\Phi_{a,-1}} \\ \ket{\Phi_{a,0}} \\
\ket{\Phi_{a,1}} \\ \dots
\end{pmatrix} = 
\underbrace{\begin{pmatrix}
	\dots & \\
	& A &0 - \omega \,  \mathbf{1} & A\\
	&& A & 0 & A\\
	&& & A &0 + \omega \,  \mathbf{1} &A\\
	&&& && &\dots
	\end{pmatrix}}_{\i \mathfrak{L}^{(0)}} 
\begin{pmatrix}
\dots \\
\ket{\Phi_{a,-1}} \\ \ket{\Phi_{a,0}} \\
\ket{\Phi_{a,1}} \\ \dots
\end{pmatrix}
\end{align}
which is solved by the eigenvalues $\Omega^{(0)}_{an}=0+n\omega$ (with a four-fold degeneracy)
and the corresponding eigenvectors
\begin{align}
{\kket{\phi^{(0)}_{an}}} = \sum_{k\in \mathbb{Z}} J_{k-n}\left(-\frac{2A}{\omega}\right) \ket{\phi^{(0)}_a} \otimes \kket{k},\\
{\bbra{\tilde\phi^{(0)}_{an}}} = \sum_{k\in \mathbb{Z}}   \bbra{k}\otimes \bra{\tilde\phi^{(0)}_a}J_{k-n}\left(-\frac{2A}{\omega}\right).
\label{eq:ev-nonpert}
\end{align}
Here the vectors $\ket{\phi^{(0)}_a}$ are any basis of the four components
on which the matrix $A$ acts, and we have used that $A^\dagger=A$. Note that the matrix-valued Bessel function can be defined via its power series
\begin{align}
J_k(X) = \sum_{r=0}^\infty \frac{(-1)^r (\frac{X}{2})^{2r+k}}{(k+r)!r!}.
\end{align}
Since in our case the matrix $A$ is hermitian and can be diagonalised, $A=U D U^\dagger$,  it can be seen  from the power series that it also
holds $J_k(-2A/\omega)=U J_k(-2D/\omega) U^\dagger$, i.e.~we can reduce the matrix Bessel function to applying the scalar Bessel function to the eigenvalues,
which gives a practical procedure to evaluate this function.

With this transformation, it holds 
\begin{align}
%\begin{split}
\i \mathfrak{L}^{(0)} {\kket{\phi^{(0)}_{an}}} =&\sum_{k\in \mathbb{Z}} \left[A J_{k-1-n}\left(-\frac{2A}{\omega}\right)
+ (0+k \omega \mathbf{1}) J_{k-n}\left(-\frac{2A}{\omega}\right)+
A J_{k+1-n}\left(-\frac{2A}{\omega}\right) \right]\ket{\phi^{(0)}_{a}} \otimes \kket{k}.
%\end{split}
\end{align}
Like in the scalar case at this step one may use the identity $x[J_{k-1}(x)+J_{k+1}(x)]= 2k J_{k}(x)$ but with $x$ as a matrix, which 
is true because $x$ is diaganolisable [the matrix identity can then be shown by transforming to the diagonal basis].

We find a high-frequency expansion by performing non-hermitian time independent perturbation theory
in extended Hilbert space by writing
\begin{align}
\i \mathfrak{L}=\i \mathfrak{L}^{(0)}+\lambda \i \mathfrak{L}^{(B)} \text{ with } \i \mathfrak{L}^{(B)} = B \otimes \sum_{k\in \mathbb{Z}}\kket{k}\bbra{k}.
\end{align}
Here $\lambda$ is a perturbative parameter which we introduce by hand to keep track of the order of the expansion. We will set $\lambda=1$
in the end. The order of $\lambda$ simply counts the powers of $B$, so strictly speaking we are performing  an expansion in powers of $B$.
\subsection{First order expansion}
Note that since the eigenvalues of $\i \mathfrak{L}^{(0)}$ are degenerate we have to perform degenerate perturbation theory.
To this end we first need to diagonalise the matrix 
\begin{align}
&{\bbra{\tilde{\phi}^{(0)}_{bn}}} \i \mathfrak{L}^{(B)} {\kket{\phi^{(0)}_{an}}}
= \bra{\tilde{\phi}^{(0)}_{b}}\sum_{k\in \mathbb{Z}} J_{k-n}\left(-\frac{2A}{\omega}\right)B J_{k-n}\left(-\frac{2A}{\omega}\right)\ket{\phi^{(0)}_{a}}
\equiv \delta_{ab} \, \Omega^{(1)}_{an},
\label{eq:def-evec-0}
\end{align}
which fixes the left and right eigenvectors, $ \bra{\tilde{\phi}^{(0)}_{a}}$ and $\ket{\phi^{(0)}_{a}}$, that should be used in Eq.~\eqref{eq:ev-nonpert}.
However, as we will see in the following, to obtain the effective generator in first order of $\lambda$ these eigenvectors do not have to be computed explicitly.
Also this gives the first correction to the eigenvalues $\Omega_{an}= \Omega^{(0)}_{an} + \Omega^{(1)}_{an}$.
Note that these eigenvectors are independent of the branch $n$ since we can shift the index $k\rightarrow k+n$ in the summation.
In the following it will be quite useful to introduce the definition
\begin{align}
\tilde{B}_n &= \tilde{B}_n(A, \omega)= \sum_{k\in \mathbb{Z}} J_{k-n}\left(-\frac{2A}{\omega}\right)B J_{k}\left(-\frac{2A}{\omega}\right).
%\tilde{B}_n &= \sum_{k\in \mathbb{Z}} \tilde{B}^k_n.
\end{align}
Let us emphasize that it therefore holds by definition
\begin{align}
\tilde{B}_0 = \sum_a \Omega^{(1)}_{an} \ket{\phi^{(0)}_{a}}  \bra{\tilde{\phi}^{(0)}_{a}}.
\label{eq:spectral}
\end{align}

This concludes the discussion of the first order in $\lambda$ already, because for reconstructing the generator we
need to consider the $n=0$ branch only where $\Omega^{(0)}_{a0}=0$ and therefore corrections
to the eigenvectors do not contribute to the first order in $\lambda$. Hence
\begin{equation}
\mathcal{K}_{0} = \sum_a -\i (\Omega^{(0)}_{a0} + \lambda \Omega^{(1)}_{a0})  \ket{\phi^{(0)}_{a}(0)}\bra{\tilde\phi^{(0)}_a(0)} + \mathcal{O}(\lambda^2).
\end{equation}
Here
\begin{align}
\ket{\phi^{(0)}_{a}(0)}&=\sum_{k\in \mathbb{Z}} \bbrakket{k}{\phi^{(0)}_{a0}} = \sum_{k\in \mathbb{Z}}J_{k}\left(-\frac{2A}{\omega}\right)\ket{\phi^{(0)}_{a}}=\ket{\phi^{(0)}_{a}},
\end{align}
where  we have used $\sum_{k\in \mathbb{Z}}J_{k}(x)=1$ which again also holds in the matrix case.
Therefore it holds  that
\begin{equation}
\mathcal{K}_{0} = \lambda \sum_a -\i \Omega^{(1)}_{a0}  \ket{\phi^{(0)}_{a}}\bra{\tilde\phi^{(0)}_a}+ \mathcal{O}(\lambda^2).
\end{equation}
which is %by definition 
(up to a factor of $-\i$) the spectral representation, Eq.~\eqref{eq:spectral}, of the matrix that we encountered in Eq.~\eqref{eq:def-evec-0}.
In conclusion, the first order term in the expansion is
\begin{equation}
\mathcal{K}_{0} =-\i \lambda \tilde{B}_0+ \mathcal{O}(\lambda^2).
\end{equation}
In Appendix \ref{sec:app-two-level system-expl}, we  compute this term analytically for the driven two-level system system, Eq.~\eqref{eq:Lindblad-two-level system}, and observe that this first order expansion
reproduces exactly the result that we obtained from the lowest order Magnus expansion in the rotating frame, Eq.~\eqref{eq:FloqLind-rotframe}.
Therefore the effective generator is a valid Lindbladian
for all $(E, \omega)$.

\subsection{Second order expansion}
\begin{figure}
	\includegraphics{Phases-2nd-Order}
	\caption{Distance to  Markovianity $\mu_\mathrm{min}$  of the candidate  $\mathcal{K}_{0}$  for the Floquet-Lindbladian obtained from second order expansion
		for the same model and parameter $\gamma=0.01$ as in Fig.~\ref{fig:phases-exact}. Note that the Floquet-Lindbladian that we obtain in first order of the expansion
		is a valid Lindbladian for all parameters $(E, \omega)$.}
	\label{fig:second-order}
\end{figure}

The first correction to the eigenvectors reads
\begin{align}
{\kket{\phi^{(1)}_{an}}} &= \sum_{m\neq n, b} {\kket{\phi^{(0)}_{bm}}} \frac{{\bbra{\tilde{\phi}^{(0)}_{bm}}} \i \mathfrak{L}^{(B)} {\kket{\phi^{(0)}_{an}}}}{(n-m)\omega}
%&=\sum_{m\neq n} \sum_{k\in \mathbb{Z}}  \frac{J_{k-m}\left(-\frac{2A}{\omega}\right)B J_{k-n}\left(-\frac{2A}{\omega}\right)}{(n-m)\omega} \ket{\phi^{(0)}_{a}}\otimes \kket{k}.
=\sum_{m\neq n}\sum_{k\in \mathbb{Z}}  J_{k-m}\left(-\frac{2A}{\omega}\right) \frac{\tilde B_{m-n}}{(n-m)\omega} \ket{\phi^{(0)}_{a}}\otimes \kket{k}.
\end{align}
Note that since the denominator is independent of index $b$, in the second step we have used $\sum_b \ket{\phi^{(0)}_{b}}\bra{\tilde{\phi}^{(0)}_{b}}=\mathbf{1}$.
Similarly, 
\begin{align}
{\bbra{\tilde\phi^{(1)}_{an}}} %=\sum_{m\neq n} \sum_{k\in \mathbb{Z}}  \bbra{k} \otimes  \bra{\tilde\phi^{(0)}_{a}} \frac{J_{k-n}\left(-\frac{2A}{\omega}\right)B J_{k-m}\left(-\frac{2A}{\omega}\right)}{(n-m)\omega}.
=\sum_{m\neq n}  \bbra{k} \otimes  \bra{\tilde\phi^{(0)}_{a}} \frac{\tilde B_{n-m}}{(n-m)\omega} J_{k-m}\left(-\frac{2A}{\omega}\right).
\end{align}
Also we need the second order of the eigenvalues,
\begin{align}
\Omega^{(2)}_{an} &= \sum_{m\neq n, b} \frac{ \bbra{\tilde{\phi}^{(0)}_{an}} \i \mathfrak{L}^{(B)} {\kket{\phi^{(0)}_{bm}}} {\bbra{\tilde{\phi}^{(0)}_{bm}}} \i \mathfrak{L}^{(B)} {\kket{\phi^{(0)}_{an}}}}{(n-m)\omega}
= \sum_{m\neq n} \frac{ \bra{\tilde{\phi}^{(0)}_{a}} \tilde B_{n-m} \tilde B_{m-n}{\ket{\phi^{(0)}_{a}}}}{(n-m)\omega}.
\end{align}
Altogether, we find that in second order of $\lambda$ the candidate for the Floquet-Lindbladian reads 
\begin{align}
\mathcal{K}_{0} =&-\i \lambda \tilde{B}_0 -\i  \lambda^2 \sum_a \left[ \Omega^{(2)}_{a0}  \ket{\phi^{(0)}_{a}}\bra{\tilde\phi^{(0)}_a} \right.\\
& +\left. \Omega^{(1)}_{a0}  \biggl(\ket{\phi^{(1)}_{a}(0)}\bra{\tilde\phi^{(0)}_a} + \ket{\phi^{(0)}_{a}}\bra{\tilde\phi^{(1)}_a(0)}\biggr)\right]+ \mathcal{O}(\lambda^3).
\end{align}
Let us evaluate 
\begin{align}
\ket{\phi^{(1)}_{a}(0)}&=\sum_{k\in \mathbb{Z}} \bbrakket{k}{\phi^{(1)}_{a0}} = - \sum_{m\neq 0} \frac{\tilde B_{m}}{m\omega}\ket{\phi^{(0)}_{a}},
\end{align}
where we again use $\sum_{k\in \mathbb{Z}}J_{k}(x)=1$.
Together with the spectral identity, Eq.~\eqref{eq:spectral}, this yields
\begin{align}
\mathcal{K}_{0} =-\i \lambda \tilde{B}_0 +\i  \lambda^2  &\sum_{m\neq 0} \left[ \sum_{a}  \ket{\phi^{(0)}_{a}} \frac{ \bra{\tilde{\phi}^{(0)}_{a}} \tilde B_{-m} \tilde B_{m}{\ket{\phi^{(0)}_{a}}}}{m\omega} \bra{\tilde\phi^{(0)}_a} %\right.\\
%& + \left.  
+\frac{\tilde B_{m}}{m\omega} \tilde B_0 + \tilde B_0  \frac{\tilde B_{-m}}{m\omega} \right]+ \mathcal{O}(\lambda^3).
\end{align}
We have numerically computed this order for our driven two-level system model and find that it almost reproduces the result
that we obtained by the first order Magnus expansion in the rotating frame Eq.~\eqref{eq:FloqLind-rotframe-ord1}, where
the small difference of both results stems from the fact that in Eq.~\eqref{eq:FloqLind-rotframe-ord1} we have higher orders in $\gamma$.

It is intriguing to see that this perturbative expansion in the static part of the Lindbladian
exactly reproduces the results of the Magnus expansion in the rotating frame. This is remarkable, because in general
it is very hard to compute higher order terms in the Magnus expansion (e.g.~in Appendix \ref{sec:app-discrep-magnus} we
discuss that  already on the second order Magnus expansion the general expressions in terms of the Fourier components of the generator
that are presented in the literature are not consistent), while going to higher orders in the perturbative expansion
seems straight-forward. }

\appendix

\section{Finding the Floquet generator from the exact map $\mathcal{P}(T)$}
\label{sec:app-Wolf-method}

Here, we summarize the results of Ref.~\cite{SchnellEtAl18FL} concerning the question of the existence of a Floquet Lindbladian.
In the time-periodically modulated isolated system, i.e.~in our notation Eq.~\eqref{eq:tdm} with $\gamma_i(t)=0$ for all $i$, it is well known that there always exists
an effective time-independent Hamiltonian $H_F$, the Floquet Hamiltonian, such that
\begin{align}
\mathcal{P}(T) = \exp\left(-{i} \left[H_F, \cdot \right] T \right).
\end{align}
%This  Floquet Hamiltonian is subject to many theoretical studies, mainly due to the fact that it can be used as an experimental tool 
%to create interesting dynamics that would not be present in the autonomous system. This technique has been coined Floquet engineering.
%Apart from that, the notion of a Floquet Hamiltonian is also a very powerful tool in nuclear magnetic resonance spectroscopy \cite{LeskesEtAl10}.
How can one see that such a Floquet Hamiltonian $H_F$ exists?
For the coherent dynamics, the evolution operator
reduces to a unitary rotation of the density matrix
\begin{align}
\mathcal{P}(T) =U(T) \cdot  U(T)^\dagger.
\end{align}
The unitary one-cycle evolution operator $U(T)$,
\begin{align} \label{eq:timev-op-closed}
U(T) = \mathcal{T} \exp\big[-{i}\int_{0}^T\!\rd t'H(t')\big]
\end{align}
%is divisible (any root of it is a unitary operator) and 
yields a countably infinite set of Hermitian generators, $H_{U,\{x_1,...,x_N\}}$, $x_a \in \mathbb{Z}$,
$U(T) = e^{-iH_{U}T}$, parametrized by a choice of a branch of the  logarithm $\log U(T)$. 
This can be seen most easily by representing the evolution operator $U(T)$, Eq.~\eqref{eq:timev-op-closed},
in its spectral decomposition. Since it is unitary we 
may represent it as
\begin{align}
U(T) = \sum_{a=1}^N e^{-i \varepsilon_a T} P_a
\end{align}
with real numbers $\varepsilon_a$ and (Hermitian) orthogonal projectors $P_a$
onto the eigenspace $a$.
Now it becomes apparent that, when computing the logarithm of $U(T)$,
for every subspace $a$ there is a freedom to pick a branch of the complex logarithm 
giving a whole set
\begin{align}
\mathrm{log}\left[U(T)\right]_{\{x_1,...,x_N\}} = -i  \sum_{a=1}^N \left(\varepsilon_a T + 2\pi x_a \right) P_a.
\end{align}
parameterized by $N$ integer numbers $x_a \in \mathbb{Z}$.
For the corresponding Hermitian generator, 
\begin{align}
H_{U, \{x_1,...,x_N\}} =  \sum_{a=1}^N \left(\varepsilon_a  +\omega x_a \right) P_a,
\end{align}
this change of branch corresponds to a redefinition of the `energy' $\varepsilon_a \rightarrow \varepsilon_a + \omega x_a$,  where $\omega=2\pi/T$ is the driving frequency. That means, the `energies' $\varepsilon_a$ are only defined 
up to integer multiples of $ \omega$, which is 
why they are typically  referred to as \emph{quasi-energies}. Note that in the case of the coherent dynamics, any of 
these generators can be chosen as Floquet Hamiltonian $H_F$,
since all of the generators $H_{U, \{x_1,...,x_N\}}$ are Hermitian.  This choice can be made, e.g, by using the principal branch, $\forall x_s \equiv 0$, or the branch 
closest to the time-averaged Hamiltonian $\overline{H(t)}$. 

Since $\mathcal{P}(T)$ is a hermiticity-preserving map, its spectrum is invariant under complex  
conjugation. Thus, its $N^2$ eigenvalues are either real or appear as complex conjugated pairs (we denote 
the numbers of real eigenvalues and complex pairs by $n_r$ and $n_c$, respectively). %number of these pairs $n_c$).
The Jordan normal form of the map $\mathcal{P}(T)$ can thus be represented as
\begin{align}
\mathcal{P}(T) =  \sum_{r=1}^{n_r} \lambda_r P_r +  \sum_{c=1}^{n_c}  \left(\lambda_c P_c +  \lambda_c^* P_{c*}\right),
\label{eq:map-jordan-form}
\end{align}
where $\lambda_r$ are the real eigenvalues, $\lambda_c, \lambda_c^*$ the pairs of complex eigenvalues, and $P_x$ the corresponding (not necessarily Hermitian)
orthogonal projectors on the corresponding subspaces.

Again, due to the nature of the complex logarithm, the Floquet generator $\mathcal{K}$ in Eq.~\eqref{eq:generator-cand} is not uniquely defined, but for every branch 
of the logarithm we get a different operator. 
A straight-forward procedure to test whether a given candidate $\mathcal{K}$ is a valid Lindblad generator is the Markovianity test proposed by Wolf \emph{et al.}\
in Refs.~\cite{WolfEtAl08,Cubitt2012}, which is based on two conditions:
(i) The operator $\mathcal{K}$  must perserve Hermiticity, i.e.
$$\mathcal{K}\sigma=\mathcal{K}\sigma^\dagger$$ for all $\sigma \in L(\mathcal{H})$ that are Hermitian, $\sigma=\sigma^\dagger$.
%Therefore the spectrum of superoperator $\mathcal{K}$ should be invariant under complex conjugation. 
%This can be checked by making sure that the eigenvalues of $\mathcal{P}_F$ do not coincide with the negative real axis.
(ii)  For the second test,  the operator $\mathcal{K}$ has to be
\textit{conditionally completely positive} \cite{WolfEtAl08}, i.e.~it has to fulfill
\begin{align}
\Phi_{\perp} \mathcal{K}^\Gamma\Phi_{\perp} \geq 0.
\label{eq:test-cond-comp}
\end{align}
Here $\Phi_{\perp} = \id - \ket{\Phi}\bra{\Phi}$ is the projector on the orthorgonal complement of the maximally entangled state 
$\ket{\Phi}=\sum_{i=1}^N \left(\ket{i} \otimes \ket{i}\right)/\sqrt{N}$ with $\lbrace\ket{i}\rbrace$ denoting the canonical basis of ${\mathcal{H}}$. 
Moreover, $\mathcal{K}^\Gamma = N (\mathcal{K} \otimes \id)[\ket{\Phi}\bra{\Phi}] \in L(\mathcal{H}^2)$ is the Choi matrix of $\mathcal{K}$.
 If one of the branches of the operator logarithm obeys both conditions it can be called Floquet Lindblaian $\mathcal{L}_F$. 
Already here the contrast with the unitary case becomes apparent: it is not guaranteed that such branch exists and if it exists, the other branches do typically not provide a Lindbladian Floquet generator as well.

%There is no need to inspect the different branches to check 
Condition (i) simply demands that the spectrum of
the candidate $\mathcal{K} $ has to be invariant under complex conjugation. This means, in turn, that the 
spectrum of the map $\mathcal{P}(T)$ should not contain negative real eigenvalues $\lambda_r =- \vert\lambda_r\vert$ (strictly speaking, there must be no negative eigenvalues of odd degeneracy). That is, because if one would set
the logarithm of such an occasion e.g.~to $\log(\lambda_r) = i\pi +  \log(\vert\lambda_r\vert)$, the spectrum is not invariant under conjugation anymore. In this case there is no Floquet Lindbladian. 
%Therefore, condition (i) requires that the spectrum of  for the Floquet-Lindbladian  is invariant under complex conjugation, 

If $\mathcal{P}(T)$ has no negative real eigenvalues, we find that we may represent the family of all
candidates $\mathcal{K}_{ \{x_1,...,x_{n_c}\}} $ as
\begin{align}
\mathcal{K}_{ \{x_1,...,x_{n_c}\}} = \mathcal{K}_0+ i \omega  \sum_{c=1}^{n_c}  x_c \left( P_c -   P_{c*}\right),
\label{eq:candidates-LF}
\end{align}
where $\mathcal{K}_0$ is the generator that follows from the principal branch of the logarithm of $\mathcal{P}(T)$.
We have the freedom to pick integer numbers $\mathbf{x}=\{x_c\}\in \mathbb{Z}^{n_c}$ that determine the branch 
of the logarithm for every pair of complex eigenvalues. Note that
for the isolated system all eigenvalues of $\mathcal{P}(T)$ lie on the
unit circle, therefore all eigenvalues of $\mathcal{K}$ are purely imaginary (or zero).
In the isolated system, with the freedom in Eq.~\eqref{eq:candidates-LF}  we recover that the eigenvalues of the Floquet Hamiltonian $H_F$,
the quasi-energies, are only defined up to multiples of the driving frequency $\omega$,
so all branches lead to a valid Lindbladian evolution. 
For the open system, typically only a few, sometimes even none of the branches lead to a generator that
 is of Lindblad form.

For that, we need to check condition (ii), which is more complicated and involves properties of the eigenelements of the Floquet map. As coined in Refs.~\cite{WolfEtAl08,Cubitt2012}, by plugging the candidates, Eq.~\eqref{eq:candidates-LF}, into the test for conditional complete positivity, Eq.~\eqref{eq:test-cond-comp}, it comes in handy to define a set of $n_c+1$ Hermitian 
matrices %$\{V_0,...,V_{n_c}\}$ obtained from the spectral projectors of $\mathcal{P}(T)$,
\begin{equation}
	V_0 = \Phi_{\perp} \mathcal{K}_0^\Gamma\Phi_{\perp}, \quad V_c = i \omega \Phi_{\perp} (P_c - P_c*)^\Gamma\Phi_{\perp}, \quad c = 1, \dots, n_c.
\end{equation}
The condition is 
fulfilled, if there is  a set of $n_c$ integers, $\mathbf{x} \in \mathbb{Z}^{n_c}$, such that
\begin{equation}
%\begin{align}\label{eq:cond2}
V_{\mathbf{x}}  =  V_0 + \sum_{c=1}^{n_c} x_c V_c \geq 0.
\label{eq:Fl-matrix-cond}
\end{equation} 
%\end{align}
%he test has an important computational aspect. 
%At a first glance, to test this condition,  we
%have to inspect all branches, i.e.,\ a countably infinite number of combinations of $n_c$ integers.

%Fortunately, the situation is not that hopeless because finding the solution %(or proving its absence) 
%for this equation is related to two known programing problems \cite{Ramana1995,Khachiyan2000}. 
%When $\{\mathbf{x}\}$ ranges over $\mathbb{R}^{n_c}$, the condition $V_{\sum}=0$ outshapes either  zero or a finite volume in which $V_{\sum}$ is positive semidefinite. 
%In the former case there is evidently no Floquet Lindbladian.  In the latter case, the volume is enclosed by a convex body called spectrahedron \cite{spect}. 
%To check whether the spectrahedron contains an integer point is a problem of
%polynomial complexity with respect to $\mathrm{max}\{|x^0_1|,...,|x^0_{n_c}| \}$, where $\{\mathbf{x}^0 \} \subset \mathbb{R}^{n_c}$ is the solution set of $V_{\sum}=0$.

Finally, when the test is successful for one branch, the Floquet Lindbladian $\mathcal{L}_F$ is found, and we can extract
from it the corresponding time-independent Hamiltonian and jump operators. 
%This decomposition of a Lindbladian into Hamiltonian and dissipative parts is not unique. However, it becomes so if we assume that all operators are traceless. The procedure is given in Appendix~\ref{sec:app-extract-lindbl}. 

\section{Discrepancy to the Magnus expansions presented in the literature}
\label{sec:app-discrep-magnus}
Here, we  discuss a discrepancy in the general expressions of the second order of the Magnus expansion (in terms of the Fourier components of the generator) that are presented in Refs.~\cite{HaddadfarshiEtAl15,LeskesEtAl10}. One should therefore be cautious when using these expressions.

As it was shown in the literature \cite{HaddadfarshiEtAl15,LeskesEtAl10},  
by plugging the Fourier expansion, Eq.~\eqref{eq:lindbl-four}, into the conventional Magnus expansion \cite{BlanesEtAl09} one finds on the lowest orders
\begin{align}
\mathcal{K}^{(1)} &= \mathcal{L}_0, \\
\mathcal{K}^{(2)} &=  \ \sum_{n=1}^{\infty} \frac{\left[\mathcal{L}_n, \mathcal{L}_{-n}\right]+ \left[\mathcal{L}_0, \mathcal{L}_n - \mathcal{L}_{-n}\right]}{n \omega}.
\end{align}
However on third order  there is a discrepancy between the results in the different works. In Ref.~\cite{HaddadfarshiEtAl15}
it is presented
\begin{align}
\begin{split}
\mathcal{K}^{(3)}_\mathrm{FCM} = & \sum_{n\neq 0}\sum_{m\neq0} \left(\frac{\left[\left[\mathcal{L}_{n}, \mathcal{L}_{-n}\right], \mathcal{L}_m\right]}{2nm\omega^2} - \frac{\left[\mathcal{L}_n, \left[\mathcal{L}_{m}, \mathcal{L}_{-n-m}\right]\right]}{3nm\omega^2} \right)\\
&-  \sum_{n\neq 0}\sum_{m\neq0, m\neq n } \frac{\left[\mathcal{L}_n, \left[\mathcal{L}_{0}, \mathcal{L}_{m}\right]\right]}{2nm\omega^2}\\
&+ \sum_{n= 1}^{\infty} \sum_{m\neq0, m\neq -n } \frac{\left[\left[\mathcal{L}_{n}, \mathcal{L}_{m}\right] + \left[\mathcal{L}_{-n}, \mathcal{L}_{-m}\right], \mathcal{L}_0 \right]}{2 n (n+m)\omega^2} ,
\end{split}
\end{align}
while in Ref.~\cite{LeskesEtAl10} it was found
\begin{align}
\begin{split}
\mathcal{K}^{(3)}_\mathrm{LMV} = & -\sum_{n\neq 0}\sum_{m\neq0} \left(\frac{\left[\mathcal{L}_m, \left[\mathcal{L}_{-m}, \mathcal{L}_{n}\right]\right]}{nm\omega^2} + \frac{\left[\mathcal{L}_m, \left[\mathcal{L}_{n}, \mathcal{L}_{0}\right]\right]}{2nm \omega^2} \right) \\
&-  \sum_{n\neq 0}\sum_{m\neq0, m\neq n } \frac{\left[\mathcal{L}_m, \left[\mathcal{L}_{n-m}, \mathcal{L}_{-n}\right]\right]}{3nm\omega^2}\\
&+\sum_{n\neq 0}  \left(\frac{\left[\mathcal{L}_0, \left[\mathcal{L}_{0}, \mathcal{L}_{n}\right]\right]}{2 n^2 \omega^2} - \frac{\left[\mathcal{L}_n, \left[\mathcal{L}_{0}, \mathcal{L}_{-n}\right]\right]}{2 n^2 \omega^2} \right),
\end{split}
\end{align}
where we have adapted the expression to our notation for the dissipative Floquet system.
Here, by $n\neq 0$ we denote the sum over $n \in \mathbb{Z}\setminus \lbrace 0 \rbrace$.

Note that with these expressions for our two-level system model with $\varphi=0$ we find
\begin{align}
\mathcal{K}^{(3)}_\mathrm{FCM} &= \frac{1}{\omega^2} \left[\mathcal{L}_0, \left[\mathcal{L}_{0}, \mathcal{L}_{1}\right]\right] +  \frac{1}{3\omega^2} \left[\mathcal{L}_1, \left[\mathcal{L}_{0}, \mathcal{L}_{1}\right]\right],\\
\mathcal{K}^{(3)}_\mathrm{LMV} &= \frac{1}{\omega^2} \left[\mathcal{L}_0, \left[\mathcal{L}_{0}, \mathcal{L}_{1}\right]\right] - \frac{1}{\omega^2} \left[\mathcal{L}_1, \left[\mathcal{L}_{0}, \mathcal{L}_{1}\right]\right],
\end{align}
which differ by the prefactors of both terms from the direct calculation
\begin{align}
\mathcal{K}^{(3)} &=  \frac{2}{\omega^2}\left[\mathcal{L}_0, \left[\mathcal{L}_{0}, \mathcal{L}_{1}\right]\right] - \frac{1}{\omega^2} \left[\mathcal{L}_1, \left[\mathcal{L}_{0}, \mathcal{L}_{1}\right]\right]. 
\label{eq:K2-qub-corr}
\end{align}
This is worrisome because the result of the direct calculation was obtained in the same way, but  for a 
special choice of the driving, so in principle all expressions should coincide.

However, in Ref.~\cite{LeskesEtAl10} another expression 
for the second order term is presented. This expression  was obtained by performing the Floquet-Magnus expansion, yielding an effective Hamiltonian/generator in the rotated basis (the basis rotation $D_F$ is unitary, if the dynamics is coherent)
\begin{align}
\Lambda(t) = D_F e^{\bar{\mathcal{L}}_F t} D_F^{-1} \equiv e^{{\mathcal{L}}_F t} .
\end{align}
The Floquet Lindbladian ${\mathcal{L}}_F$ can then be obtained in second order in $1/\omega$ by finding $\bar{\mathcal{L}}_F$ up to second order  combined with
the second order of the expansion of the rotation matrix 
\begin{align}
D_F= \exp\left[i (S^{(1)}/\omega +S^{(2)}/\omega^2)\right].
\end{align} 
%Note that for the 
%Floquet-Lindblad case which we discuss here it does not suffice to check whether the operator $\bar{\mathcal{L}}_F$ is completely positive. The evolution is
%only physical if the full combination of the three operators is completely positive. In the open system $D_F$ is a superoperator that is not necessarily only a rotation anymore.

With this identification it is found
\begin{align}
\mathcal{K}^{(1)'} &= \mathcal{K}^{(1)}, \quad \mathcal{K}^{(2)'} =  \mathcal{K}^{(2)}, \\
 \mathcal{K}^{(3)'}_\mathrm{LMV} &=  \mathcal{K}^{(3)}_\mathrm{LMV}-\sum_{n\neq 0}\sum_{m\neq0} \frac{\left[\mathcal{L}_0, \left[\mathcal{L}_{m}, \mathcal{L}_{n-m}\right]\right]}{nm\omega^2}
\end{align}
and argued that the difference between the both expressions is due to approximations in the derivation of the Floquet-Magnus expansion \cite{LeskesEtAl10}.

Interestingly, in our case of the driven two-level system, by calculating 
\begin{align}
\mathcal{K}^{(3)'}_\mathrm{LMV} =  \frac{2}{\omega^2}\left[\mathcal{L}_0, \left[\mathcal{L}_{0}, \mathcal{L}_{1}\right]\right] - \frac{1}{\omega^2} \left[\mathcal{L}_1, \left[\mathcal{L}_{0}, \mathcal{L}_{1}\right]\right]
\end{align}
we recover the expression in Eq.~\eqref{eq:K2-qub-corr} that we found by directly performing the conventional Magnus expansion. 
We therefore expect that there could be a small error in the direct derivation of $\mathcal{K}^{(3)}_\mathrm{LMV} $
via the Magnus expansion
and that it maybe also holds that $ \mathcal{K}^{(3)}_\mathrm{LMV} =  \mathcal{K}^{(3)'}_\mathrm{LMV}$. 

As a result, the only expression that could be correct is   $\mathcal{K}^{(3)'}_\mathrm{LMV}$.

\section{Degenerate perturbation theory in extended space for the dissipative system}
\label{sec:app-deg-perttheory}
For the coherent system, it was shown \cite{EckardtAnisimovas15} that a high-frequency expansion can be derived from a canonical van-Vleck degenerate perturbation theory in the extended Hilbert space. Here we list the steps that are necessary to generalize this ansatz to the open system.

To this end, let us suppose that we may divide the quasienergy superoperator in the following fashion
\begin{align}
\bar{\mathcal{Q}} = \bar{\mathcal{Q}}_0 + \lambda \bar{\mathcal{V}}
\end{align}
where the spectrum of the operator $\bar{\mathcal{Q}}_0$ is known.
Note that since the system is dissipative, we need to consider the 
right eigenvectors 
\begin{align}
\bar{\mathcal{Q}}_0 \kket{a, m} = \Omega^{(0)}_{a,m}\kket{a, m}
\end{align}
as well as the left eigenvectors 
\begin{align}
\bbra{\tilde a, m} \bar{\mathcal{Q}}_0 = \bbra{\tilde a, m} \Omega^{(0)}_{a,m}
\end{align}
since for non-hermitian operators these will differ in general. Here we split the photon 
index $m$ from the eigenindex, since the spectrum will obey
\begin{align}
\Omega^{(0)}_{a,m+n} = \Omega^{(0)}_{a,m}  + n\omega.
\end{align}
It holds the orthogonality relation
\begin{align}
\bbrakket{\tilde a, m}{b, n} = \delta_{ab} \delta_{mn}.
\label{eq:left-right-orthog}
\end{align}
Note that even though we denote the eigenvectors as ket- and bra-vectors, they are actually
density matrices, so e.g.~in Eq.~\eqref{eq:left-right-orthog} the inner product that  is occurring is
actually relying on the Frobenius inner product 
\begin{align}
(A, B)_F  = \mathrm{tr}(A^\dagger B) .
\end{align}
Let us elaborate a bit on this point. The eigenvectors have the form
\begin{align}
\kket{a, m} &\equiv \begin{pmatrix}
\dots \\
{\Phi_{a,m,-1}} \\ {\Phi_{a,m,0}} \\
{\Phi_{a,m,1}} \\ \dots
\end{pmatrix}, \\
 \bbra{\tilde a, m} &\equiv \begin{pmatrix}
\dots &
{\tilde \Phi_{a,m,-1}} & {\tilde \Phi_{a,m,0}} &
{\tilde \Phi_{a,m,1}} & \dots
\end{pmatrix},
\end{align}
As we show in Appendix~\ref{sec:app-matrep-two-level system} as an example for the two-level system,
it is possible to map density matrices $\Phi_{ij}$ (here, $i,j$ are the matrix indices)  in the $N$-dimensional Hilbert space $\mathcal{H}$
onto $N^2$-dimensional vectors $\ket{\Phi} = \ket{\Phi_{11}, \dots \Phi_{1N}, \Phi_{21}, \dots, \Phi_{NN}}$. Then, superoperators are just (non-hermitian) matrices 
of shape $N^2 \times N^2$. We can then use standard linear algebra to
diagonalize the matrix representation of the superoperator.
For this matrix we find eigenvectors $\ket{\Phi_b}, \bra{\tilde \Phi_a}$ fulfilling 
$\braket{\tilde \Phi_a}{\Phi_b}= \delta_{ab}$. Translating 
it back to density matrices we find
\begin{align}
\delta_{ab}= \braket{\tilde \Phi_a}{\Phi_b} = \sum_{i,j}  \left(\tilde\Phi_a\right)^{*}_{ij} \left(\Phi_b\right)_{ij}= \mathrm{tr}({\tilde \Phi_a}^\dagger{\Phi_b}) = ({\tilde \Phi_a},{\Phi_b})_F.
\end{align}
Therefore, the inner product in the extended Hilbert space, Eq.~\eqref{eq:left-right-orthog}, reads 
\begin{align}
\bbrakket{\tilde a, m}{b, n} = \sum_{k} ({\tilde \Phi_{a,m,k}},\Phi_{b,n,k})_F.
\end{align}

%\twocolumngrid

Remarkably, using this language, one is able to generalize the perturbative
procedure that was found in Ref.~\cite{EckardtAnisimovas15}.
The aim is to find a transformation to the new basis states of the perturbed problem,
\begin{align}
\kket{a, m}_B =  \bar{\mathcal{D}}\kket{a, m}, \qquad \ _B \bbra{\tilde a, m} =  \bbra{\tilde a, m} \bar{\mathcal{D}}^{-1},
\end{align}
such that in the transformed basis the quasi-energy operator is block diagonal,
\begin{align}
\ _B \bbra{\tilde b, m} \bar{\mathcal{Q}} \kket{a, n}_B = 0,  \quad \forall  m \neq n.
\end{align}
It is clear that the left eigenvectors have to transform with $\bar{\mathcal{D}}^{-1}$, because also in the transformed basis,
it has to hold$\ _B\bbrakket{\tilde a, m}{b, n}_B = \delta_{ab} \delta_{mn}$. 

Now, like in the coherent case \cite{EckardtAnisimovas15}, we can separate the block-diagonal part 
of this equation
\begin{align}
\left[\bar{\mathcal{D}}^{-1}(\bar{\mathcal{Q}}_0 + \lambda \bar{\mathcal{V}}_D + \lambda \bar{\mathcal{V}}_X) \bar{\mathcal{D}}\right]_D &= \bar{\mathcal{Q}}_0  +  \bar{\mathcal{W}}_D,
\label{eq:expansion-diag}
\intertext{from the block-off-diagonal
	part}
\left[\bar{\mathcal{D}}^{-1}(\bar{\mathcal{Q}}_0 + \lambda \bar{\mathcal{V}}_D + \lambda \bar{\mathcal{V}}_X) \bar{\mathcal{D}}\right]_X &= 0.
\label{eq:expansion-offdiag}
\end{align}
with some block diagonal operator $\bar{\mathcal{W}} = \bar{\mathcal{W}}_D$. Here, we use the convention
\begin{align}
\bar{\mathcal{A}}_D = \sum_{m} \bar{\mathcal{P}}_m \bar{\mathcal{A}}\bar{\mathcal{P}}_m, \qquad  \bar{\mathcal{A}}_X = \sum_{m\neq n} \bar{\mathcal{P}}_m \bar{\mathcal{A}}\bar{\mathcal{P}}_n
\end{align}
with projector $\bar{\mathcal{P}}_m=\sum_a\kket{a,m}\bbra{\tilde a,m}$. By representing the rotation as 
\begin{align}
\bar{\mathcal{D}}= \exp(\bar{\mathcal{G}}_X) \quad \text{ it directly follows} \quad \bar{\mathcal{D}}^{-1}= \exp(-\bar{\mathcal{G}}_X).
\end{align}
Here the rotation $\bar{\mathcal{G}}= \bar{\mathcal{G}}_X$ is chosen such that it does
not affect the blocks with the same photon number $m$.
We then can expand the operators
\begin{align}
\bar{\mathcal{G}}_X = \sum_{n=1}^{\infty} \lambda^n\bar{\mathcal{G}}^{(n)}_X, \qquad \bar{\mathcal{W}}_D = \sum_{n=1}^{\infty} \lambda^n\bar{\mathcal{W}}^{(n)}_D,
\end{align}
plug this into Eq.~\eqref{eq:expansion-diag} and Eq.~\eqref{eq:expansion-offdiag}, sort it by orders of $\lambda$
and find exactly the same expressions as in Appendix C of Ref.~\cite{EckardtAnisimovas15}.
Let us just present the first nontrivial order $\propto \lambda^1$, where it has to hold
\begin{align}
\bar{\mathcal{W}}^{(1)}_D= \bar{\mathcal{V}}_D,\quad  \text{ as well as } \quad  \left[\bar{\mathcal{G}}^{(1)}_X, \bar{\mathcal{Q}}_0\right]= \bar{\mathcal{V}}_X.
\end{align}
Very similar to the coherent case, the occurring commutators $\left[\bar{\mathcal{G}}^{(n)}_X, \bar{\mathcal{Q}}_0\right]$ may be unraveled 
by taking matrix elements of the form
\begin{align}
\bbra{\tilde a, m} \left[\bar{\mathcal{G}}^{(1)}_X, \bar{\mathcal{Q}}_0\right] \kket{b, n} 
&= (\Omega_{a,m}-\Omega_{b,n})\bbra{\tilde a, m} \bar{\mathcal{G}}^{(1)}_X \kket{b, n}\\
&= \bbra{\tilde a, m} \bar{\mathcal{V}}_X \kket{b, n},
\end{align}
with $m \neq n$. Therefore, we see that the argumentation for the closed system can be
directly translated to the open system by replacing the real quasienergies $\varepsilon^{(0)}_{a,m}$
with the complex eigenvalues $\Omega^{(0)}_{a,m}$, the bra-vectos $\bbra{a, m}$
with left eigenvectors $\bbra{\tilde a, m}$ and the rotation $\bar{{U}}$ with $\bar{\mathcal{D}}$ as well as $\bar{{U}}^\dagger$ with $\bar{\mathcal{D}}^{-1}$.

Thus, like in the coherent case, we may find a high-frequency expansion of the superoperator by taking 
\begin{align}
\mathcal{Q}_0 = -i \partial_t, \quad \text{ such that } \quad \mathcal{Q}_0 \kket{a, m} = m \omega \kket{a, m}
\end{align}
and with the natural basis $\kket{a, m}$. Note that $\mathcal{Q}_0$
is hermitian, therefore the left eigenvectors are just $\bbra{a, m}$.

\section{Commutator of two general two-level system Lindblad superoperators}
\label{sec:app-comm-lind}
%Given are two arbitrary Lindbladians $\mathcal{L}^{(1)}$ and $\mathcal{L}^{(2)}$ for a two-level system system. 
%We find a general expression for their commutator.
Here we derive general expressions for the commutator of two arbitrary Lindbladians $\mathcal{L}^{(1)}$ and $\mathcal{L}^{(2)}$ for a two-level system system. 

The Lindbladians  $\mathcal{L}^{(1)}$ and $\mathcal{L}^{(2)}$ can be represented as 
\begin{align}
\mathcal{L}^{(i)}= -i [H^{(i)}, \cdot] + \sum_{nm} d^{(i)}_{nm} \, \left(\sigma_n \cdot \sigma_m - \frac{1}{2} \left\lbrace\sigma_m \sigma_n, \cdot \right\rbrace \right),
\end{align}
where the indices $n,m$ in the following run over $1,2,3$.
Their commutator therefore reads 
\begin{align}
\begin{split}
\left[\mathcal{L}^{(1)}, \mathcal{L}^{(2)}\right]& = - \left[H^{(1)}, \left[H^{(2)}, \cdot \right]  \right] +\left[H^{(2)}, \left[H^{(1)}, \cdot \right]  \right] \\
- i \sum_{nm} d_{nm}^{(1)}& \left( \sigma_n \left[H^{(2)}, \cdot \right] \sigma_m - \frac{1}{2}\left\lbrace\sigma_m \sigma_n,  \left[H^{(2)}, \cdot \right] \right\rbrace \right. \\
 & \left. -  \left[H^{(2)}, \sigma_n \cdot \sigma_m \right]  +  \frac{1}{2}\left[H^{(2)},   \left\lbrace\sigma_m \sigma_n, \cdot \right\rbrace \right] \right)\\
 + i \sum_{nm} d_{nm}^{(2)} &\left( \sigma_n \left[H^{(1)}, \cdot \right] \sigma_m - \frac{1}{2}\left\lbrace\sigma_m \sigma_n,  \left[H^{(1)}, \cdot \right] \right\rbrace \right.  \\
 & \left.-  \left[H^{(1)}, \sigma_n \cdot \sigma_m \right]  +  \frac{1}{2}\left[H^{(1)},  \left\lbrace\sigma_m \sigma_n, \cdot \right\rbrace \right] \right)\\
 +\sum_{nm,kl} (d_{nm}^{(1)}& d_{kl}^{(2)} -d_{kl}^{(1)} d_{nm}^{(2)} ) \left[ \sigma_n \left(\sigma_k \cdot \sigma_l - \frac{1}{2} \left\lbrace\sigma_l \sigma_k, \cdot \right\rbrace \right) \sigma_m \right.\\ 
 & \left. - \frac{1}{2}\left\lbrace\sigma_m \sigma_n,  \sigma_k \cdot \sigma_l - \frac{1}{2} \left\lbrace\sigma_l \sigma_k, \cdot \right\rbrace  \right\rbrace \right].
\end{split}
\end{align}
This can be simplified to read
\begin{align}
\begin{split}
\left[\mathcal{L}^{(1)}, \mathcal{L}^{(2)}\right] &= -i \left[H^\mathrm{coh}, \cdot \right] + i \sum_{nm} d_{nm}^{(1)} \left(  \left[H^{(2)}, \sigma_n \right] \cdot \sigma_m \right. \\
&\left.+ \sigma_n \cdot \left[H^{(2)}, \sigma_m \right]  - \frac{1}{2}\left\lbrace \left[H^{(2)},\sigma_m \sigma_n\right],  \cdot  \right\rbrace  \right)\\
 - i \sum_{nm} d_{nm}^{(2)} &\left(  \left[H^{(1)}, \sigma_n \right] \cdot \sigma_m + \sigma_n \cdot \left[H^{(1)}, \sigma_m \right]  - \frac{1}{2}\left\lbrace \left[H^{(1)},\sigma_m \sigma_n\right],  \cdot  \right\rbrace  \right)\\
+\sum_{nm,kl} (d_{nm}^{(1)}& d_{kl}^{(2)}-d_{kl}^{(1)} d_{nm}^{(2)} ) \left[ \sigma_n \left(\sigma_k \cdot \sigma_l - \frac{1}{2} \left\lbrace\sigma_l \sigma_k, \cdot \right\rbrace \right) \sigma_m \right. \\ 
& \left. - \frac{1}{2}\left\lbrace\sigma_m \sigma_n,  \sigma_k \cdot \sigma_l - \frac{1}{2} \left\lbrace\sigma_l \sigma_k, \cdot \right\rbrace  \right\rbrace \right].
\end{split}
\label{eq:lind-commu-step1}
\end{align}
with resulting Hamiltonian due to the coherent parts
\begin{align}
H^\mathrm{coh} = -i \left[H^{(1)}, H^{(2)} \right]=2 \sum_{kql} \varepsilon_{kql} h^{(1)}_k h^{(2)}_q \sigma_l.
\end{align}
In the last step we have represented the Hamiltonians in the Pauli basis, 
\begin{align}
H^{(i)} = h^{(i)}_0 \mathbf{1} + \sum_k h^{(i)}_k \sigma_k.
\end{align}
Note that the first three lines of Eq.~\eqref{eq:lind-commu-step1} are already in Lindblad form. The third line, however, needs more work, but one can show that it can be brought
to Lindblad form
\begin{align}
\begin{split}
\sum_{nm,kl}& (d_{nm}^{(1)} d_{kl}^{(2)}-d_{kl}^{(1)} d_{nm}^{(2)} ) \left[ \sigma_n \left(\sigma_k \cdot \sigma_l - \frac{1}{2} \left\lbrace\sigma_l \sigma_k, \cdot \right\rbrace \right) \sigma_m \right. \\ &\left. - \frac{1}{2}\left\lbrace\sigma_m \sigma_n,  \sigma_k \cdot \sigma_l - \frac{1}{2} \left\lbrace\sigma_l \sigma_k, \cdot \right\rbrace  \right\rbrace \right]
\end{split}\\
&=-i \left[H^\mathrm{diss}, \cdot \right] + \sum_{mn} d^{\mathrm{diss}}_{mn} \, \left(\sigma_m \cdot \sigma_n - \frac{1}{2} \left\lbrace\sigma_m \sigma_n, \cdot \right\rbrace \right)
\end{align}
with resulting hamiltonian due to the dissipative parts,
\begin{align}
H^\mathrm{diss} = -2 \sum_{nmkq} \varepsilon_{nmq} \mathrm{Re}(d^{(1)}_{nk})\mathrm{Re}(d^{(2)}_{mk}) \sigma_q,
\end{align}
as well as
\begin{align}
d^\mathrm{diss}_{nm} = 2 i \sum_{k} \mathrm{Im}(d_{nk}^{(1)} d_{mk}^{(2)}-d_{mk}^{(1)} d_{nk}^{(2)} ).
\end{align}
Therefore, in total the commutator reads
\begin{align}
\begin{split}
\left[\mathcal{L}^{(1)}, \mathcal{L}^{(2)}\right] &= -i \left[H^\mathrm{coh}+H^\mathrm{diss} , \cdot \right] \\&+ \sum_{nm} (d^\mathrm{c-d}_{nm} + d^\mathrm{diss}_{nm}) \, \left[\sigma_n \cdot \sigma_m - \frac{1}{2} \left\lbrace\sigma_m \sigma_n, \cdot \right\rbrace \right]
\end{split}
\end{align}
where we have also evaluated the terms coming from the mixed coherent and dissipative terms 
\begin{align}
\begin{split}
d^\mathrm{c-d}_{nm} = 2 \sum_{kl} &\left[ \left( d_{lm}^{(1)}h^{(2)}_k - d_{lm}^{(2)}h^{(1)}_k \right) \varepsilon_{knl} \right. \\ 
&\left. + \left(d_{nl}^{(1)}h^{(2)}_k  - d_{nl}^{(2)}h^{(1)}_k \right) \varepsilon_{kml}\right].
\end{split}
\end{align}

\section{Matrix representation of the most general two-level system Lindbladian} 
\label{sec:app-matrep-two-level system}

For the two-level system the Hilbert space is $\mathcal{H}=\mathbb{C}^2$. Under the identification
\begin{align}
\varrho = \begin{pmatrix}
a & b  \\
c & d
\end{pmatrix}
\quad \rightarrow \quad
\ket{\varrho} = \begin{pmatrix}
a \\ b \\
c \\ d
\end{pmatrix}
\end{align}
we may represent density matrices as vectors and superoperators as matrices.
Here we %extract the corresponding Hamiltonian and the Dissipators we 
provide an explicit
translation table of the superoperator into matrix notation  for the most general static two-level system Lindbladian. 

The most general Lindbladian has the form
\begin{align}
\mathcal{L}= -i \left[\sum_k h_k  \sigma_k, \cdot\right] + \sum_{mn} d_{mn} \, \left(\sigma_m \cdot \sigma_n - \frac{1}{2} \lbrace\sigma_n \sigma_m, \cdot \rbrace \right)
\end{align}
with coefficient matrix  
\begin{align}
{d} = \begin{pmatrix}
a & d+i e & f+i g\\
d-i e & b& s+i t\\
f-i g & s+i t & c
\end{pmatrix}.
\end{align}
After some algebra one finds its matrix form as
\onecolumngrid
\begin{align}
%\scriptsize
\mathcal{L}= \begin{pmatrix}
-a-b-2e&i h_1 -h_2+f+i s & -i h_1 -h_2+f-i s& a+b-2e\\
i h_1 + h_2+f-i s -2i g -2t & -2 i h_3 -a-b-2c & a-b-2i d & -i h_1 - h_2-f+i s -2i g -2t \\
-i h_1 + h_2+f+i s +2i g -2t & a-b+2i d & 2 i h_3 -a-b-2c &  i h_1 - h_2-f-i s +2i g -2t \\
a+b+2e & -i h_1 +h_2-f-i s & i h_1 +h_2-f+i s & -a -b +2e
\end{pmatrix}.
\label{eq:L-general}
\end{align}
\twocolumngrid

\section{Fourier components of the superoperator generating the rotating frame transformation}
\label{sec:app-rotfr-fourier}
Here we prove Eq.~\eqref{eq:Lambda_n-expl} which provides an explicit expression of the extended-space superoperator $\bar{\Lambda}$ generating the (generalized) rotating frame transformation  for an operator of the form of Eq.~\eqref{eq:scalar-drivingterm-Lindbl}.
%\begin{align}
%\mathcal{L}_d(t) = \phi(t) \mathcal{L}_d',  \quad \text{ with scalar function } \quad \phi(t) = \sum_{m\neq 0} e^{im\omega t} \chi_m.
%\end{align}

By definition
\begin{align}
\Lambda_n = \frac{1}{T} \int_0^T \mathrm{d}t e^{-in\omega t} \exp\left({\int_0^t \mathrm{d}t' \mathcal{L}_d(t')}\right). 
\end{align}
We can further evaluate this expression if we assume that, like  for our model system,
it holds that 
\begin{align}
\mathcal{L}_d(t) = \phi(t) \mathcal{L}_d'
\end{align}
with some periodic scalar function $\phi(t) = \sum_{m\neq 0} e^{im\omega t} \phi_m$. Then we may evaluate 
\begin{align}
&\int_0^t \mathrm{d}t' \mathcal{L}_d(t') = \chi(t) \mathcal{L}_d' 
\intertext{ with }  
&\chi(t)= \int_0^t \mathrm{d}t'  \phi(t') =  \sum_{m\neq 0}  \frac{e^{im\omega t}-1}{im\omega} \phi_m .
\end{align}
We may rewrite $e^{im\omega t}-1=\cos(m\omega t)-1 + i \sin(m\omega t)$.
This gives
\begin{align}
\begin{split}
\Lambda_n = \frac{1}{T} \int_0^T \mathrm{d}t e^{-in\omega t} &\exp\left( \sum_{m\neq 0} \frac{\sin(m\omega t)}{m\omega} \phi_m \mathcal{L}_d' \right.\\
& \left.+ \sum_{m\neq 0} \frac{\cos(m\omega t)-1}{im\omega} \phi_m \mathcal{L}_d'\right)
\end{split}\\
\begin{split}
=  \frac{1}{T} \int_0^T \mathrm{d}t e^{-in\omega t} \prod_{m\neq 0} & \exp\left(  \frac{\sin(m\omega t)}{m\omega} \phi_m \mathcal{L}_d'\right) \\
&\exp\left( \frac{\cos(m\omega t)-1}{im\omega} \phi_m \mathcal{L}_d'\right)
\end{split}
%\exp\left( i \sum_{m\neq 0} \frac{ \varphi_m}{m\omega} \mathcal{L}_d'\right). 
\label{eq:Lambda_n-epl1}
\end{align}
We may now represent $\mathcal{L}_d'$ using its spectral decomposition
\begin{align}
\mathcal{L}_d' = \sum_a \lambda_a \kket{\Phi^{(d)}_{a}} \bbra{\tilde \Phi^{(d)}_a}
\end{align}
and may use the Bessel functions of first kind $J_n$ to evaluate
\begin{align}
f^{(m)}_n(x) &= \frac{1}{T} \int_0^T \mathrm{d}t e^{-in\omega t + ix \sin(m\omega t)} \\ 
&=  \frac{1}{T} \int_0^T \mathrm{d}t e^{-in\omega t} \sum_{k\in\mathbb{Z}} J_k(x) e^{ikm\omega t} \\
&= 
\left\{\begin{array}{cc} 
J_{n/m}(x) &\text{ if } n = k m, k \in \mathbb{Z}\\
0  &  \text{ else. }
\end{array} \right.
\end{align}
Similarly, with the modified Bessel functions of first kind $I_n$ we find
\begin{align}
g^{(m)}_n(x) &= \frac{1}{T} e^{-x}\int_0^T \mathrm{d}t e^{-in\omega t + x \cos(m\omega t)}\\
&= 
\left\{\begin{array}{cc} 
e^{-x} I_{n/m}(x) &\text{ if } n = k m, k \in \mathbb{Z}\\
0  &  \text{ else. }
\end{array} \right.
\end{align}
Note that in Eq.~\eqref{eq:Lambda_n-epl1} occurs the Fourier transform of  a product of the functions that we transformed above, which gives rise to a relatively involved structure.
A compact form can be obtained  in extended Hilbert space  where it holds
\begin{align}
\bar{ \Lambda} &= \sum_a \prod_{m\neq0} \bar{f}^{(m)}\left(\frac{\phi_m \lambda_a}{im\omega}\right)\bar{g}^{(m)}\left(\frac{\phi_m \lambda_a}{im\omega}\right)\kket{\Phi^{(d)}_{a}} \bbra{\tilde \Phi^{(d)}_a} \\
&= \prod_{m\neq0} \bar{f}^{(m)}\left(\frac{\phi_m \mathcal{L}_d'}{im\omega}\right)\bar{g}^{(m)}\left(\frac{\phi_m \mathcal{L}_d'}{im\omega}\right).
\end{align}

\section{Explicit calculation of the perturbative expansion in extended space for the driven-dissipative two-level system}
\label{sec:app-two-level system-expl}

Instead of the explicit rotating-frame transformation on the level of the superoperator, as presented in Sec.~\eqref{sec:rotfr-Ln-expl} for the driven-dissipative two-level system, here we calculate the components $\tilde{\mathcal{L}}_n$ in matrix representation 
by using Eq.~\eqref{eq:L_n-matrx-expl}. This matrix representation can be used to evaluate the Floquet-Magnus expansion numerically.

For our model system, by using Eq.~\eqref{eq:L-general} we find the matrix representations
\begin{align}
A&=i \mathcal{L}_1  = i \mathcal{L}_{-1}= \frac{E}{2} \begin{pmatrix}
0 & - 1 & \phantom{-}1 & 0  \\
-1 & 0 & 0 & \phantom{-}1 \\
\phantom{-}1&  0 & 0 & -1 \\
0 & \phantom{-}1 & -1 & 0
\end{pmatrix}
 \intertext{ and }
\mathcal{L}_0 &=  \begin{pmatrix}
-4  \gamma & 0 & 0 & 0  \\
0 & -i - 2 \gamma & 0 & 0 \\
0&  0 & i - 2  \gamma & 0 \\
4 \gamma & 0 & 0 & 0
\end{pmatrix}.
\end{align}
We start by diagonalizing the Hermitian matrix $A$. One can show that $A=U D U^\dagger$
with
\begin{align}
U= \frac{1}{2} \begin{pmatrix}
-1 & 0 & \sqrt{2} & -1  \\
-1 & \sqrt{2} & 0 & \phantom{-}1 \\
\phantom{-}1&  \sqrt{2} & 0 & -1 \\
\phantom{-}1 & 0 & \sqrt{2} & \phantom{-}1
\end{pmatrix}
\text{ and } 
D=  \begin{pmatrix}
-E & 0 & 0 & 0  \\
0 & 0 & 0 & 0 \\
0&  0 & 0 & 0 \\
0 & 0 & 0 & E
\end{pmatrix} .
\end{align}
%As we discuss in the main text, 
As can be seen from the power series of $J_k$
it holds that $J_k(-2A/\omega)=U J_k(-2D/\omega) U^\dagger$ yielding
\begin{align}
J_k\left(-\frac{2A}{\omega}\right) = \frac{1}{2}  \begin{pmatrix}
\phantom{-}a_k & \phantom{-}c_k & -c_k & \phantom{-}b_k  \\
\phantom{-}c_k &  \phantom{-}a_k & \phantom{-}b_k & -c_k \\
-c_k &  \phantom{-}b_k & \phantom{-}a_k & \phantom{-}c_k \\
\phantom{-}b_k & -c_k & \phantom{-}c_k & \phantom{-}a_k
\end{pmatrix}(z).
\end{align}
where we set $z=2E/\omega$ and define the functions
\begin{align}
a_k(z)&=e_k J_k(z) +\delta_{k0},\\
b_k(z)&=-e_k J_k(z)+\delta_{k0},\\
c_k(z)&=o_k J_k(z)
\end{align}
Here we have used that $J_k(0)=\delta_{k0}$, $J_k(-z)=(-1)^k J_k(z)$ and the definitions
\begin{align}
e_k =\left\lbrace \begin{array}{cc} 1, & k \text{ even,}\\ 0, & k \text{ odd,} \end{array} \right. \text{ and } o_k =\left\lbrace \begin{array}{cc} 0, &  k \text{ even,}\\ 1, & k \text{ odd.} \end{array} \right.
\end{align}
With this, we evaluate
%\onecolumngrid
\begin{align}
\begin{split}
&\mathcal{L}_0 J_k\left(-\frac{2A}{\omega}\right) =\\&   \begin{pmatrix}
-4\gamma a_k & -4\gamma c_k &  -4\gamma c_k & 4\gamma b_k  \\
(-i-2\gamma)c_k &  (-i-2\gamma)a_k & (-i-2\gamma)b_k & (i+2\gamma)c_k \\
(-i+2\gamma)c_k &  (i-2\gamma)b_k & (i-2\gamma)a_k & (i-2\gamma)c_k \\
4\gamma a_k & 4\gamma c_k & -4\gamma c_k &  4\gamma b_k
\end{pmatrix}(z),
\end{split}
\end{align}
and
\onecolumngrid
\begin{align}
\begin{split}
\tilde{\mathcal{L}}_n &= \sum_{k\in\mathbb Z}J_{k-n}\left(-\frac{2A}{\omega}\right)  {\mathcal{L}}_0 J_k\left(-\frac{2A}{\omega}\right) \\&= - \gamma  \sum_{k\in\mathbb Z} J_{k-n}\left(z\right)  J_{k}\left(z\right)  \begin{pmatrix}
\phantom{-}e_n p_k & \phantom{-}o_n q_k & -o_n q_k  & {-}e_n p_k  \\
\phantom{-}o_n p_k & \phantom{-}e_n q_k & -e_n q_k  & {-}o_n p_k \\
{-}o_n p_k & {-}e_n q_k &  \phantom{-}e_n q_k  &  \phantom{-}o_n p_k  \\
{-}e_n p_k & {-}o_n q_k & \phantom{-}o_n q_k  & \phantom{-}e_n p_k 
\end{pmatrix}- \gamma  \delta_{n0}  \begin{pmatrix}
0 & 0 & 0  & 0  \\
0 & 1 & 1  & 0\\
0 & 1 & 1  & 0  \\
0 & 0 & 0  & 0
\end{pmatrix}\\
&+  \frac{1}{2}  \begin{pmatrix}
-4 \gamma e_n  J_{0}  &  -io_n  J_0 &  -io_n J_0 & -4 \gamma e_n J_0\\
o_n  (-4 \gamma J_{0} -i J_{n}) & -ie_n (J_{0}+ J_{n})  & -ie_n (J_{0}- J_{n})  & o_n (-4 \gamma J_{0} +i J_{n}) \\
o_n (4 \gamma J_{0} + J_{n})   & -ie_n (-J_{0}+ J_{n})  & -ie_n(-J_{0}- J_{n})  & o_n  (4 \gamma J_{0} +i J_{n})   \\
4 \gamma  e_n J_{0}  & io_n J_{0}  &  i o_n J_{0}  & 4 \gamma   e_n J_{0}
\end{pmatrix}\left(z\right),
\end{split}
\end{align}
with $p_k=2e_k+o_k$, as well as $q_k=2o_k+e_k$. Therefore, we finally 
find the representation of the zeroth order expansion 
\begin{align}\begin{split}
\mathcal K^{(1)} = \tilde{\mathcal{L}}_0 = 
\begin{pmatrix}
- \gamma [2J_0+2f+g] & 0 & 0  &  - \gamma [2J_0-2f-g]   \\
0 & -i J_0- \gamma [1+f+2g] & - \gamma [1-f-2g]   & 0\\
0 & - \gamma [1-f-2g] & i J_0- \gamma [1+f+2g]  & 0  \\
\gamma [2J_0+2f+g]  & 0 & 0  &  \gamma [2J_0-2f-g] 
\end{pmatrix}(z)
\end{split}
\end{align}
where we define $f(z)=\sum_{k\in \mathbb{Z}} e_k J_k(z)^2$ as well as $g(z)=\sum_{k\in \mathbb{Z}} o_k J_k(z)^2$. Note that it holds,
\begin{align}
f(z) + g(z) = \sum_{k\in \mathbb{Z}} J_k(z)^2=1,
\end{align}
which allows to express $\mathcal{K}^{(1)}$ in terms of $J_0(z)$ and $g(z)$ only 
\begin{align}\begin{split}
\mathcal K^{(1)} = 
\begin{pmatrix}
- \gamma [2J_0+2-g] & 0 & 0  &  - \gamma [2J_0-2+g]   \\
0 & -i J_0- \gamma [2+g] &  \gamma g   & 0\\
0 &  \gamma g & i J_0- \gamma [2+g]  & 0  \\
\gamma [2J_0+2-g]  & 0 & 0  &  \gamma [2J_0-2+g] 
\end{pmatrix}(z).
\end{split}
\end{align}
\twocolumngrid

By comparing the matrix representation $\mathcal K^{(1)} $
to the most general form of the two-level system Lindbladian, Eq.~\eqref{eq:L-general}, we find the Hamiltonian
and the dissipator matrix,
\begin{widetext}
\begin{align}
%\begin{split}
&\mathcal K^{(0)}=\mathcal{L}(H, {d}),   \text{ with } H = \frac{J_0(z)}{2} \sigma_z 
%\\ 
%&
\text{ and }{d}=\gamma \begin{pmatrix}
1 & i J_0(z)& 0   \\
-i  J_0(z)&  1-g(z) & 0 \\
0 & 0 & g(z) 
\end{pmatrix}.
%\end{split}
\end{align}
\end{widetext}
Note that this is exactly the same result that we obtained %on lowest order of the Magnus expansion in the rotating frame, 
in Eq.~\eqref{eq:FloqLind-rotframe}.
To see this, we use the Bessel function identity $J_n(y+z)= \sum_{k \in \mathbb{Z}} J_k(y) J_{n-k}(z)$ to rewrite
\begin{align} 
J_0(2z) &= \sum_{k \in \mathbb{Z}} J_k(z) J_{-k}(z) = \sum_{k \in \mathbb{Z}} (-1)^k J_k(z)^2 \\ \nonumber
&=  \sum_{k \in \mathbb{Z}} e_k J_k(z)^2  -  \sum_{k \in \mathbb{Z}} o_k J_k(z)^2 = f(z)-g(z).
\end{align}
Together with $f(z)+g(z)=1$ we find that
\begin{align}
g(z) = \frac{1}{2} \left[ 1- J_0(2z) \right].
\end{align}

 \bibliography{mybib,mybib2}

\begin{thebibliography}{70}%
\makeatletter
\providecommand \@ifxundefined [1]{%
 \@ifx{#1\undefined}
}%
\providecommand \@ifnum [1]{%
 \ifnum #1\expandafter \@firstoftwo
 \else \expandafter \@secondoftwo
 \fi
}%
\providecommand \@ifx [1]{%
 \ifx #1\expandafter \@firstoftwo
 \else \expandafter \@secondoftwo
 \fi
}%
\providecommand \natexlab [1]{#1}%
\providecommand \enquote  [1]{``#1''}%
\providecommand \bibnamefont  [1]{#1}%
\providecommand \bibfnamefont [1]{#1}%
\providecommand \citenamefont [1]{#1}%
\providecommand \href@noop [0]{\@secondoftwo}%
\providecommand \href [0]{\begingroup \@sanitize@url \@href}%
\providecommand \@href[1]{\@@startlink{#1}\@@href}%
\providecommand \@@href[1]{\endgroup#1\@@endlink}%
\providecommand \@sanitize@url [0]{\catcode `\\12\catcode `\$12\catcode
  `\&12\catcode `\#12\catcode `\^12\catcode `\_12\catcode `\%12\relax}%
\providecommand \@@startlink[1]{}%
\providecommand \@@endlink[0]{}%
\providecommand \url  [0]{\begingroup\@sanitize@url \@url }%
\providecommand \@url [1]{\endgroup\@href {#1}{\urlprefix }}%
\providecommand \urlprefix  [0]{URL }%
\providecommand \Eprint [0]{\href }%
\providecommand \doibase [0]{http://dx.doi.org/}%
\providecommand \selectlanguage [0]{\@gobble}%
\providecommand \bibinfo  [0]{\@secondoftwo}%
\providecommand \bibfield  [0]{\@secondoftwo}%
\providecommand \translation [1]{[#1]}%
\providecommand \BibitemOpen [0]{}%
\providecommand \bibitemStop [0]{}%
\providecommand \bibitemNoStop [0]{.\EOS\space}%
\providecommand \EOS [0]{\spacefactor3000\relax}%
\providecommand \BibitemShut  [1]{\csname bibitem#1\endcsname}%
\let\auto@bib@innerbib\@empty
%</preamble>
\bibitem [{\citenamefont {Aidelsburger}\ \emph {et~al.}(2011)\citenamefont
  {Aidelsburger}, \citenamefont {Atala}, \citenamefont {Nascimb{\`e}ne},
  \citenamefont {Trotzky}, \citenamefont {Chen},\ and\ \citenamefont
  {Bloch}}]{AidelsburgerEtAl11}%
  \BibitemOpen
  \bibfield  {author} {\bibinfo {author} {\bibfnamefont {M.}~\bibnamefont
  {Aidelsburger}}, \bibinfo {author} {\bibfnamefont {M.}~\bibnamefont {Atala}},
  \bibinfo {author} {\bibfnamefont {S.}~\bibnamefont {Nascimb{\`e}ne}},
  \bibinfo {author} {\bibfnamefont {S.}~\bibnamefont {Trotzky}}, \bibinfo
  {author} {\bibfnamefont {Y.-A.}\ \bibnamefont {Chen}}, \ and\ \bibinfo
  {author} {\bibfnamefont {I.}~\bibnamefont {Bloch}},\ }\href {\doibase
  10.1103/PhysRevLett.107.255301} {\bibfield  {journal} {\bibinfo  {journal}
  {Phys. Rev. Lett.}\ }\textbf {\bibinfo {volume} {107}},\ \bibinfo {pages}
  {255301} (\bibinfo {year} {2011})}\BibitemShut {NoStop}%
\bibitem [{\citenamefont {Struck}\ \emph {et~al.}(2013)\citenamefont {Struck},
  \citenamefont {Weinberg}, \citenamefont {{\"O}lschl{\"a}ger}, \citenamefont
  {Windpassinger}, \citenamefont {Simonet}, \citenamefont {Sengstock},
  \citenamefont {H{\"o}ppner}, \citenamefont {Hauke}, \citenamefont {Eckardt},
  \citenamefont {Lewenstein},\ and\ \citenamefont {Mathey}}]{StruckEtAl13}%
  \BibitemOpen
  \bibfield  {author} {\bibinfo {author} {\bibfnamefont {J.}~\bibnamefont
  {Struck}}, \bibinfo {author} {\bibfnamefont {M.}~\bibnamefont {Weinberg}},
  \bibinfo {author} {\bibfnamefont {C.}~\bibnamefont {{\"O}lschl{\"a}ger}},
  \bibinfo {author} {\bibfnamefont {P.}~\bibnamefont {Windpassinger}}, \bibinfo
  {author} {\bibfnamefont {J.}~\bibnamefont {Simonet}}, \bibinfo {author}
  {\bibfnamefont {K.}~\bibnamefont {Sengstock}}, \bibinfo {author}
  {\bibfnamefont {R.}~\bibnamefont {H{\"o}ppner}}, \bibinfo {author}
  {\bibfnamefont {P.}~\bibnamefont {Hauke}}, \bibinfo {author} {\bibfnamefont
  {A.}~\bibnamefont {Eckardt}}, \bibinfo {author} {\bibfnamefont
  {M.}~\bibnamefont {Lewenstein}}, \ and\ \bibinfo {author} {\bibfnamefont
  {L.}~\bibnamefont {Mathey}},\ }\href@noop {} {\bibfield  {journal} {\bibinfo
  {journal} {Nat.\ Phys.}\ }\textbf {\bibinfo {volume} {9}},\ \bibinfo {pages}
  {738} (\bibinfo {year} {2013})}\BibitemShut {NoStop}%
\bibitem [{\citenamefont {{Gregor Jotzu AND Michael Messer AND R{\'e}mi
  Desbuquois}}\ \emph {et~al.}(2014)\citenamefont {{Gregor Jotzu AND Michael
  Messer AND R{\'e}mi Desbuquois}}, \citenamefont {Greif},\ and\ \citenamefont
  {Esslinger}}]{JotzuEtAl14}%
  \BibitemOpen
  \bibfield  {author} {\bibinfo {author} {\bibfnamefont {T.~U.}\ \bibnamefont
  {{Gregor Jotzu AND Michael Messer AND R{\'e}mi Desbuquois}}, \bibfnamefont
  {Martin~Lebrat}}, \bibinfo {author} {\bibfnamefont {D.}~\bibnamefont
  {Greif}}, \ and\ \bibinfo {author} {\bibfnamefont {T.}~\bibnamefont
  {Esslinger}},\ }\href@noop {} {\bibfield  {journal} {\bibinfo  {journal}
  {Nature}\ }\textbf {\bibinfo {volume} {515}},\ \bibinfo {pages} {237}
  (\bibinfo {year} {2014})}\BibitemShut {NoStop}%
\bibitem [{\citenamefont {Aidelsburger}\ \emph {et~al.}(2015)\citenamefont
  {Aidelsburger}, \citenamefont {Lohse}, \citenamefont {Schweizer},
  \citenamefont {Atala}, \citenamefont {Barreiro}, \citenamefont
  {Nascimb{\`e}ne}, \citenamefont {Cooper}, \citenamefont {Bloch},\ and\
  \citenamefont {Goldman}}]{AidelsburgerEtAl15}%
  \BibitemOpen
  \bibfield  {author} {\bibinfo {author} {\bibfnamefont {M.}~\bibnamefont
  {Aidelsburger}}, \bibinfo {author} {\bibfnamefont {M.}~\bibnamefont {Lohse}},
  \bibinfo {author} {\bibfnamefont {C.}~\bibnamefont {Schweizer}}, \bibinfo
  {author} {\bibfnamefont {M.}~\bibnamefont {Atala}}, \bibinfo {author}
  {\bibfnamefont {J.~T.}\ \bibnamefont {Barreiro}}, \bibinfo {author}
  {\bibfnamefont {S.}~\bibnamefont {Nascimb{\`e}ne}}, \bibinfo {author}
  {\bibfnamefont {N.~R.}\ \bibnamefont {Cooper}}, \bibinfo {author}
  {\bibfnamefont {I.}~\bibnamefont {Bloch}}, \ and\ \bibinfo {author}
  {\bibfnamefont {N.}~\bibnamefont {Goldman}},\ }\href@noop {} {\bibfield
  {journal} {\bibinfo  {journal} {Nat.\ Phys.}\ }\textbf {\bibinfo {volume}
  {1}},\ \bibinfo {pages} {162} (\bibinfo {year} {2015})}\BibitemShut {NoStop}%
\bibitem [{\citenamefont {Fl{\"a}schner}\ \emph {et~al.}(2016)\citenamefont
  {Fl{\"a}schner}, \citenamefont {Rem}, \citenamefont {Tarnowski},
  \citenamefont {Vogel}, \citenamefont {L{\"u}hmann}, \citenamefont
  {Sengstock},\ and\ \citenamefont {Weitenberg}}]{Flaeschner16}%
  \BibitemOpen
  \bibfield  {author} {\bibinfo {author} {\bibfnamefont {N.}~\bibnamefont
  {Fl{\"a}schner}}, \bibinfo {author} {\bibfnamefont {B.~S.}\ \bibnamefont
  {Rem}}, \bibinfo {author} {\bibfnamefont {M.}~\bibnamefont {Tarnowski}},
  \bibinfo {author} {\bibfnamefont {D.}~\bibnamefont {Vogel}}, \bibinfo
  {author} {\bibfnamefont {D.-S.}\ \bibnamefont {L{\"u}hmann}}, \bibinfo
  {author} {\bibfnamefont {K.}~\bibnamefont {Sengstock}}, \ and\ \bibinfo
  {author} {\bibfnamefont {C.}~\bibnamefont {Weitenberg}},\ }\href {\doibase
  10.1126/science.aad4568} {\bibfield  {journal} {\bibinfo  {journal}
  {Science}\ }\textbf {\bibinfo {volume} {352}},\ \bibinfo {pages} {1091}
  (\bibinfo {year} {2016})}\BibitemShut {NoStop}%
\bibitem [{\citenamefont {Eckardt}(2017)}]{Eckardt17}%
  \BibitemOpen
  \bibfield  {author} {\bibinfo {author} {\bibfnamefont {A.}~\bibnamefont
  {Eckardt}},\ }\href {\doibase 10.1103/RevModPhys.89.011004} {\bibfield
  {journal} {\bibinfo  {journal} {Rev. Mod. Phys.}\ }\textbf {\bibinfo {volume}
  {89}},\ \bibinfo {pages} {011004} (\bibinfo {year} {2017})}\BibitemShut
  {NoStop}%
\bibitem [{\citenamefont {Tarnowski}\ \emph {et~al.}(2019)\citenamefont
  {Tarnowski}, \citenamefont {{\"U}nal}, \citenamefont {Fl{\"a}schner},
  \citenamefont {Rem}, \citenamefont {Eckardt}, \citenamefont {Sengstock},\
  and\ \citenamefont {Weitenberg}}]{TarnowskiEtAl19}%
  \BibitemOpen
  \bibfield  {author} {\bibinfo {author} {\bibfnamefont {M.}~\bibnamefont
  {Tarnowski}}, \bibinfo {author} {\bibfnamefont {F.~N.}\ \bibnamefont
  {{\"U}nal}}, \bibinfo {author} {\bibfnamefont {N.}~\bibnamefont
  {Fl{\"a}schner}}, \bibinfo {author} {\bibfnamefont {B.~S.}\ \bibnamefont
  {Rem}}, \bibinfo {author} {\bibfnamefont {A.}~\bibnamefont {Eckardt}},
  \bibinfo {author} {\bibfnamefont {K.}~\bibnamefont {Sengstock}}, \ and\
  \bibinfo {author} {\bibfnamefont {C.}~\bibnamefont {Weitenberg}},\
  }\href@noop {} {\bibfield  {journal} {\bibinfo  {journal} {Nature
  Communications}\ }\textbf {\bibinfo {volume} {10}},\ \bibinfo {pages} {1}
  (\bibinfo {year} {2019})}\BibitemShut {NoStop}%
\bibitem [{\citenamefont {Viebahn}\ \emph {et~al.}(2021)\citenamefont
  {Viebahn}, \citenamefont {Minguzzi}, \citenamefont {Sandholzer},
  \citenamefont {Walter}, \citenamefont {Sajnani}, \citenamefont {G{\"o}rg},\
  and\ \citenamefont {Esslinger}}]{ViebahnEtAl21}%
  \BibitemOpen
  \bibfield  {author} {\bibinfo {author} {\bibfnamefont {K.}~\bibnamefont
  {Viebahn}}, \bibinfo {author} {\bibfnamefont {J.}~\bibnamefont {Minguzzi}},
  \bibinfo {author} {\bibfnamefont {K.}~\bibnamefont {Sandholzer}}, \bibinfo
  {author} {\bibfnamefont {A.-S.}\ \bibnamefont {Walter}}, \bibinfo {author}
  {\bibfnamefont {M.}~\bibnamefont {Sajnani}}, \bibinfo {author} {\bibfnamefont
  {F.}~\bibnamefont {G{\"o}rg}}, \ and\ \bibinfo {author} {\bibfnamefont
  {T.}~\bibnamefont {Esslinger}},\ }\href {\doibase 10.1103/PhysRevX.11.011057}
  {\bibfield  {journal} {\bibinfo  {journal} {Phys. Rev. X}\ }\textbf {\bibinfo
  {volume} {11}},\ \bibinfo {pages} {011057} (\bibinfo {year}
  {2021})}\BibitemShut {NoStop}%
\bibitem [{\citenamefont {Houck}\ \emph {et~al.}(2012)\citenamefont {Houck},
  \citenamefont {T{\"u}reci},\ and\ \citenamefont {Koch}}]{Koch2012}%
  \BibitemOpen
  \bibfield  {author} {\bibinfo {author} {\bibfnamefont {A.~A.}\ \bibnamefont
  {Houck}}, \bibinfo {author} {\bibfnamefont {H.~E.}\ \bibnamefont
  {T{\"u}reci}}, \ and\ \bibinfo {author} {\bibfnamefont {J.}~\bibnamefont
  {Koch}},\ }\href@noop {} {\bibfield  {journal} {\bibinfo  {journal} {Nature
  Phys.}\ }\textbf {\bibinfo {volume} {8}},\ \bibinfo {pages} {292} (\bibinfo
  {year} {2012})}\BibitemShut {NoStop}%
\bibitem [{\citenamefont {Georgescu}\ \emph {et~al.}(2014)\citenamefont
  {Georgescu}, \citenamefont {Ashhab},\ and\ \citenamefont {Nori}}]{Nori2014}%
  \BibitemOpen
  \bibfield  {author} {\bibinfo {author} {\bibfnamefont {I.~M.}\ \bibnamefont
  {Georgescu}}, \bibinfo {author} {\bibfnamefont {S.}~\bibnamefont {Ashhab}}, \
  and\ \bibinfo {author} {\bibfnamefont {F.}~\bibnamefont {Nori}},\ }\href
  {\doibase 10.1103/revmodphys.86.153} {\bibfield  {journal} {\bibinfo
  {journal} {Rev. Mod. Phys.}\ }\textbf {\bibinfo {volume} {86}},\ \bibinfo
  {pages} {153} (\bibinfo {year} {2014})}\BibitemShut {NoStop}%
\bibitem [{\citenamefont {Breuer}\ \emph {et~al.}(2000)\citenamefont {Breuer},
  \citenamefont {Huber},\ and\ \citenamefont {Petruccione}}]{BreuerEtAl00}%
  \BibitemOpen
  \bibfield  {author} {\bibinfo {author} {\bibfnamefont {H.-P.}\ \bibnamefont
  {Breuer}}, \bibinfo {author} {\bibfnamefont {W.}~\bibnamefont {Huber}}, \
  and\ \bibinfo {author} {\bibfnamefont {F.}~\bibnamefont {Petruccione}},\
  }\href@noop {} {\bibfield  {journal} {\bibinfo  {journal} {Phys.\ Rev.\ E}\
  }\textbf {\bibinfo {volume} {61}},\ \bibinfo {pages} {4883} (\bibinfo {year}
  {2000})}\BibitemShut {NoStop}%
\bibitem [{\citenamefont {Ketzmerick}\ and\ \citenamefont
  {Wustmann}(2010)}]{KetzmerickWustmann10}%
  \BibitemOpen
  \bibfield  {author} {\bibinfo {author} {\bibfnamefont {R.}~\bibnamefont
  {Ketzmerick}}\ and\ \bibinfo {author} {\bibfnamefont {W.}~\bibnamefont
  {Wustmann}},\ }\href@noop {} {\bibfield  {journal} {\bibinfo  {journal}
  {Phys.\ Rev.\ E}\ }\textbf {\bibinfo {volume} {82}},\ \bibinfo {pages}
  {021114} (\bibinfo {year} {2010})}\BibitemShut {NoStop}%
\bibitem [{\citenamefont {Vorberg}\ \emph {et~al.}(2013)\citenamefont
  {Vorberg}, \citenamefont {Wustmann}, \citenamefont {Ketzmerick},\ and\
  \citenamefont {Eckardt}}]{VorbergEtAl13}%
  \BibitemOpen
  \bibfield  {author} {\bibinfo {author} {\bibfnamefont {D.}~\bibnamefont
  {Vorberg}}, \bibinfo {author} {\bibfnamefont {W.}~\bibnamefont {Wustmann}},
  \bibinfo {author} {\bibfnamefont {R.}~\bibnamefont {Ketzmerick}}, \ and\
  \bibinfo {author} {\bibfnamefont {A.}~\bibnamefont {Eckardt}},\ }\href
  {\doibase 10.1103/PhysRevLett.111.240405} {\bibfield  {journal} {\bibinfo
  {journal} {Phys. Rev. Lett.}\ }\textbf {\bibinfo {volume} {111}},\ \bibinfo
  {pages} {240405} (\bibinfo {year} {2013})}\BibitemShut {NoStop}%
\bibitem [{\citenamefont {Shirai}\ \emph {et~al.}(2015)\citenamefont {Shirai},
  \citenamefont {Mori},\ and\ \citenamefont {Miyashita}}]{ShiraiEtAl14}%
  \BibitemOpen
  \bibfield  {author} {\bibinfo {author} {\bibfnamefont {T.}~\bibnamefont
  {Shirai}}, \bibinfo {author} {\bibfnamefont {T.}~\bibnamefont {Mori}}, \ and\
  \bibinfo {author} {\bibfnamefont {S.}~\bibnamefont {Miyashita}},\ }\href
  {\doibase 10.1103/PhysRevE.91.030101} {\bibfield  {journal} {\bibinfo
  {journal} {Phys. Rev. E}\ }\textbf {\bibinfo {volume} {91}},\ \bibinfo
  {pages} {030101} (\bibinfo {year} {2015})}\BibitemShut {NoStop}%
\bibitem [{\citenamefont {Seetharam}\ \emph {et~al.}(2015)\citenamefont
  {Seetharam}, \citenamefont {Bardyn}, \citenamefont {Lindner}, \citenamefont
  {Rudner},\ and\ \citenamefont {Refael}}]{SeetharamEtAl15}%
  \BibitemOpen
  \bibfield  {author} {\bibinfo {author} {\bibfnamefont {K.~I.}\ \bibnamefont
  {Seetharam}}, \bibinfo {author} {\bibfnamefont {C.-E.}\ \bibnamefont
  {Bardyn}}, \bibinfo {author} {\bibfnamefont {N.~H.}\ \bibnamefont {Lindner}},
  \bibinfo {author} {\bibfnamefont {M.~S.}\ \bibnamefont {Rudner}}, \ and\
  \bibinfo {author} {\bibfnamefont {G.}~\bibnamefont {Refael}},\ }\href
  {\doibase 10.1103/PhysRevX.5.041050} {\bibfield  {journal} {\bibinfo
  {journal} {Phys. Rev. X}\ }\textbf {\bibinfo {volume} {5}},\ \bibinfo {pages}
  {041050} (\bibinfo {year} {2015})}\BibitemShut {NoStop}%
\bibitem [{\citenamefont {Dehghani}\ \emph {et~al.}(2015)\citenamefont
  {Dehghani}, \citenamefont {Oka},\ and\ \citenamefont
  {Mitra}}]{DehghaniEtAl15}%
  \BibitemOpen
  \bibfield  {author} {\bibinfo {author} {\bibfnamefont {H.}~\bibnamefont
  {Dehghani}}, \bibinfo {author} {\bibfnamefont {T.}~\bibnamefont {Oka}}, \
  and\ \bibinfo {author} {\bibfnamefont {A.}~\bibnamefont {Mitra}},\ }\href
  {\doibase 10.1103/PhysRevB.91.155422} {\bibfield  {journal} {\bibinfo
  {journal} {Phys. Rev. B}\ }\textbf {\bibinfo {volume} {91}},\ \bibinfo
  {pages} {155422} (\bibinfo {year} {2015})}\BibitemShut {NoStop}%
\bibitem [{\citenamefont {Iadecola}\ \emph {et~al.}(2015)\citenamefont
  {Iadecola}, \citenamefont {Neupert},\ and\ \citenamefont
  {Chamon}}]{IadecolaEtAl15}%
  \BibitemOpen
  \bibfield  {author} {\bibinfo {author} {\bibfnamefont {T.}~\bibnamefont
  {Iadecola}}, \bibinfo {author} {\bibfnamefont {T.}~\bibnamefont {Neupert}}, \
  and\ \bibinfo {author} {\bibfnamefont {C.}~\bibnamefont {Chamon}},\ }\href
  {\doibase 10.1103/PhysRevB.91.235133} {\bibfield  {journal} {\bibinfo
  {journal} {Phys. Rev. B}\ }\textbf {\bibinfo {volume} {91}},\ \bibinfo
  {pages} {235133} (\bibinfo {year} {2015})}\BibitemShut {NoStop}%
\bibitem [{\citenamefont {Vorberg}\ \emph {et~al.}(2015)\citenamefont
  {Vorberg}, \citenamefont {Wustmann}, \citenamefont {Schomerus}, \citenamefont
  {Ketzmerick},\ and\ \citenamefont {Eckardt}}]{VorbergEtAl15}%
  \BibitemOpen
  \bibfield  {author} {\bibinfo {author} {\bibfnamefont {D.}~\bibnamefont
  {Vorberg}}, \bibinfo {author} {\bibfnamefont {W.}~\bibnamefont {Wustmann}},
  \bibinfo {author} {\bibfnamefont {H.}~\bibnamefont {Schomerus}}, \bibinfo
  {author} {\bibfnamefont {R.}~\bibnamefont {Ketzmerick}}, \ and\ \bibinfo
  {author} {\bibfnamefont {A.}~\bibnamefont {Eckardt}},\ }\href {\doibase
  10.1103/PhysRevE.92.062119} {\bibfield  {journal} {\bibinfo  {journal} {Phys.
  Rev. E}\ }\textbf {\bibinfo {volume} {92}},\ \bibinfo {pages} {062119}
  (\bibinfo {year} {2015})}\BibitemShut {NoStop}%
\bibitem [{\citenamefont {Shirai}\ \emph {et~al.}(2016)\citenamefont {Shirai},
  \citenamefont {Thingna}, \citenamefont {Mori}, \citenamefont {Denisov},
  \citenamefont {H{\"a}nggi},\ and\ \citenamefont {Miyashita}}]{ShiraiEtAl16}%
  \BibitemOpen
  \bibfield  {author} {\bibinfo {author} {\bibfnamefont {T.}~\bibnamefont
  {Shirai}}, \bibinfo {author} {\bibfnamefont {J.}~\bibnamefont {Thingna}},
  \bibinfo {author} {\bibfnamefont {T.}~\bibnamefont {Mori}}, \bibinfo {author}
  {\bibfnamefont {S.}~\bibnamefont {Denisov}}, \bibinfo {author} {\bibfnamefont
  {P.}~\bibnamefont {H{\"a}nggi}}, \ and\ \bibinfo {author} {\bibfnamefont
  {S.}~\bibnamefont {Miyashita}},\ }\href@noop {} {\bibfield  {journal}
  {\bibinfo  {journal} {New J.\ Phys.}\ }\textbf {\bibinfo {volume} {18}},\
  \bibinfo {pages} {053008} (\bibinfo {year} {2016})}\BibitemShut {NoStop}%
\bibitem [{\citenamefont {Letscher}\ \emph {et~al.}(2017)\citenamefont
  {Letscher}, \citenamefont {Thomas}, \citenamefont {Niederpr{\"u}m},
  \citenamefont {Fleischhauer},\ and\ \citenamefont {Ott}}]{LetscherEtAl17}%
  \BibitemOpen
  \bibfield  {author} {\bibinfo {author} {\bibfnamefont {F.}~\bibnamefont
  {Letscher}}, \bibinfo {author} {\bibfnamefont {O.}~\bibnamefont {Thomas}},
  \bibinfo {author} {\bibfnamefont {T.}~\bibnamefont {Niederpr{\"u}m}},
  \bibinfo {author} {\bibfnamefont {M.}~\bibnamefont {Fleischhauer}}, \ and\
  \bibinfo {author} {\bibfnamefont {H.}~\bibnamefont {Ott}},\ }\href {\doibase
  10.1103/PhysRevX.7.021020} {\bibfield  {journal} {\bibinfo  {journal} {Phys.
  Rev. X}\ }\textbf {\bibinfo {volume} {7}},\ \bibinfo {pages} {021020}
  (\bibinfo {year} {2017})}\BibitemShut {NoStop}%
\bibitem [{\citenamefont {Schnell}\ \emph {et~al.}(2018)\citenamefont
  {Schnell}, \citenamefont {Ketzmerick},\ and\ \citenamefont
  {Eckardt}}]{SchnellEtAl18}%
  \BibitemOpen
  \bibfield  {author} {\bibinfo {author} {\bibfnamefont {A.}~\bibnamefont
  {Schnell}}, \bibinfo {author} {\bibfnamefont {R.}~\bibnamefont {Ketzmerick}},
  \ and\ \bibinfo {author} {\bibfnamefont {A.}~\bibnamefont {Eckardt}},\ }\href
  {\doibase 10.1103/PhysRevE.97.032136} {\bibfield  {journal} {\bibinfo
  {journal} {Phys. Rev. E}\ }\textbf {\bibinfo {volume} {97}},\ \bibinfo
  {pages} {032136} (\bibinfo {year} {2018})}\BibitemShut {NoStop}%
\bibitem [{\citenamefont {Chong}\ \emph {et~al.}(2018)\citenamefont {Chong},
  \citenamefont {Kim}, \citenamefont {Kim}, \citenamefont {Yoon}, \citenamefont
  {Kang},\ and\ \citenamefont {An}}]{ChongEtAl2018}%
  \BibitemOpen
  \bibfield  {author} {\bibinfo {author} {\bibfnamefont {K.~O.}\ \bibnamefont
  {Chong}}, \bibinfo {author} {\bibfnamefont {J.-R.}\ \bibnamefont {Kim}},
  \bibinfo {author} {\bibfnamefont {J.}~\bibnamefont {Kim}}, \bibinfo {author}
  {\bibfnamefont {S.}~\bibnamefont {Yoon}}, \bibinfo {author} {\bibfnamefont
  {S.}~\bibnamefont {Kang}}, \ and\ \bibinfo {author} {\bibfnamefont
  {K.}~\bibnamefont {An}},\ }\href@noop {} {\bibfield  {journal} {\bibinfo
  {journal} {Nature Communications Physics}\ }\textbf {\bibinfo {volume} {1}},\
  \bibinfo {pages} {25} (\bibinfo {year} {2018})}\BibitemShut {NoStop}%
\bibitem [{\citenamefont {Qin}\ \emph {et~al.}(2018)\citenamefont {Qin},
  \citenamefont {Schnell}, \citenamefont {Sengstock}, \citenamefont
  {Weitenberg}, \citenamefont {Eckardt},\ and\ \citenamefont
  {Hofstetter}}]{QinEtAl18}%
  \BibitemOpen
  \bibfield  {author} {\bibinfo {author} {\bibfnamefont {T.}~\bibnamefont
  {Qin}}, \bibinfo {author} {\bibfnamefont {A.}~\bibnamefont {Schnell}},
  \bibinfo {author} {\bibfnamefont {K.}~\bibnamefont {Sengstock}}, \bibinfo
  {author} {\bibfnamefont {C.}~\bibnamefont {Weitenberg}}, \bibinfo {author}
  {\bibfnamefont {A.}~\bibnamefont {Eckardt}}, \ and\ \bibinfo {author}
  {\bibfnamefont {W.}~\bibnamefont {Hofstetter}},\ }\href {\doibase
  10.1103/PhysRevA.98.033601} {\bibfield  {journal} {\bibinfo  {journal} {Phys.
  Rev. A}\ }\textbf {\bibinfo {volume} {98}},\ \bibinfo {pages} {033601}
  (\bibinfo {year} {2018})}\BibitemShut {NoStop}%
\bibitem [{\citenamefont {Schnell}\ \emph {et~al.}(2020)\citenamefont
  {Schnell}, \citenamefont {Eckardt},\ and\ \citenamefont
  {Denisov}}]{SchnellEtAl18FL}%
  \BibitemOpen
  \bibfield  {author} {\bibinfo {author} {\bibfnamefont {A.}~\bibnamefont
  {Schnell}}, \bibinfo {author} {\bibfnamefont {A.}~\bibnamefont {Eckardt}}, \
  and\ \bibinfo {author} {\bibfnamefont {S.}~\bibnamefont {Denisov}},\ }\href
  {\doibase 10.1103/PhysRevB.101.100301} {\bibfield  {journal} {\bibinfo
  {journal} {Phys. Rev. B}\ }\textbf {\bibinfo {volume} {101}},\ \bibinfo
  {pages} {100301} (\bibinfo {year} {2020})}\BibitemShut {NoStop}%
\bibitem [{\citenamefont {Haddadfarshi}\ \emph {et~al.}(2015)\citenamefont
  {Haddadfarshi}, \citenamefont {Cui},\ and\ \citenamefont
  {Mintert}}]{HaddadfarshiEtAl15}%
  \BibitemOpen
  \bibfield  {author} {\bibinfo {author} {\bibfnamefont {F.}~\bibnamefont
  {Haddadfarshi}}, \bibinfo {author} {\bibfnamefont {J.}~\bibnamefont {Cui}}, \
  and\ \bibinfo {author} {\bibfnamefont {F.}~\bibnamefont {Mintert}},\ }\href
  {\doibase 10.1103/PhysRevLett.114.130402} {\bibfield  {journal} {\bibinfo
  {journal} {Phys. Rev. Lett.}\ }\textbf {\bibinfo {volume} {114}},\ \bibinfo
  {pages} {130402} (\bibinfo {year} {2015})}\BibitemShut {NoStop}%
\bibitem [{\citenamefont {Reimer}\ \emph {et~al.}(2018)\citenamefont {Reimer},
  \citenamefont {Pedersen}, \citenamefont {Tanger}, \citenamefont
  {Pletyukhov},\ and\ \citenamefont {Gritsev}}]{ReimerEtAl18}%
  \BibitemOpen
  \bibfield  {author} {\bibinfo {author} {\bibfnamefont {V.}~\bibnamefont
  {Reimer}}, \bibinfo {author} {\bibfnamefont {K.~G.~L.}\ \bibnamefont
  {Pedersen}}, \bibinfo {author} {\bibfnamefont {N.}~\bibnamefont {Tanger}},
  \bibinfo {author} {\bibfnamefont {M.}~\bibnamefont {Pletyukhov}}, \ and\
  \bibinfo {author} {\bibfnamefont {V.}~\bibnamefont {Gritsev}},\ }\href
  {\doibase 10.1103/PhysRevA.97.043851} {\bibfield  {journal} {\bibinfo
  {journal} {Phys. Rev. A}\ }\textbf {\bibinfo {volume} {97}},\ \bibinfo
  {pages} {043851} (\bibinfo {year} {2018})}\BibitemShut {NoStop}%
\bibitem [{\citenamefont {Restrepo}\ \emph {et~al.}(2016)\citenamefont
  {Restrepo}, \citenamefont {Cerrillo}, \citenamefont {Bastidas}, \citenamefont
  {Angelakis},\ and\ \citenamefont {Brandes}}]{RestrepoEtAl17}%
  \BibitemOpen
  \bibfield  {author} {\bibinfo {author} {\bibfnamefont {S.}~\bibnamefont
  {Restrepo}}, \bibinfo {author} {\bibfnamefont {J.}~\bibnamefont {Cerrillo}},
  \bibinfo {author} {\bibfnamefont {V.~M.}\ \bibnamefont {Bastidas}}, \bibinfo
  {author} {\bibfnamefont {D.~G.}\ \bibnamefont {Angelakis}}, \ and\ \bibinfo
  {author} {\bibfnamefont {T.}~\bibnamefont {Brandes}},\ }\href {\doibase
  10.1103/PhysRevLett.117.250401} {\bibfield  {journal} {\bibinfo  {journal}
  {Phys. Rev. Lett.}\ }\textbf {\bibinfo {volume} {117}},\ \bibinfo {pages}
  {250401} (\bibinfo {year} {2016})}\BibitemShut {NoStop}%
\bibitem [{\citenamefont {Dai}\ \emph {et~al.}(2016)\citenamefont {Dai},
  \citenamefont {Shi},\ and\ \citenamefont {Yi}}]{DaiEtAl16}%
  \BibitemOpen
  \bibfield  {author} {\bibinfo {author} {\bibfnamefont {C.~M.}\ \bibnamefont
  {Dai}}, \bibinfo {author} {\bibfnamefont {Z.~C.}\ \bibnamefont {Shi}}, \ and\
  \bibinfo {author} {\bibfnamefont {X.~X.}\ \bibnamefont {Yi}},\ }\href
  {\doibase 10.1103/PhysRevA.93.032121} {\bibfield  {journal} {\bibinfo
  {journal} {Phys. Rev. A}\ }\textbf {\bibinfo {volume} {93}},\ \bibinfo
  {pages} {032121} (\bibinfo {year} {2016})}\BibitemShut {NoStop}%
\bibitem [{\citenamefont {Dai}\ \emph {et~al.}(2017)\citenamefont {Dai},
  \citenamefont {Li}, \citenamefont {Wang},\ and\ \citenamefont
  {Yi}}]{DaiEtAl2017}%
  \BibitemOpen
  \bibfield  {author} {\bibinfo {author} {\bibfnamefont {C.}~\bibnamefont
  {Dai}}, \bibinfo {author} {\bibfnamefont {H.}~\bibnamefont {Li}}, \bibinfo
  {author} {\bibfnamefont {W.}~\bibnamefont {Wang}}, \ and\ \bibinfo {author}
  {\bibfnamefont {X.}~\bibnamefont {Yi}},\ }\href@noop {} {\bibfield  {journal}
  {\bibinfo  {journal} {arXiv:1707.05030}\ } (\bibinfo {year}
  {2017})}\BibitemShut {NoStop}%
\bibitem [{\citenamefont {Hotz}\ and\ \citenamefont
  {Schaller}(2021)}]{HotzSchaller21}%
  \BibitemOpen
  \bibfield  {author} {\bibinfo {author} {\bibfnamefont {R.}~\bibnamefont
  {Hotz}}\ and\ \bibinfo {author} {\bibfnamefont {G.}~\bibnamefont
  {Schaller}},\ }\href@noop {} {\enquote {\bibinfo {title} {Coarse-graining
  master equation for periodically driven systems},}\ } (\bibinfo {year}
  {2021}),\ \Eprint {http://arxiv.org/abs/2102.03063} {arXiv:2102.03063
  [quant-ph]} \BibitemShut {NoStop}%
\bibitem [{\citenamefont {Szczygielski}(2021)}]{Szczygielski21}%
  \BibitemOpen
  \bibfield  {author} {\bibinfo {author} {\bibfnamefont {K.}~\bibnamefont
  {Szczygielski}},\ }\href {\doibase 10.1016/j.laa.2020.09.005} {\bibfield
  {journal} {\bibinfo  {journal} {Linear Algebra and its Applications}\
  }\textbf {\bibinfo {volume} {609}},\ \bibinfo {pages} {176} (\bibinfo {year}
  {2021})}\BibitemShut {NoStop}%
\bibitem [{\citenamefont {Gunderson}\ \emph {et~al.}(2021)\citenamefont
  {Gunderson}, \citenamefont {Muldoon}, \citenamefont {Murch},\ and\
  \citenamefont {Joglekar}}]{GundersonEtAl21}%
  \BibitemOpen
  \bibfield  {author} {\bibinfo {author} {\bibfnamefont {J.}~\bibnamefont
  {Gunderson}}, \bibinfo {author} {\bibfnamefont {J.}~\bibnamefont {Muldoon}},
  \bibinfo {author} {\bibfnamefont {K.~W.}\ \bibnamefont {Murch}}, \ and\
  \bibinfo {author} {\bibfnamefont {Y.~N.}\ \bibnamefont {Joglekar}},\ }\href
  {\doibase 10.1103/PhysRevA.103.023718} {\bibfield  {journal} {\bibinfo
  {journal} {Phys. Rev. A}\ }\textbf {\bibinfo {volume} {103}},\ \bibinfo
  {pages} {023718} (\bibinfo {year} {2021})}\BibitemShut {NoStop}%
\bibitem [{\citenamefont {Mizuta}\ \emph {et~al.}(2021)\citenamefont {Mizuta},
  \citenamefont {Takasan},\ and\ \citenamefont {Kawakami}}]{MizutaEtAl21}%
  \BibitemOpen
  \bibfield  {author} {\bibinfo {author} {\bibfnamefont {K.}~\bibnamefont
  {Mizuta}}, \bibinfo {author} {\bibfnamefont {K.}~\bibnamefont {Takasan}}, \
  and\ \bibinfo {author} {\bibfnamefont {N.}~\bibnamefont {Kawakami}},\ }\href
  {\doibase 10.1103/PhysRevA.103.L020202} {\bibfield  {journal} {\bibinfo
  {journal} {Phys. Rev. A}\ }\textbf {\bibinfo {volume} {103}},\ \bibinfo
  {pages} {L020202} (\bibinfo {year} {2021})}\BibitemShut {NoStop}%
\bibitem [{\citenamefont {Eckardt}\ and\ \citenamefont
  {Anisimovas}(2015)}]{EckardtAnisimovas15}%
  \BibitemOpen
  \bibfield  {author} {\bibinfo {author} {\bibfnamefont {A.}~\bibnamefont
  {Eckardt}}\ and\ \bibinfo {author} {\bibfnamefont {E.}~\bibnamefont
  {Anisimovas}},\ }\href@noop {} {\bibfield  {journal} {\bibinfo  {journal}
  {New J.\ Phys.}\ }\textbf {\bibinfo {volume} {17}},\ \bibinfo {pages}
  {093039} (\bibinfo {year} {2015})}\BibitemShut {NoStop}%
\bibitem [{\citenamefont {Weinberg}\ \emph {et~al.}(2015)\citenamefont
  {Weinberg}, \citenamefont {{\"O}lschl{\"a}ger}, \citenamefont {Str{\"a}ter},
  \citenamefont {Prelle}, \citenamefont {Eckardt}, \citenamefont {Sengstock},\
  and\ \citenamefont {Simonet}}]{WeinbergEtAl15}%
  \BibitemOpen
  \bibfield  {author} {\bibinfo {author} {\bibfnamefont {M.}~\bibnamefont
  {Weinberg}}, \bibinfo {author} {\bibfnamefont {C.}~\bibnamefont
  {{\"O}lschl{\"a}ger}}, \bibinfo {author} {\bibfnamefont {C.}~\bibnamefont
  {Str{\"a}ter}}, \bibinfo {author} {\bibfnamefont {S.}~\bibnamefont {Prelle}},
  \bibinfo {author} {\bibfnamefont {A.}~\bibnamefont {Eckardt}}, \bibinfo
  {author} {\bibfnamefont {K.}~\bibnamefont {Sengstock}}, \ and\ \bibinfo
  {author} {\bibfnamefont {J.}~\bibnamefont {Simonet}},\ }\href {\doibase
  10.1103/PhysRevA.92.043621} {\bibfield  {journal} {\bibinfo  {journal} {Phys.
  Rev. A}\ }\textbf {\bibinfo {volume} {92}},\ \bibinfo {pages} {043621}
  (\bibinfo {year} {2015})}\BibitemShut {NoStop}%
\bibitem [{\citenamefont {Bilitewski}\ and\ \citenamefont
  {Cooper}(2015)}]{BilitewskiCooper15b}%
  \BibitemOpen
  \bibfield  {author} {\bibinfo {author} {\bibfnamefont {T.}~\bibnamefont
  {Bilitewski}}\ and\ \bibinfo {author} {\bibfnamefont {N.~R.}\ \bibnamefont
  {Cooper}},\ }\href {\doibase 10.1103/PhysRevA.91.063611} {\bibfield
  {journal} {\bibinfo  {journal} {Phys. Rev. A}\ }\textbf {\bibinfo {volume}
  {91}},\ \bibinfo {pages} {063611} (\bibinfo {year} {2015})}\BibitemShut
  {NoStop}%
\bibitem [{\citenamefont {Wintersperger}\ \emph {et~al.}(2020)\citenamefont
  {Wintersperger}, \citenamefont {Bukov}, \citenamefont {N\"ager},
  \citenamefont {Lellouch}, \citenamefont {Demler}, \citenamefont {Schneider},
  \citenamefont {Bloch}, \citenamefont {Goldman},\ and\ \citenamefont
  {Aidelsburger}}]{WinterspergerEtAl20}%
  \BibitemOpen
  \bibfield  {author} {\bibinfo {author} {\bibfnamefont {K.}~\bibnamefont
  {Wintersperger}}, \bibinfo {author} {\bibfnamefont {M.}~\bibnamefont
  {Bukov}}, \bibinfo {author} {\bibfnamefont {J.}~\bibnamefont {N\"ager}},
  \bibinfo {author} {\bibfnamefont {S.}~\bibnamefont {Lellouch}}, \bibinfo
  {author} {\bibfnamefont {E.}~\bibnamefont {Demler}}, \bibinfo {author}
  {\bibfnamefont {U.}~\bibnamefont {Schneider}}, \bibinfo {author}
  {\bibfnamefont {I.}~\bibnamefont {Bloch}}, \bibinfo {author} {\bibfnamefont
  {N.}~\bibnamefont {Goldman}}, \ and\ \bibinfo {author} {\bibfnamefont
  {M.}~\bibnamefont {Aidelsburger}},\ }\href {\doibase
  10.1103/PhysRevX.10.011030} {\bibfield  {journal} {\bibinfo  {journal} {Phys.
  Rev. X}\ }\textbf {\bibinfo {volume} {10}},\ \bibinfo {pages} {011030}
  (\bibinfo {year} {2020})}\BibitemShut {NoStop}%
\bibitem [{\citenamefont {Blanes}\ \emph {et~al.}(2009)\citenamefont {Blanes},
  \citenamefont {Casas}, \citenamefont {Oteo},\ and\ \citenamefont
  {Ros}}]{BlanesEtAl09}%
  \BibitemOpen
  \bibfield  {author} {\bibinfo {author} {\bibfnamefont {S.}~\bibnamefont
  {Blanes}}, \bibinfo {author} {\bibfnamefont {F.}~\bibnamefont {Casas}},
  \bibinfo {author} {\bibfnamefont {J.~A.}\ \bibnamefont {Oteo}}, \ and\
  \bibinfo {author} {\bibfnamefont {J.}~\bibnamefont {Ros}},\ }\href@noop {}
  {\bibfield  {journal} {\bibinfo  {journal} {Physics Reports}\ }\textbf
  {\bibinfo {volume} {470}},\ \bibinfo {pages} {151} (\bibinfo {year}
  {2009})}\BibitemShut {NoStop}%
\bibitem [{\citenamefont {Ikeda}\ \emph {et~al.}(2021)\citenamefont {Ikeda},
  \citenamefont {Chinzei},\ and\ \citenamefont {Sato}}]{IkedaEtAl21}%
  \BibitemOpen
  \bibfield  {author} {\bibinfo {author} {\bibfnamefont {T.~N.}\ \bibnamefont
  {Ikeda}}, \bibinfo {author} {\bibfnamefont {K.}~\bibnamefont {Chinzei}}, \
  and\ \bibinfo {author} {\bibfnamefont {M.}~\bibnamefont {Sato}},\ }\href@noop
  {} {\bibfield  {journal} {\bibinfo  {journal} {arXiv preprint
  arXiv:2107.07911}\ } (\bibinfo {year} {2021})}\BibitemShut {NoStop}%
\bibitem [{\citenamefont {Breuer}\ \emph {et~al.}(2009)\citenamefont {Breuer},
  \citenamefont {Laine},\ and\ \citenamefont {Piilo}}]{BreuerEtAl09}%
  \BibitemOpen
  \bibfield  {author} {\bibinfo {author} {\bibfnamefont {H.-P.}\ \bibnamefont
  {Breuer}}, \bibinfo {author} {\bibfnamefont {E.-M.}\ \bibnamefont {Laine}}, \
  and\ \bibinfo {author} {\bibfnamefont {J.}~\bibnamefont {Piilo}},\ }\href
  {\doibase 10.1103/PhysRevLett.103.210401} {\bibfield  {journal} {\bibinfo
  {journal} {Phys. Rev. Lett.}\ }\textbf {\bibinfo {volume} {103}},\ \bibinfo
  {pages} {210401} (\bibinfo {year} {2009})}\BibitemShut {NoStop}%
\bibitem [{\citenamefont {Rivas}\ \emph {et~al.}(2014)\citenamefont {Rivas},
  \citenamefont {Huelga},\ and\ \citenamefont {Plenio}}]{RivasEtAl2014}%
  \BibitemOpen
  \bibfield  {author} {\bibinfo {author} {\bibfnamefont {{\'{A}}.}~\bibnamefont
  {Rivas}}, \bibinfo {author} {\bibfnamefont {S.~F.}\ \bibnamefont {Huelga}}, \
  and\ \bibinfo {author} {\bibfnamefont {M.~B.}\ \bibnamefont {Plenio}},\
  }\href {\doibase 10.1088/0034-4885/77/9/094001} {\bibfield  {journal}
  {\bibinfo  {journal} {Reports on Progress in Physics}\ }\textbf {\bibinfo
  {volume} {77}},\ \bibinfo {pages} {094001} (\bibinfo {year}
  {2014})}\BibitemShut {NoStop}%
\bibitem [{\citenamefont {Breuer}\ \emph {et~al.}(2016)\citenamefont {Breuer},
  \citenamefont {Laine}, \citenamefont {Piilo},\ and\ \citenamefont
  {Vacchini}}]{BreuerEtAl16RMP}%
  \BibitemOpen
  \bibfield  {author} {\bibinfo {author} {\bibfnamefont {H.-P.}\ \bibnamefont
  {Breuer}}, \bibinfo {author} {\bibfnamefont {E.-M.}\ \bibnamefont {Laine}},
  \bibinfo {author} {\bibfnamefont {J.}~\bibnamefont {Piilo}}, \ and\ \bibinfo
  {author} {\bibfnamefont {B.}~\bibnamefont {Vacchini}},\ }\href {\doibase
  10.1103/RevModPhys.88.021002} {\bibfield  {journal} {\bibinfo  {journal}
  {Rev. Mod. Phys.}\ }\textbf {\bibinfo {volume} {88}},\ \bibinfo {pages}
  {021002} (\bibinfo {year} {2016})}\BibitemShut {NoStop}%
\bibitem [{\citenamefont {de~Vega}\ and\ \citenamefont
  {Alonso}(2017)}]{DeVegaAlonsoRMP17}%
  \BibitemOpen
  \bibfield  {author} {\bibinfo {author} {\bibfnamefont {I.}~\bibnamefont
  {de~Vega}}\ and\ \bibinfo {author} {\bibfnamefont {D.}~\bibnamefont
  {Alonso}},\ }\href {\doibase 10.1103/RevModPhys.89.015001} {\bibfield
  {journal} {\bibinfo  {journal} {Rev. Mod. Phys.}\ }\textbf {\bibinfo {volume}
  {89}},\ \bibinfo {pages} {015001} (\bibinfo {year} {2017})}\BibitemShut
  {NoStop}%
\bibitem [{\citenamefont {Wolf}\ and\ \citenamefont {Cirac}(2008)}]{Wolf2008}%
  \BibitemOpen
  \bibfield  {author} {\bibinfo {author} {\bibfnamefont {M.~M.}\ \bibnamefont
  {Wolf}}\ and\ \bibinfo {author} {\bibfnamefont {J.~I.}\ \bibnamefont
  {Cirac}},\ }\href@noop {} {\bibfield  {journal} {\bibinfo  {journal}
  {Communications in Mathematical Physics}\ }\textbf {\bibinfo {volume}
  {279}},\ \bibinfo {pages} {147} (\bibinfo {year} {2008})}\BibitemShut
  {NoStop}%
\bibitem [{\citenamefont {Addis}\ \emph {et~al.}(2014)\citenamefont {Addis},
  \citenamefont {Bylicka}, \citenamefont {Chru\ifmmode \acute{s}\else
  \'{s}\fi{}ci\ifmmode~\acute{n}\else \'{n}\fi{}ski},\ and\ \citenamefont
  {Maniscalco}}]{AddisEtAl14}%
  \BibitemOpen
  \bibfield  {author} {\bibinfo {author} {\bibfnamefont {C.}~\bibnamefont
  {Addis}}, \bibinfo {author} {\bibfnamefont {B.}~\bibnamefont {Bylicka}},
  \bibinfo {author} {\bibfnamefont {D.}~\bibnamefont {Chru\ifmmode
  \acute{s}\else \'{s}\fi{}ci\ifmmode~\acute{n}\else \'{n}\fi{}ski}}, \ and\
  \bibinfo {author} {\bibfnamefont {S.}~\bibnamefont {Maniscalco}},\ }\href
  {\doibase 10.1103/PhysRevA.90.052103} {\bibfield  {journal} {\bibinfo
  {journal} {Phys. Rev. A}\ }\textbf {\bibinfo {volume} {90}},\ \bibinfo
  {pages} {052103} (\bibinfo {year} {2014})}\BibitemShut {NoStop}%
\bibitem [{\citenamefont {Siudzi{\'{n}}ska}\ and\ \citenamefont
  {Chru{\'{s}}ci{\'{n}}ski}(2020)}]{SiudzinskaChruscinski2020}%
  \BibitemOpen
  \bibfield  {author} {\bibinfo {author} {\bibfnamefont {K.}~\bibnamefont
  {Siudzi{\'{n}}ska}}\ and\ \bibinfo {author} {\bibfnamefont {D.}~\bibnamefont
  {Chru{\'{s}}ci{\'{n}}ski}},\ }\href {\doibase 10.1088/1751-8121/aba7f2}
  {\bibfield  {journal} {\bibinfo  {journal} {Journal of Physics A:
  Mathematical and Theoretical}\ }\textbf {\bibinfo {volume} {53}},\ \bibinfo
  {pages} {375305} (\bibinfo {year} {2020})}\BibitemShut {NoStop}%
\bibitem [{\citenamefont {Breuer}\ and\ \citenamefont
  {Petruccione}(2002)}]{BreuerPetruccione}%
  \BibitemOpen
  \bibfield  {author} {\bibinfo {author} {\bibfnamefont {H.}~\bibnamefont
  {Breuer}}\ and\ \bibinfo {author} {\bibfnamefont {F.}~\bibnamefont
  {Petruccione}},\ }\href@noop {} {\emph {\bibinfo {title} {The Theory of Open
  Quantum Systems}}}\ (\bibinfo  {publisher} {Oxford University Press},\
  \bibinfo {address} {Oxford \& New York},\ \bibinfo {year} {2002})\BibitemShut
  {NoStop}%
\bibitem [{\citenamefont {Gorini}\ \emph {et~al.}(1976)\citenamefont {Gorini},
  \citenamefont {Kossakowski},\ and\ \citenamefont {Sudarshan}}]{GoriniEtAl76}%
  \BibitemOpen
  \bibfield  {author} {\bibinfo {author} {\bibfnamefont {V.}~\bibnamefont
  {Gorini}}, \bibinfo {author} {\bibfnamefont {A.}~\bibnamefont {Kossakowski}},
  \ and\ \bibinfo {author} {\bibfnamefont {E.~C.~G.}\ \bibnamefont
  {Sudarshan}},\ }\href {\doibase 10.1063/1.522979} {\bibfield  {journal}
  {\bibinfo  {journal} {J. Math. Phys.}\ }\textbf {\bibinfo {volume} {17}},\
  \bibinfo {pages} {821} (\bibinfo {year} {1976})}\BibitemShut {NoStop}%
\bibitem [{\citenamefont {Lindblad}(1976)}]{Lindblad1976}%
  \BibitemOpen
  \bibfield  {author} {\bibinfo {author} {\bibfnamefont {G.}~\bibnamefont
  {Lindblad}},\ }\href@noop {} {\bibfield  {journal} {\bibinfo  {journal}
  {Comm. Math. Phys.}\ }\textbf {\bibinfo {volume} {48}},\ \bibinfo {pages}
  {119} (\bibinfo {year} {1976})}\BibitemShut {NoStop}%
\bibitem [{\citenamefont {Holevo}(2012)}]{Holevo2012}%
  \BibitemOpen
  \bibfield  {author} {\bibinfo {author} {\bibfnamefont {A.~S.}\ \bibnamefont
  {Holevo}},\ }\href@noop {} {\emph {\bibinfo {title} {Quantum systems,
  channels, information: a mathematical introduction}}},\ Vol.~\bibinfo
  {volume} {16}\ (\bibinfo  {publisher} {Walter de Gruyter},\ \bibinfo {year}
  {2012})\BibitemShut {NoStop}%
\bibitem [{\citenamefont {Szczygielski}(2014)}]{szczygielski2014}%
  \BibitemOpen
  \bibfield  {author} {\bibinfo {author} {\bibfnamefont {K.}~\bibnamefont
  {Szczygielski}},\ }\href@noop {} {\bibfield  {journal} {\bibinfo  {journal}
  {J. Math. Phys.}\ }\textbf {\bibinfo {volume} {55}},\ \bibinfo {pages}
  {083506} (\bibinfo {year} {2014})}\BibitemShut {NoStop}%
\bibitem [{\citenamefont {Hartmann}\ \emph {et~al.}(2017)\citenamefont
  {Hartmann}, \citenamefont {Poletti}, \citenamefont {Ivanchenko},
  \citenamefont {Denisov},\ and\ \citenamefont {H{\"a}nggi}}]{HartmannEtAl17}%
  \BibitemOpen
  \bibfield  {author} {\bibinfo {author} {\bibfnamefont {M.}~\bibnamefont
  {Hartmann}}, \bibinfo {author} {\bibfnamefont {D.}~\bibnamefont {Poletti}},
  \bibinfo {author} {\bibfnamefont {M.}~\bibnamefont {Ivanchenko}}, \bibinfo
  {author} {\bibfnamefont {S.}~\bibnamefont {Denisov}}, \ and\ \bibinfo
  {author} {\bibfnamefont {P.}~\bibnamefont {H{\"a}nggi}},\ }\href
  {http://stacks.iop.org/1367-2630/19/i=8/a=083011} {\bibfield  {journal}
  {\bibinfo  {journal} {New J.\ Phys.}\ }\textbf {\bibinfo {volume} {19}},\
  \bibinfo {pages} {083011} (\bibinfo {year} {2017})}\BibitemShut {NoStop}%
\bibitem [{Note1()}]{Note1}%
  \BibitemOpen
  \bibinfo {note} {Note that, despite the fact that the Floquet generator of an
  isolated system is Hermitian and can, thus, be considered a Floquet \protect
  \emph {Hamiltonian}, its properties can be rather different from those of a
  Hamiltonian of an undriven system. Namely, for generic (interacting
  non-integrable) systems, the eigenstates of the Floquet Hamiltonian are
  expected to be superpositions of states at all energies corresponding to
  infinite-temperature ensembles in the sense of eigenstate thermalisation
  \cite {LazaridesEtAl14b,DAlessioRigol14}.}\BibitemShut {Stop}%
\bibitem [{\citenamefont {Wolf}\ \emph {et~al.}(2008)\citenamefont {Wolf},
  \citenamefont {Eisert}, \citenamefont {Cubitt},\ and\ \citenamefont
  {Cirac}}]{WolfEtAl08}%
  \BibitemOpen
  \bibfield  {author} {\bibinfo {author} {\bibfnamefont {M.~M.}\ \bibnamefont
  {Wolf}}, \bibinfo {author} {\bibfnamefont {J.}~\bibnamefont {Eisert}},
  \bibinfo {author} {\bibfnamefont {T.~S.}\ \bibnamefont {Cubitt}}, \ and\
  \bibinfo {author} {\bibfnamefont {J.~I.}\ \bibnamefont {Cirac}},\ }\href@noop
  {} {\bibfield  {journal} {\bibinfo  {journal} {Phys. Rev. Lett.}\ }\textbf
  {\bibinfo {volume} {101}} (\bibinfo {year} {2008})}\BibitemShut {NoStop}%
\bibitem [{\citenamefont {Chakraborty}\ and\ \citenamefont {Chru\ifmmode
  \acute{s}\else \'{s}\fi{}ci\ifmmode~\acute{n}\else
  \'{n}\fi{}ski}(2019)}]{ChakrabortyEtAl19}%
  \BibitemOpen
  \bibfield  {author} {\bibinfo {author} {\bibfnamefont {S.}~\bibnamefont
  {Chakraborty}}\ and\ \bibinfo {author} {\bibfnamefont {D.}~\bibnamefont
  {Chru\ifmmode \acute{s}\else \'{s}\fi{}ci\ifmmode~\acute{n}\else
  \'{n}\fi{}ski}},\ }\href {\doibase 10.1103/PhysRevA.99.042105} {\bibfield
  {journal} {\bibinfo  {journal} {Phys. Rev. A}\ }\textbf {\bibinfo {volume}
  {99}},\ \bibinfo {pages} {042105} (\bibinfo {year} {2019})}\BibitemShut
  {NoStop}%
\bibitem [{\citenamefont {Chru\ifmmode \acute{s}\else
  \'{s}\fi{}ci\ifmmode~\acute{n}\else \'{n}\fi{}ski}\ \emph
  {et~al.}(2011)\citenamefont {Chru\ifmmode \acute{s}\else
  \'{s}\fi{}ci\ifmmode~\acute{n}\else \'{n}\fi{}ski}, \citenamefont
  {Kossakowski},\ and\ \citenamefont {Rivas}}]{ChruscinskiEtAl11}%
  \BibitemOpen
  \bibfield  {author} {\bibinfo {author} {\bibfnamefont {D.}~\bibnamefont
  {Chru\ifmmode \acute{s}\else \'{s}\fi{}ci\ifmmode~\acute{n}\else
  \'{n}\fi{}ski}}, \bibinfo {author} {\bibfnamefont {A.}~\bibnamefont
  {Kossakowski}}, \ and\ \bibinfo {author} {\bibfnamefont {A.}~\bibnamefont
  {Rivas}},\ }\href {\doibase 10.1103/PhysRevA.83.052128} {\bibfield  {journal}
  {\bibinfo  {journal} {Phys. Rev. A}\ }\textbf {\bibinfo {volume} {83}},\
  \bibinfo {pages} {052128} (\bibinfo {year} {2011})}\BibitemShut {NoStop}%
\bibitem [{\citenamefont {Chru\ifmmode \acute{s}\else
  \'{s}\fi{}ci\ifmmode~\acute{n}\else \'{n}\fi{}ski}\ and\ \citenamefont
  {Maniscalco}(2014)}]{ChruscinskiManiscalco14}%
  \BibitemOpen
  \bibfield  {author} {\bibinfo {author} {\bibfnamefont {D.}~\bibnamefont
  {Chru\ifmmode \acute{s}\else \'{s}\fi{}ci\ifmmode~\acute{n}\else
  \'{n}\fi{}ski}}\ and\ \bibinfo {author} {\bibfnamefont {S.}~\bibnamefont
  {Maniscalco}},\ }\href {\doibase 10.1103/PhysRevLett.112.120404} {\bibfield
  {journal} {\bibinfo  {journal} {Phys. Rev. Lett.}\ }\textbf {\bibinfo
  {volume} {112}},\ \bibinfo {pages} {120404} (\bibinfo {year}
  {2014})}\BibitemShut {NoStop}%
\bibitem [{\citenamefont {Rivas}\ \emph {et~al.}(2010)\citenamefont {Rivas},
  \citenamefont {Huelga},\ and\ \citenamefont {Plenio}}]{RivasEtAl10}%
  \BibitemOpen
  \bibfield  {author} {\bibinfo {author} {\bibfnamefont {{\'A}.}~\bibnamefont
  {Rivas}}, \bibinfo {author} {\bibfnamefont {S.~F.}\ \bibnamefont {Huelga}}, \
  and\ \bibinfo {author} {\bibfnamefont {M.~B.}\ \bibnamefont {Plenio}},\
  }\href {\doibase 10.1103/PhysRevLett.105.050403} {\bibfield  {journal}
  {\bibinfo  {journal} {Phys. Rev. Lett.}\ }\textbf {\bibinfo {volume} {105}},\
  \bibinfo {pages} {050403} (\bibinfo {year} {2010})}\BibitemShut {NoStop}%
\bibitem [{\citenamefont {Grossmann}\ \emph {et~al.}(1991)\citenamefont
  {Grossmann}, \citenamefont {Dittrich}, \citenamefont {Jung},\ and\
  \citenamefont {H{\"a}nggi}}]{GrossmannEtAl91}%
  \BibitemOpen
  \bibfield  {author} {\bibinfo {author} {\bibfnamefont {F.}~\bibnamefont
  {Grossmann}}, \bibinfo {author} {\bibfnamefont {T.}~\bibnamefont {Dittrich}},
  \bibinfo {author} {\bibfnamefont {P.}~\bibnamefont {Jung}}, \ and\ \bibinfo
  {author} {\bibfnamefont {P.}~\bibnamefont {H{\"a}nggi}},\ }\href@noop {}
  {\bibfield  {journal} {\bibinfo  {journal} {Phys.\ Rev.\ Lett.}\ }\textbf
  {\bibinfo {volume} {67}},\ \bibinfo {pages} {516} (\bibinfo {year}
  {1991})}\BibitemShut {NoStop}%
\bibitem [{\citenamefont {Eckardt}\ \emph {et~al.}(2005)\citenamefont
  {Eckardt}, \citenamefont {Weiss},\ and\ \citenamefont
  {Holthaus}}]{EckardtEtAl05b}%
  \BibitemOpen
  \bibfield  {author} {\bibinfo {author} {\bibfnamefont {A.}~\bibnamefont
  {Eckardt}}, \bibinfo {author} {\bibfnamefont {C.}~\bibnamefont {Weiss}}, \
  and\ \bibinfo {author} {\bibfnamefont {M.}~\bibnamefont {Holthaus}},\
  }\href@noop {} {\bibfield  {journal} {\bibinfo  {journal} {Phys.\ Rev.\
  Lett.}\ }\textbf {\bibinfo {volume} {95}},\ \bibinfo {pages} {260404}
  (\bibinfo {year} {2005})}\BibitemShut {NoStop}%
\bibitem [{\citenamefont {Eckardt}\ \emph {et~al.}(2009)\citenamefont
  {Eckardt}, \citenamefont {Holthaus}, \citenamefont {Lignier}, \citenamefont
  {Zenesini}, \citenamefont {Ciampini}, \citenamefont {Morsch},\ and\
  \citenamefont {Arimondo}}]{EckardtEtAl09}%
  \BibitemOpen
  \bibfield  {author} {\bibinfo {author} {\bibfnamefont {A.}~\bibnamefont
  {Eckardt}}, \bibinfo {author} {\bibfnamefont {M.}~\bibnamefont {Holthaus}},
  \bibinfo {author} {\bibfnamefont {H.}~\bibnamefont {Lignier}}, \bibinfo
  {author} {\bibfnamefont {A.}~\bibnamefont {Zenesini}}, \bibinfo {author}
  {\bibfnamefont {D.}~\bibnamefont {Ciampini}}, \bibinfo {author}
  {\bibfnamefont {O.}~\bibnamefont {Morsch}}, \ and\ \bibinfo {author}
  {\bibfnamefont {E.}~\bibnamefont {Arimondo}},\ }\href@noop {} {\bibfield
  {journal} {\bibinfo  {journal} {Phys.\ Rev.\ A}\ }\textbf {\bibinfo {volume}
  {79}},\ \bibinfo {pages} {013611} (\bibinfo {year} {2009})}\BibitemShut
  {NoStop}%
\bibitem [{\citenamefont {Moan}\ and\ \citenamefont
  {Niesen}(2008)}]{MoanNiesen2008}%
  \BibitemOpen
  \bibfield  {author} {\bibinfo {author} {\bibfnamefont {P.~C.}\ \bibnamefont
  {Moan}}\ and\ \bibinfo {author} {\bibfnamefont {J.}~\bibnamefont {Niesen}},\
  }\href@noop {} {\bibfield  {journal} {\bibinfo  {journal} {Foundations of
  Computational Mathematics}\ }\textbf {\bibinfo {volume} {8}},\ \bibinfo
  {pages} {291} (\bibinfo {year} {2008})}\BibitemShut {NoStop}%
\bibitem [{\citenamefont {Bastidas}\ \emph {et~al.}(2018)\citenamefont
  {Bastidas}, \citenamefont {Kyaw}, \citenamefont {Tangpanitanon},
  \citenamefont {Romero}, \citenamefont {Kwek},\ and\ \citenamefont
  {Angelakis}}]{BastidasEtAl18}%
  \BibitemOpen
  \bibfield  {author} {\bibinfo {author} {\bibfnamefont {V.~M.}\ \bibnamefont
  {Bastidas}}, \bibinfo {author} {\bibfnamefont {T.~H.}\ \bibnamefont {Kyaw}},
  \bibinfo {author} {\bibfnamefont {J.}~\bibnamefont {Tangpanitanon}}, \bibinfo
  {author} {\bibfnamefont {G.}~\bibnamefont {Romero}}, \bibinfo {author}
  {\bibfnamefont {L.-C.}\ \bibnamefont {Kwek}}, \ and\ \bibinfo {author}
  {\bibfnamefont {D.~G.}\ \bibnamefont {Angelakis}},\ }\href {\doibase
  10.1088/1367-2630/aadcbd} {\bibfield  {journal} {\bibinfo  {journal} {New
  Journal of Physics}\ }\textbf {\bibinfo {volume} {20}},\ \bibinfo {pages}
  {093004} (\bibinfo {year} {2018})}\BibitemShut {NoStop}%
\bibitem [{\citenamefont {Scopa}\ \emph {et~al.}(2019)\citenamefont {Scopa},
  \citenamefont {Landi}, \citenamefont {Hammoumi},\ and\ \citenamefont
  {Karevski}}]{ScopaEtAl19}%
  \BibitemOpen
  \bibfield  {author} {\bibinfo {author} {\bibfnamefont {S.}~\bibnamefont
  {Scopa}}, \bibinfo {author} {\bibfnamefont {G.~T.}\ \bibnamefont {Landi}},
  \bibinfo {author} {\bibfnamefont {A.}~\bibnamefont {Hammoumi}}, \ and\
  \bibinfo {author} {\bibfnamefont {D.}~\bibnamefont {Karevski}},\ }\href
  {\doibase 10.1103/PhysRevA.99.022105} {\bibfield  {journal} {\bibinfo
  {journal} {Phys. Rev. A}\ }\textbf {\bibinfo {volume} {99}},\ \bibinfo
  {pages} {022105} (\bibinfo {year} {2019})}\BibitemShut {NoStop}%
\bibitem [{\citenamefont {Kohler}\ \emph {et~al.}(1997)\citenamefont {Kohler},
  \citenamefont {Dittrich}, \citenamefont {H{\"a}nggi},\ and\ \citenamefont
  {Dittrich}}]{KohlerEtAl97}%
  \BibitemOpen
  \bibfield  {author} {\bibinfo {author} {\bibfnamefont {S.}~\bibnamefont
  {Kohler}}, \bibinfo {author} {\bibfnamefont {T.}~\bibnamefont {Dittrich}},
  \bibinfo {author} {\bibfnamefont {P.}~\bibnamefont {H{\"a}nggi}}, \ and\
  \bibinfo {author} {\bibfnamefont {T.}~\bibnamefont {Dittrich}},\ }\href@noop
  {} {\bibfield  {journal} {\bibinfo  {journal} {Phys.\ Rev.\ E}\ }\textbf
  {\bibinfo {volume} {55}},\ \bibinfo {pages} {300} (\bibinfo {year}
  {1997})}\BibitemShut {NoStop}%
\bibitem [{\citenamefont {Grifoni}\ and\ \citenamefont
  {H{\"a}nggi}(1998)}]{GrifoniHaenggi98}%
  \BibitemOpen
  \bibfield  {author} {\bibinfo {author} {\bibfnamefont {M.}~\bibnamefont
  {Grifoni}}\ and\ \bibinfo {author} {\bibfnamefont {P.}~\bibnamefont
  {H{\"a}nggi}},\ }\href@noop {} {\bibfield  {journal} {\bibinfo  {journal}
  {Phys.\ Rep.}\ }\textbf {\bibinfo {volume} {304}},\ \bibinfo {pages} {229}
  (\bibinfo {year} {1998})}\BibitemShut {NoStop}%
\bibitem [{\citenamefont {Cubitt}\ \emph {et~al.}(2012)\citenamefont {Cubitt},
  \citenamefont {Eisert},\ and\ \citenamefont {Wolf}}]{Cubitt2012}%
  \BibitemOpen
  \bibfield  {author} {\bibinfo {author} {\bibfnamefont {T.~S.}\ \bibnamefont
  {Cubitt}}, \bibinfo {author} {\bibfnamefont {J.}~\bibnamefont {Eisert}}, \
  and\ \bibinfo {author} {\bibfnamefont {M.~M.}\ \bibnamefont {Wolf}},\ }\href
  {\doibase 10.1007/s00220-011-1402-y} {\bibfield  {journal} {\bibinfo
  {journal} {Comm.\ Math.\ Phys.}\ }\textbf {\bibinfo {volume} {310}},\
  \bibinfo {pages} {383} (\bibinfo {year} {2012})}\BibitemShut {NoStop}%
\bibitem [{\citenamefont {Leskes}\ \emph {et~al.}(2010)\citenamefont {Leskes},
  \citenamefont {Madhu},\ and\ \citenamefont {Vega}}]{LeskesEtAl10}%
  \BibitemOpen
  \bibfield  {author} {\bibinfo {author} {\bibfnamefont {M.}~\bibnamefont
  {Leskes}}, \bibinfo {author} {\bibfnamefont {P.}~\bibnamefont {Madhu}}, \
  and\ \bibinfo {author} {\bibfnamefont {S.}~\bibnamefont {Vega}},\ }\href
  {\doibase 10.1016/j.pnmrs.2010.06.002} {\bibfield  {journal} {\bibinfo
  {journal} {Progress in Nuclear Magnetic Resonance Spectroscopy}\ }\textbf
  {\bibinfo {volume} {57}},\ \bibinfo {pages} {345} (\bibinfo {year}
  {2010})}\BibitemShut {NoStop}%
\bibitem [{\citenamefont {Lazarides}\ \emph {et~al.}(2014)\citenamefont
  {Lazarides}, \citenamefont {Das},\ and\ \citenamefont
  {Moessner}}]{LazaridesEtAl14b}%
  \BibitemOpen
  \bibfield  {author} {\bibinfo {author} {\bibfnamefont {A.}~\bibnamefont
  {Lazarides}}, \bibinfo {author} {\bibfnamefont {A.}~\bibnamefont {Das}}, \
  and\ \bibinfo {author} {\bibfnamefont {R.}~\bibnamefont {Moessner}},\ }\href
  {\doibase 10.1103/PhysRevE.90.012110} {\bibfield  {journal} {\bibinfo
  {journal} {Phys. Rev. E}\ }\textbf {\bibinfo {volume} {90}},\ \bibinfo
  {pages} {012110} (\bibinfo {year} {2014})}\BibitemShut {NoStop}%
\bibitem [{\citenamefont {D'Alessio}\ and\ \citenamefont
  {Rigol}(2014)}]{DAlessioRigol14}%
  \BibitemOpen
  \bibfield  {author} {\bibinfo {author} {\bibfnamefont {L.}~\bibnamefont
  {D'Alessio}}\ and\ \bibinfo {author} {\bibfnamefont {M.}~\bibnamefont
  {Rigol}},\ }\href {\doibase 10.1103/PhysRevX.4.041048} {\bibfield  {journal}
  {\bibinfo  {journal} {Phys. Rev. X}\ }\textbf {\bibinfo {volume} {4}},\
  \bibinfo {pages} {041048} (\bibinfo {year} {2014})}\BibitemShut {NoStop}%
\end{thebibliography}%
\end{document}